\documentclass[structabstract]{aa}
\usepackage{verbatim}
\usepackage{graphicx}
\usepackage{txfonts}
\usepackage{longtable}
\usepackage{lscape}
\newcommand{\GALEX}{$GALEX$\,}
\newcommand{\GOODS}{$GOODS$\,}
\newcommand{\HST}{$HST$\,}
\newcommand{\SDSS}{$SDSS$\,}
\newcommand{\ANG}{$ANG$\,}
\newcommand{\NGS}{$NGS$\,}
\newcommand{\mb}{$M_{b}$\,}
\newcommand{\mstar}{$M_{\star}$\,}
\newcommand{\msun}{$M_{\odot}$\,}
\newcommand{\bb}{$B_{435}$\,}
\newcommand{\vv}{$V_{606}$\,}
\newcommand{\ii}{$i_{775}$\,}
\newcommand{\zz}{$z_{850}$\,}
\newcommand{\NUV}{$NUV$\,}
\newcommand{\NUVrest}{$NUV_{RF}$\,}
\newcommand{\FUV}{$FUV$\,}
\newcommand{\us}{$u'$\,}
\newcommand{\gs}{$g'$\,}
\newcommand{\rs}{$r'$\,}
\newcommand{\is}{$i'$\,}
\newcommand{\zs}{$z'$\,}
\newcommand{\B}{$B$\,}
\newcommand{\Brest}{$B_{RF}$\,}
\newcommand{\magarcsq}{$mag/arcsec^{2}$\,}
\newcommand{\FWHM}{$FWHM$\,}
\newcommand{\FWHMkpc}{$FWHM_{kpc}$\,}
\newcommand{\PSF}{$PSF$\,}
\newcommand{\mur}{$\mu-r$\,}
\newcommand{\Reff}{$R_{eff}$\,}
\newcommand{\Rtwenty}{$R_{20}$\,}
\newcommand{\Rfifty}{$R_{50}$\,}
\newcommand{\Reighty}{$R_{80}$\,}
\newcommand{\Rbr}{$R_{Br}$\,}
\newcommand{\C}{$C_{42}$\,}
\newcommand{\midz}{0.5$\leq z \leq$0.8\,}
\newcommand{\farz}{0.8$< z \leq$1.1\,}
\newcommand{\Rp}{$R_P$\,}
\newcommand{\Roskar}{Ro\v{s}kar\,}
\newcommand{\R}{$R$\,}
\newcommand{\ResFIFTY}{$\Gamma=$0.5 kpc, $FWHM_{kpc}=$1.5}

\newcommand{\medReffB}{$\tilde R_{eff}(B)$\,}

\newcommand{\Dmax}{$D_{max}$\,}

\begin{document}

   \title{Radial Distribution of Near-UV Flux in Disc Galaxies in the range 0$\lesssim$z$\lesssim$1}

   \author{Ruym\'an Azzollini,\inst{1}
          John E. Beckman\inst{1,2}
          \and
	  Ignacio Trujillo\inst{1,3}
	  }

   \institute{Instituto de Astrof\'isica de Canarias,\\
              C/V\'ia L\'actea s/n, E38205, La Laguna, S/C de Tenerife, Spain\\
              \email{ruyman@iac.es; jeb@iac.es; trujillo@iac.es}\\
         \and
             Consejo Superior de Investigaciones Cient\'ificas,\\
	     C/Serrano, 117. Madrid E-28006, Spain.\\
	 \and
	     Ram\'on y Cajal Fellow
             }

   \date{}

 
  \abstract
 {Abridged} 
   {The goal of this paper is to quantify the changes on the SF distribution within the disc galaxies in the last $\sim$8 Gyr. We use as a proxy for the SF radial profile the Near-UV surface brightness distributions, allowing suitably for extinction.}
   {We compare the effective radii (\Reff) and concentration of the flux distribution in the rest-frame Near-UV for a sample of 270 galaxies in the range $0<z<1$. This radial distribution is compared to that measured in the rest-frame \B-band, which traces older stellar populations. The analysis is performed using deep, high resolution, multi-band images from  \GALEX, \SDSS, and HST/ACS - \GOODS-South.}
   {The relation \Reff (\NUV)-\mstar suffers a moderate change between $z\sim$1 and $z\sim$0, at a fixed stellar mass of $10^{10}$ \msun galaxies increase their effective radii by a factor 1.18$\pm$0.06. The ratio \Reff(\NUV)/\Reff(\B) has increased by $\sim$10\% in the same period. Median profiles in \NUV show signs of truncation at \R$\sim$\Reff, and median colour profiles (\NUV-\B) show a minimum (a ``bluest'' point) also around \R$\sim$1-1.5\Reff. The surface brightness of the discs (at \Reff) has decreased by $\sim$80\% in \NUV, and by $\sim$60\% in \B since $z\sim$1. The distributions of \NUV flux are more compact at z$\sim$1 than nowadays, in terms of the fraction of flux enclosed in a specific radius (in kpc). The central brightness in \B and \NUV has increased with respect to the brightness level of the discs, both in \B and \NUV, indicating perhaps a continuous accumulation of the bulge structure since $z\sim$1, or that they have been left exposed because of the significant decline in the the star-forming activity in the discs.}
   {Our results indicate that the SF surface density has decreased dramatically in discs since $z\sim$1, and this decline has been more intense in the central parts ($\lesssim$\Reff) of the galaxies. In addition, our data suggest that the bulges/pseudo-bulges have grown in surface brightness with regard to the discs since z$\sim$1.}
\titlerunning{Radial Distribution of \NUV Flux in disc Galaxies}
\authorrunning{Azzollini, Beckman \& Trujillo}

   \maketitle
%


\section{Introduction}\label{Sec:introduction}

Observations in the last decade have shown that the global star formation rate (SFR) in galaxies has suffered a significant decrease, by roughly tenfold, since z$\sim$1 (e.g. Lilly et al. \cite{Lilly96}, Madau et al. \cite{Madau96}, \cite{Madau98}). Other observations have shown that this star-forming activity has gradually shifted to less massive galaxies, a process commonly termed as ``downsizing'' (e.g. Cowie et al. \cite{Cowie96}; Brinchmann \& Ellis \cite{Brinchmann00}). This dramatic change in the pace at which star formation (SF hereafter) has taken place, and the shift in the masses of the hosts which harbour this activity are yet to be fully understood. Also important, and pending, is to find a consistent scheme in which this evidence is related to the evolution in the mass content (e.g. Rudnick et al. \cite{Rudnick03}, Bell et al. \cite{Bell07}, P\'erez-Gonz\'alez et al. \cite{PerezGonzalez08}), luminosities (e.g. Ilbert et al. \cite{Ilbert06}, Conselice et al. \cite{Conselice05}, Rudnick et al. \cite{Rudnick03}, Wolf et al. \cite{Wolf03}) and structure (morphologies) of galaxies (e.g. Conselice et al. \cite{Conselice08}, Coe et al. \cite{Coe06}, Cassata et al. \cite{Cassata05}, Conselice \cite{Conselice03}).

An area of study which has not yet received enough attention, and could provide clarifying evidence to solve the problems posed, is that of the spatial distribution of the SF within galaxies as a function of redshift. In particular, we refer to studies of the radial distribution of SF as a function of epoch. This kind of work adds constraints on the process of galaxy evolution, as knowing where stars are being born in galaxies during the history of the universe is essential to reconstruct their evolutionary history. This article is directed at this research field, surveying the radial distribution of Near-UV (\NUV) flux, used as a tracer of SF, in a sample of disc galaxies within the redshift range 0$\lesssim$$z$$\lesssim$1.

There are several pieces of evidence which suggest that undisturbed, i.e. non interacting, disc galaxies are the most promising subjects to study the change of the spatial distribution of SF with cosmic time. First of all, these are the most numerous galaxies in the Local Universe, accounting for $\sim$70\% of the total, and their preponderance in number density seems to hold, at least, up to $z\sim$1 (Conselice et al. \cite{Conselice05}). Moreover these galaxies harbour most of the unobscured SF at $z\lesssim$1 (Bell et al. \cite{Bell05}). It has been hypothesized that the decline in global SFR between $z\sim$1 and $z\sim$0 is due to a decreased rate of major mergers (e.g. Somerville et al. \cite{Somerville01}), but there is evidence to challenge this picture. For example, Bell et al. (\cite{Bell05}) found that more than half of the intensely star-forming galaxies at z$\sim$0.7 have spiral morphologies,  without major disturbances, whereas less than 30\% are strongly interacting. They proposed alternative explanations for the decrease in global SFR, involving gas depletion and weak interactions with small satellites. In a complementary line of research, Bell et al. \cite{Bell07} explored the relation between star formation and the acquisition of stellar mass for galaxies at $z\lesssim$1. They found that the increase in stellar mass content in this time interval is consistent with the integrated SFR. Most of these stars are formed in galaxies located in the ``blue cloud'' (typically late-type galaxies with young stellar populations), but there must be some mechanism of star formation quenching by which part of these galaxies pass to the ``red sequence'' (galaxies which are passively evolving, with little or no star-forming activity, though there are cases in which they harbour significant dust-obscured SF, and which usually have early morphological types) at some point in that period. Otherwise, there would be a dramatic overproduction of blue cloud galaxies in the local universe over the observed populations. For these reasons, it is clear that undisturbed disc galaxies are an interesting target for a study of the spatial distribution of SF.

An aspect of spiral galaxies which has received much attention in recent years is the formation and evolution of their stellar discs. Recent studies of galaxies at $z>$1 in the \HST-\emph{Ultra Deep Field} have shown that most late-type galaxies at those times present morphologies which are dominated by massive clumps of stars ($10^7\lesssim$\mstar$\lesssim$ $10^9$ \msun), which some authors argue can dissolve, by dynamical friction, into exponential discs in time scales of the order of 1 Gyr (see e.g. Bournaud et al. 2007; Bournaud et al. 2008; Elmegreen et al. 2008a,b). Between $z=$1 and $z=$0 the stellar discs, which show more regular morphologies, have grown in effective radius, \Reff, measured in \B-band, so as to keep their surface mass densities roughly constant (Barden et al. \cite{Barden05}). This means, that there is no evolution in the \Reff-\mstar relation in that period, although each individual galaxy must evolve along this relation. In addition, from the study of stellar disc truncations in this range of redshifts, it seems that there is a moderate growth, by a 25-30\%, in the radii of these truncations at a given stellar mass of the galaxies (Trujillo \& Pohlen \cite{TP05}, Azzollini et al. \cite{ATB08b}). Moreover, there is growing evidence that supports the idea that this growth, at least in relatively recent times, has been inside-out. Observing how this apparent growth of the stellar discs actually takes place, by building a history of the spatial distribution of SF with epoch is vital to reach a sound knowledge of the underlying processes.

The study of the spatial distribution of SF has been so far restricted to galaxies in the nearby universe. The early work of Paul Hodge and collaborators was pioneering in this field (e.g. Hodge \cite{Hodge69a}, \cite{Hodge69b}; Hodge \& Kennicutt \cite{Hodge83}). In these early studies SF was traced by the associated H$\alpha$ emission (in HII regions) using interference filters, a technique widely used with this aim in the local universe since then. A considerable corpus of knowledge has been collected about the spatial distribution of SF in the local universe, characterizing its radial extent (e.g. Hodge \cite{Hodge69a}, Hodge \& Kennicutt \cite{Hodge83}), and studying how it relates to the morphological types of galaxies (e.g. Ho et al. \cite{Ho97b}, Shane \cite{Shane02}), the infall of new gas (e.g. Phookun et al. \cite{Phookun93}), spiral patterns (e.g. Roberts \cite{Roberts69}, Cepa \& Beckman \cite{Cepa90}, Seigar \& James \cite{Seigar02}), the influence of stellar bars (e.g. Phillips \cite{Phillips93}, Ho et al. \cite{Ho97b}), interactions with other galaxies (e.g. Kennicutt et al. \cite{Kennicutt87}, Moore et al. \cite{Moore96}), or with intra-group or intra-cluster media (e.g. Koopman \& Kenney \cite{Koopman04}).

In recent years there have also been some examples of extensive studies on the radial distribution of SF in disc galaxies, with samples of dozens and even hundreds of galaxies. In \emph{The H$\alpha$ Galaxy Survey}, James et al. (\cite{James04}) surveyed $\gtrsim$300 galaxies as imaged in H$\alpha$, and Shane (\cite{Shane02}), basing his results on these data, studied the radial distribution (through several measures of the ``concentration'') of SF in interacting galaxies, in clusters and the field, and also as a function of the Hubble type. Other studies have benefited from GALEX, such as Boissier et al. \cite{Boissier07}. They presented a study on the radial profiles in UV and IR for a sample of 43 nearby galaxies, and compared them to their distributions of gas (CO and HI), studying the radial variation of SFR and dust-absorption in the UV. In another interesting example, Mu\~noz-Mateos et al. \cite{MuñozMateos07} studied the radial profiles of specific SFR, sSFR, for a sample of 161 nearby spiral galaxies. They found a large dispersion in the slope of these profiles with a slightly positive mean value, which they interpreted as proof of a moderate inside-out disc formation. Also based on deep GALEX observations has been the recent discovery of extended UV discs (XUV discs, Thilker et al. \cite{Thilker05}, Gil de Paz \cite{GdP05}). This UV emission, which is due to SF, and found far beyond the extension of the optical disc, stands as a piece of evidence which is not easy to fit into the puzzle of disc formation.

In this work we present a study of the radial distribution of rest-frame Near-UV flux in a sample of 260 disc galaxies, extending over the redshift range 0$\lesssim$$z$$\lesssim$1. The Near-UV flux (1500$<$$\lambda$$<$2800\AA) is dominated by the contribution from young stars, with ages $\lesssim$ 100Myr (Kennicutt \cite{Kennicutt98}), and so the flux in these wavelengths could be taken, to a fair approximation, as a tracer of on-going SF. Nonetheless, dust attenuation is significant at these wavelengths, and so, the observed distribution of \NUV flux does in fact reflect the distribution of SF, as modified by extinction. In this work we explore the dust problem, although we are limited by the complexity of the issue, and lack of relevant data, particularly in the rest-frame infrared. We perform some simple tests, based on known radial profiles of dust attenuation, to quantify the impact of dust extinction on the retrieved results. Our aim is to probe the evolution in the spatial distribution of SF across stellar discs, and understand the results in the context of disc formation and evolution.

This paper is organized as follows. In section \ref{Sec:data} we describe our sample of galaxies, the data that we use and how they are handled for our study. In section \ref{Sec:analysis} we present results on the radial distribution of rest-frame Near-UV flux in these galaxies, and compare it to the distribution of flux in rest-frame \B-band. Also in this section we analyse our results, accounting for the observational effects. In section \ref{Sec:discussion} we discuss the results, relating them to the distribution of star formation, and put them in context, by relating them to other observational studies, and also to current models on disc formation. Finally, in section \ref{Sec:conclusions} we summarize our conclusions on the presented analysis.

Throughout this work, we assume a flat $\Lambda$-dominated cosmology ($\Omega_{M}$ $=$0.3, $\Omega_{\Lambda}$ $=$ 0.7, and $H_0$ $=$ 70 $km$ $s^{-1}$ $Mpc^{-1}$).

\section{Data \& Selection of Samples}\label{Sec:data}

The observational basis for this work is taken from several public databases which provide photometric and imaging data in rest-frame \NUV and other bands for large samples of galaxies in the local universe and at intermediate redshifts ($z<1$). Here we explain the criteria for selecting the objects under study in different redshift bins, and provide information on the resulting samples and the data we use for the analysis.

\subsection{Local Sample}

To build our Near UV (\NUV) sample in the local universe we have chosen as parent sample \emph{The Galex Ultraviolet Atlas of Nearby Galaxies} (\ANG hereafter, Gil de Paz et al. \cite{GdP07}). This is composed of 1034 nearby galaxies, and includes objects from the \GALEX \emph{Nearby Galaxies Survey} (\NGS; Gil de Paz et al. \cite{GdP04}; Bianchi et al. \cite{Bianchi2003a}, \cite{Bianchi2003b}), plus galaxies serendipitously found in the \NGS fields or in fields with similar or greater depth, obtained as part of other \GALEX imaging surveys, and which have optical diameters at the $\mu_B$ = 25 \magarcsq isophote larger than 1 arcminute.

To test for evolutionary changes, it is important to know to what extent this parent sample is representative of the galaxies found in the nearby universe. With this aim, we refer to a comparison presented in Gil de Paz et al. \cite{GdP07} of their sample with the \emph{Nearby Field Galaxy Survey} by Jansen et al. \cite{Jansen00} (NFGS). This survey (listing 193 galaxies) was designed to be representative of the nearby population of galaxies, as portrayed in the magnitude-limited sample of the \emph{CfA Survey} (Huchra et al. \cite{Huchra83}). The \GALEX-$ANG$ and the NFGS are sampling the same volume of the universe and represent a similar population of galaxies, according to their distributions in redshift and apparent magnitude. The distributions of morphological types are also similar in these samples. There is, though, a small difference in the luminosity distribution between samples. The most luminous spirals (-21$<$\mb$<$-20) are in moderate excess in the \ANG with respect to the $NFGS$, while the low-luminosity spirals (-19$<$\mb$<$-18) are slightly under-represented.

We define our sample in the local universe by applying the following criteria: a) it is restricted to disc galaxies; b) the objects must be of moderate inclination, to facilitate the extraction of useful radial brightness profiles, and minimize the complications in the analysis due to dust absorption; c) the objects must be bright enough for their counterparts at higher redshifts to be accessible for study with the available data; and d) we also want to have imaging data of the objects at longer wavelengths, within the optical range. With these criteria, we select the objects whose parameters are within these ranges: a) de Vaucouleurs morphological type, $T$, within 0$<$$T$$<$10 (Sa to Sm) ; b) axial-ratio $q$$>$0.5 (inclination $<60\deg$) ; c) absolute magnitude in the \B-band \mb$\leq$-19.5 mag (luminous galaxies above the completeness limit in the highest redshift bin, \farz) ; d) not classified as part of a ``multiple'' system in HyperLeda; and e) within the footprint of the Data Release 6 of the \emph{Sloan Digital Sky Survey}\footnote{http://www.sdss.org} (\SDSS, Adelman-McCarthy et al. \cite{Adelman08}). The morphological types, axial ratios and \mb are taken from the HyperLeda\footnote{http://leda.univ-lyon1.fr} database (Paturel et al. \cite{Paturel03}). 

\begin{figure}
\centering
\includegraphics[width=8cm]{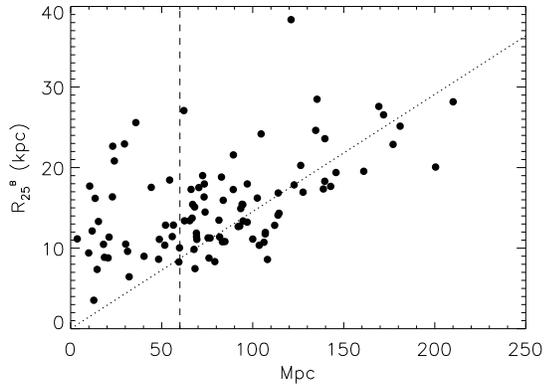}
\caption{Radii of the galaxies at the 25 \magarcsq isophote in the \B-band ($R_{25}$) against distance, for objects in the Local sample. The dotted line corresponds to the length subtended by half an arcmin as a function of distance, which is a constraint put on the selection of many of the objects in the \ANG survey. The vertical dashed line marks a distance of 60 Mpc, which is a further restriction we set on the sample to minimize the bias in apparent size inherited from the \ANG.}
\label{Fig:DisLoc}
\end{figure}

These criteria constrain our local sample to a population of 98 nearby, almost face-on disc galaxies, extending to $\sim$200 Mpc ($z\sim0.05$). In addition, we further restrict the sample to galaxies which lie at a maximum distance of 60 Mpc. That is to avoid a bias in the selection that favours the largest galaxies, due to the criteria applied to build the \ANG, our parent sample. In Fig. \ref{Fig:DisLoc} we present the distribution of physical sizes (kpc) for galaxies in the Local sample, as a function of distance (Mpc). These sizes are taken from HyperLeda, and correspond to the radius of the isophote at 25 \magarcsq in the \B-band. The dotted line marks the projected size of half an arcmin at the given distance. As mentioned above, the \ANG contains objects serendipitously found in the \NGS and other \GALEX images which have a diameter (at the 25 \magarcsq isophote) equal to or larger than 1 arcmin. Given that roughly 70\% of the galaxies in the parent sample were included in this way, this means a selection effect by which there is a lower envelope in the distribution of sizes against distance, as observed in this figure (in Section \ref{Sec:Reff} we justify quantitatively the election of the imposed limit on distance, when the effective radii of the galaxies are discussed). After this restriction in distance ($D<$60 Mpc), our local sample is composed of 33 galaxies. 

From this set we have  discarded 2 more galaxies: NGC1042 because of sky gradients in the \SDSS-\gs image; and NGC2782 as suspect of undergoing a merger. This final sample of 31 objects will be referred to hereafter as the ``Local'' sample.

The dataset for exploring the radial distribution of flux in our Local sample is composed of images in 2 bands, which we list with their corresponding effective wavelengths ($\lambda_{eff}$) and full widths to half maximum of the transmission curves ($\Delta\lambda$): \GALEX-\NUV ($\lambda_{eff} = 2250 \AA$, $\Delta\lambda = 1000 \AA$) and \SDSS-\gs ($\lambda_{eff} =$ 4770\AA, $\Delta\lambda =$ 1370\AA). The \GALEX images have a PSF which is slightly broader for brighter sources, and variable throughout the field of view (FOV), which is circular and 1.2$\deg$ in diameter. For objects within the central 0.5$\deg$ of the \GALEX FOV, the full width at half maximum (\FWHM) of the Point Spread Function (PSF) is in the range 5.0''-5.5'' in the \NUV band. The projected pixel scale on the sky is 1.5'' per side. The \SDSS-\gs images have a median value of the \FWHM of 1.5'', and an angular scale of 0.396''/pixel.

The \ANG images have a characteristic exposure time of one orbit ($\sim$1700 sec.), while for the \SDSS images the exposure time is fixed at 54 sec. Together with other relevant parameters, in Table \ref{Tab:1} are listed 1$\sigma$ fluctuations in surface-brightness of the background in an aperture with an area of 1 $arcsec^2$, in all bands used in this work ($\sigma_{arcsec^2}$, 6th column; all magnitudes there and throughout this work are in the AB system). We also list the 1$\sigma$ fluctuations of the background in the typical area used to estimate the background, an elliptical annulus with semi-major axis equal to 1.5 times the isophotal radius of the object in the \Brest band, and width of 1 pixel ($\sigma_{back}$, 7th column). This isophotal level to which we refer is, of course, that used to detect the objects. As we can see in this table, there are substantial differences between the noise levels amongst bands. How we account for these differences in our analysis is explained in detail in section \ref{Sec:analysis}.

We could have benefited from other bands also available both at shorter and longer wavelengths to extend the analysis on the radial distribution of flux for galaxies in the Local sample. To extend to the ``blue'' side, which would be interesting, for example, to estimate dust absorptions in the UV, we could use the \GALEX-\FUV images ($\lambda_{eff} = 1520 \AA$), that are available for most of the galaxies in the \ANG surveys. On the ``red'' side, which could provide information on the distribution of yet older stars than those traced with rest-frame \B-band, \rs, \is and \zs \SDSS bands ($\lambda_{eff} \gtrsim 6000\AA$), are also available. The reason not to use them is that we are limited by the available bands at z$\sim$1 from GOODS, if we want to make a meaningful comparison of the radial distributions of flux of the objects. This is, we make this restriction in order to compare distributions of fluxes in similar rest-frame wavelengths at all redshifts. Nonetheless, we use all bands listed in Table \ref{Tab:1}, including all the \SDSS bands for the Local samples, to make estimates of \B-band luminosities and stellar masses of the galaxies, by means of SED modelling, as described in section \ref{Sec:analysis}.

\subsection{Intermediate Redshift Sample}

We survey galaxies which extend up to z$\sim$1, which corresponds to a look back time ($lbt$) of $\sim$8 Gyr, according to the most favoured $\Lambda$-CDM cosmological model. In order to have an imaging database which provides rest-frame \NUV coverage, and with usable spatial resolution and sensitivity levels, we chose the \HST-ACS imaging of the \GOODS-South\footnote{http://www.stsci.edu/ftp/science/goods/} field (Giavalisco et al. \cite{Giavalisco04}). This field, which has been intensively observed using several space and ground observatories (e.g. \HST, $Chandra$, $XMM-Newton$, $Spitzer$, VLT, etc.), subtends an area on the sky of roughly 170 square arcmin. The \HST-ACS imaging set consist of images in the $F435W$, $F606W$, $F775W$ and $F850LP$ \HST pass-bands, hereafter referred to as \bb, \vv, \ii and \zz, as is common in the literature, and for which total exposure times per pixel in the final images were of approximately 7200, 3040, 3040 and 6280 seconds, respectively. The images have an angular scale of 0.03''/pixel and the \FWHM of the PSF in \zz measures 0.09''.

We have benefited from the fact that the \GOODS-South field is one of the most thoroughly observed patches in the sky, and thus we rely amply on the work of other authors to define our sample at intermediate redshift and gather valuable data on our targets. First, the \GOODS-South field is contained in the \emph{Galaxy Evolution from Morphologies and SEDs imaging Survey} (GEMS ; Rix et al. \cite{Rix04}). In the redshift range 0.1$\leq$$z$$\leq$1, GEMS provides morphologies and structural parameters for nearly 10,000 galaxies, obtained using \HST images (Barden et al. \cite{Barden05}; McIntosh et al. \cite{McIntosh05}). Second, for many of these objects there exist photometric redshift estimates and Spectral Energy Distributions (SEDs) from COMBO-17 (\emph{Classifying Objects by Medium-Band Observations in 17 filters}; Wolf et al. \cite{Wolf01}, \cite{Wolf03}). The COMBO-17 team made this data publicly available through a catalogue with precise redshifts (with errors $\delta$$z$/(1+$z$)$\sim$0.02) for approximately 9000 galaxies down to $m_R$$<$24 mag (Wolf et al. \cite{Wolf04}), which we use. In the same data release were included rest-frame absolute magnitudes and colours (accurate to $\sim$ 0.1 mag). Finally, we have also used the stellar mass estimates published in Barden et al. \cite{Barden05}, which are taken from Borch (\cite{Borch04}), and are deduced from the COMBO-17 photometric data.

Barden et al. (2005) conducted a morphological analysis of the galaxies in the GEMS field by fitting S\'ersic $r^{1/n}$ (S\'ersic \cite{Sersic68}) profiles to the surface brightness distributions. Ravindranath et al. (\cite{Ravindranath04}) showed that using the S\'ersic index $n$ as the criterion, it is feasible to distinguish between late and early-type galaxies at intermediate redshifts. Late-types (Sab-Sdm) are defined as having $n<$2-2.5. Moreover, the morphological analysis conducted by Barden et al. provides information about the inclination of the galaxies, through the axial ratio of the isophotes, $q$. This is a parameter to take into account, since we want to study the radial distribution of the rest-frame \NUV flux. This is particularly sensitive to dust absorption, and thus to the inclination of the discs, as already mentioned in the description of the selection of the Local sample.

Our intermediate redshift sample, to which we refer throughout this paper as the High-z sample, is constructed from the collection of data just described as follows. First, we selected objects from the Barden et al. sample within the following ranges of parameters: S\'ersic index $n<$2.5 to isolate disc-dominated galaxies (Barden et al. \cite{Barden05}; Shen et al. \cite{Shen03}; Ravindranath et al. \cite{Ravindranath04}); axial ratio $q>$0.5 to select objects with inclination $<$60$\deg$ ; and \mb$\leq$-19.5 mag, as for the Local sample. Moreover, only objects within 0.5$<$z$<$1.1 were selected in order to have spectral coverage in the rest-frame \NUV band with the available images from \HST-ACS. Then, the resulting sample was matched to a photometric catalogue extracted by ourselves from the \GOODS-South \HST-ACS data (\GOODS data hereafter), using SExtractor\footnote{This catalogue was obtained by detecting sources in the \zz band which had at least 16 contiguous pixels (0.014 arcsec$^{2}$) at an isophotal level of 0.6 $\sigma_{sky}$/pixel (25.4 \magarcsq) or higher.} (Bertin \& Arnouts \cite{BA96}). The resulting sample contains 265 objects. Also in this case it was necessary to discard a number of objects because of different problems that make them ill conditioned for our analysis: a) proximity to a bright source, such as an star or galaxy, b) image artefacts, a condition that is more relevant at the borders of the images, c) the object is apparently undergoing a major merger event. After these rejections, the High-z sample is composed of 239 disc galaxies.

Furthermore, we divide the High-z sample into two subsamples, according to redshift: a) ``mid-z'' subsample, with 134 objects within \midz, and b) ``far-z'' subsample, with 105 galaxies within \farz. In the first redshift range, the \GOODS band that best traces the rest-frame \B-band is \ii, while in the second range, it is \zz. We introduce this segregation to allow for a more accurate comparison of the radial distribution of \NUV flux relative to rest-frame \B-band flux, as the changes in observing band due to redshift are less significant in this way.

\begin{table*} 
\centering
\begin{tabular}{l c c c c c c c}
\hline\hline

Band & $\lambda_{eff}$ & $\Delta\lambda$ & exposure-time & scale &
$\sigma_{arcsec^{2}}$ & $\sigma_{back}$ & \FWHM \\
     & \AA & \AA & sec. & ''/pix & \magarcsq & \magarcsq & ''\\ 
\hline                    
   \NUV$^{(a)}$ & 2250 & 1000  & $\sim$1700  &  1.5  & 27.4 & 31.0 & 5\\
   \us          & 3540 &  570  & 54          & 0.396 & 24.8 & 27.7 & 1.6\\
   g'$^{(b)}$   & 4770 & 1370  & '' 	     &   ''  & 25.7 & 28.6 & 1.5\\
   r'           & 6230 & 1370  & ''	         &   ''  & 25.2 & 28.1 & 1.4\\
\hline                  
 \bb$^{(c)}$    & 4320 &  690  & 7200        &  0.03 & 29.8 & 29.1 & 0.09\\
 \vv            & 5920 & 1580  & 3040	     &   ''  & 29.7 & 29.0 & ''\\
 \ii$^{(d)}$    & 7690 & 1020  & 3040	     &   ''  & 29.0 & 28.3 & ''\\
 \zz$^{(e)}$    & 9050 & 1270  & 6280	     &   ''  & 28.6 & 27.9 & ''\\
\end{tabular}
\caption{Relevant Parameters of the Imaging Dataset. Columns: 1) band; 2) effective wavelength of the transmission curve; 3) full width at half maximum of the transmission curve; 4) typical exposure time; 5) projected size of the pixel in the sky; 6) 1 $\sigma$ fluctuations of the background in an area of 1 arcsec$^2$; 7) 1 $\sigma$ fluctuations of the background in the typical area used to estimate the background (an elliptical annulus with semi-major axis equal to 1.5$R_P$ (Petrosian radius in the \Brest band), and width of 1 pixel); 8) \FWHM of the \PSF.\newline Notes: a) \NUVrest at $z\sim$0; b) \Brest at $z\sim$0 ; c) \NUVrest within \midz (mid-z) and \farz (far-z); d) \Brest within \midz; e) \Brest within \farz.}\label{Tab:1}       
\end{table*}

\begin{figure}
\centering
\includegraphics[width=10cm]{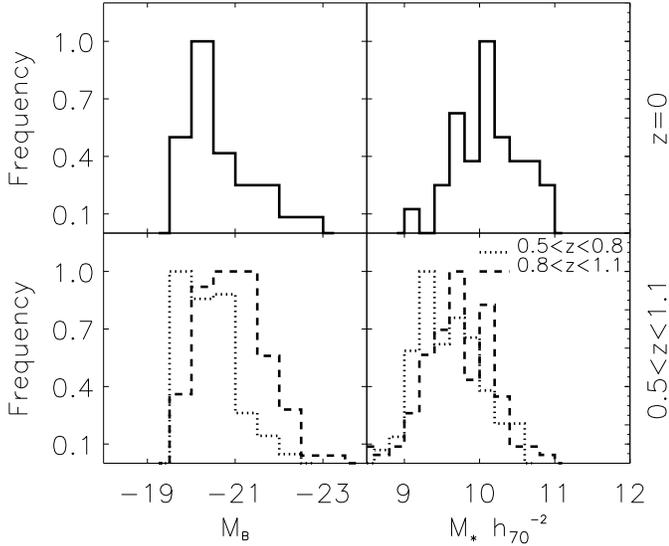}
\caption{In the upper panels are shown the distributions of Luminosities (\mb, left), and stellar masses (\mstar, right) for galaxies in the Local sample. In the lower row are the corresponding distributions for galaxies in the High-z sample: dotted line for the mid-z subsample, dashed for the far-z subsample, and continuous for the whole High-z sample.}
\label{Fig:HisLM}
\end{figure}

\begin{figure}
\centering
\includegraphics[width=4.2cm]{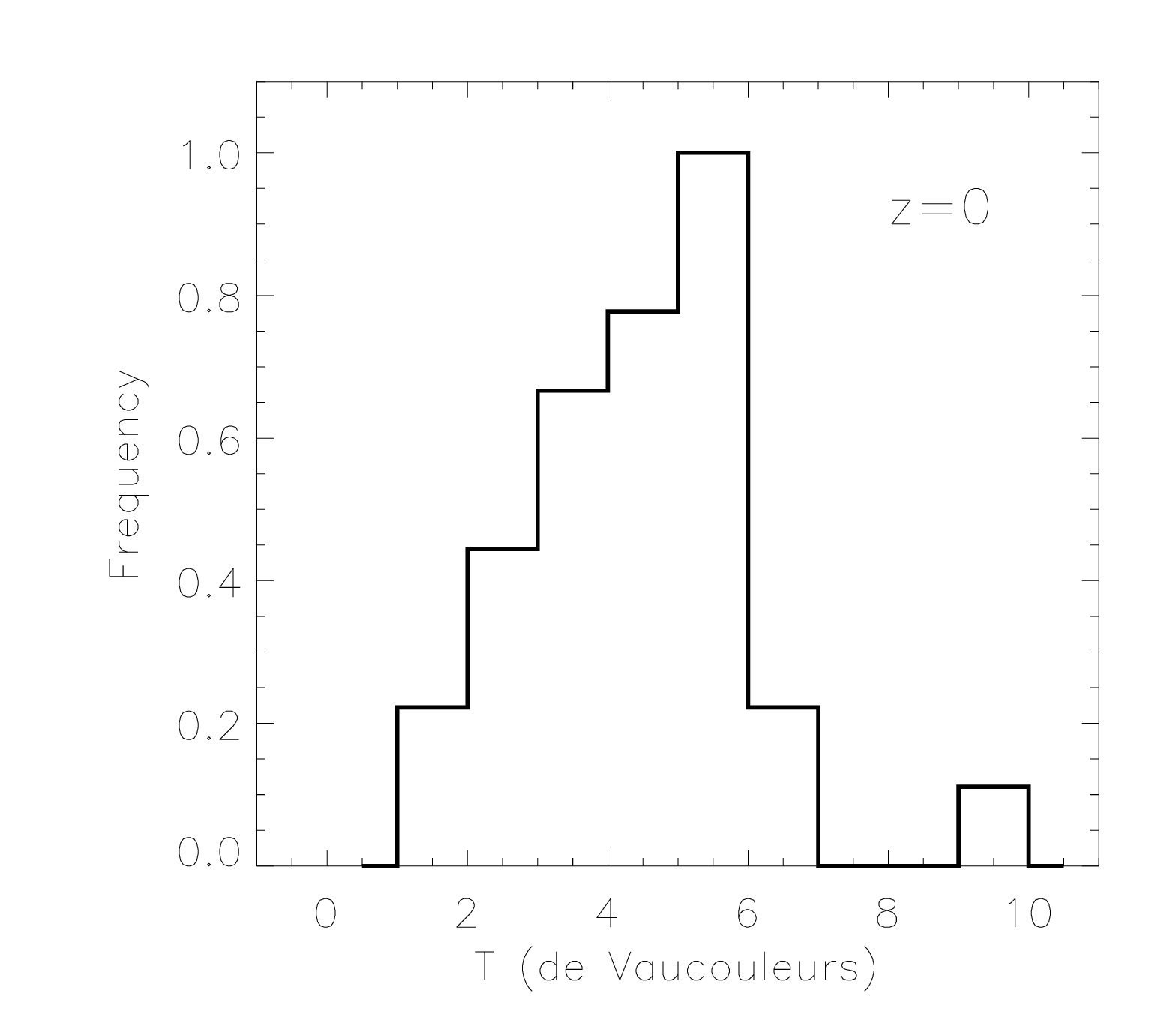}
\includegraphics[width=4.2cm]{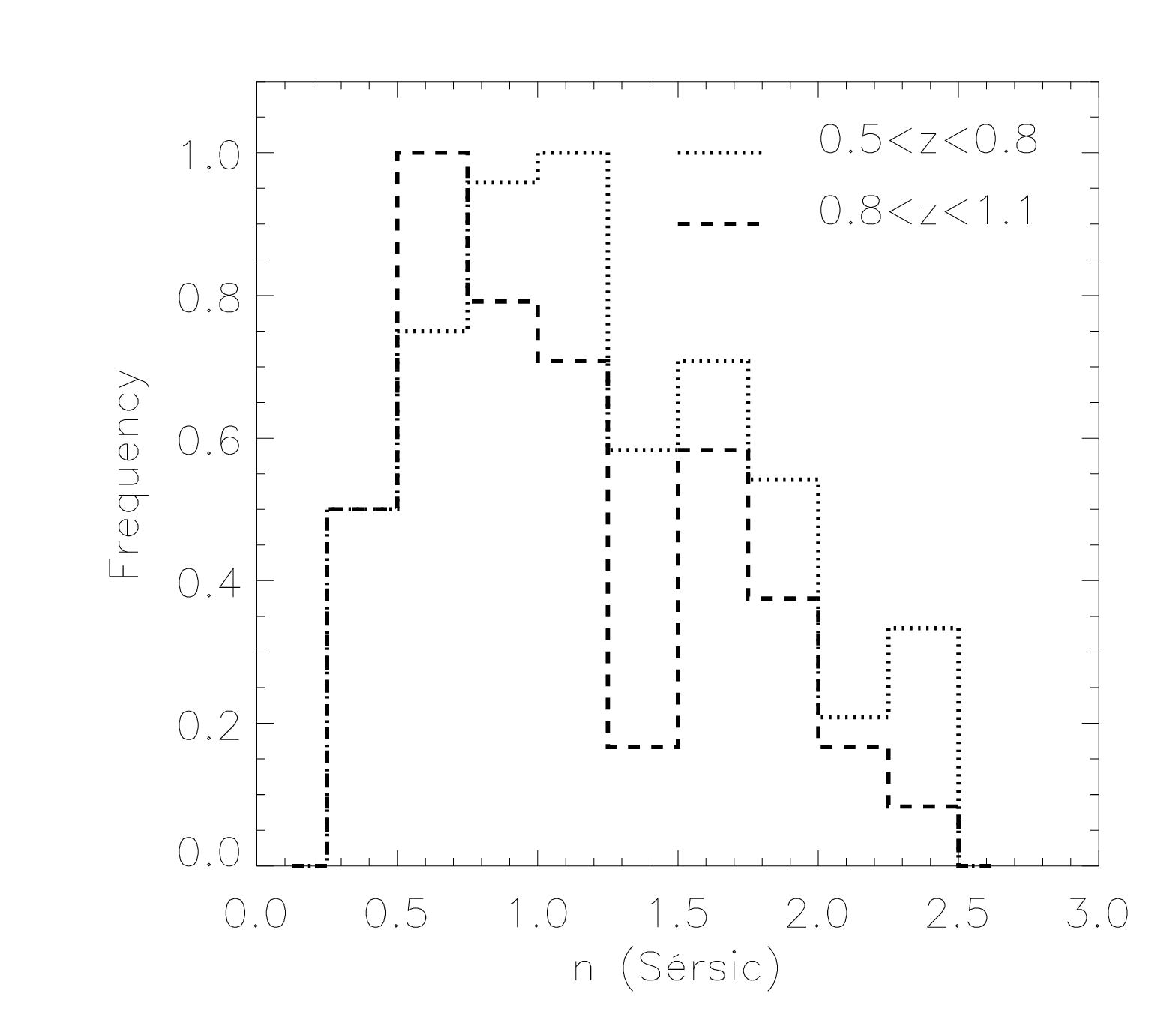}
\caption{Left: Morphological types (de Vaucouleurs) of galaxies in the Local sample. Right: S\'ersic index of the galaxies in the High-z sample (continuous line). Dotted line: mid-z subsample (\midz); dashed line: far-z subsample (\farz).}
\label{Fig:HisTS}
\end{figure}

In Fig. \ref{Fig:HisLM} we show the distributions of \B-band luminosities (\mb), and stellar masses (\mstar) for objects in the Local and High-z samples. The luminosities are quite similar amongst Local and High-z galaxies, with median values of -20.4 mag and -20.6 mag respectively. The distributions of masses are not so similar, with the Local sample extending somewhat to more massive galaxies. The Local sample has a median value of stellar mass, \mstar, of log(\mstar)$=$10.1, and the High-z sample of 9.6. These differences have relevant implications for the analysis strategy, as we are interested in comparing the radial distribution of \NUV flux amongst galaxies which are indeed comparable, so as not to confound selection effects with \emph{genuine} evolution. We explain this strategy extensively in section \ref{Sec:analysis}. The luminosities and stellar masses shown in this figure for both samples, and used throughout this work, are based on our own photometry of the objects, and obtained through model fitting of the SEDs with HyperZ, as described in section \ref{Sec:analysis}.

Another characteristic of the galaxies which is relevant to our study is the morphological type of the galaxies, as different trends have been reported in the spatial distribution of SF for different Hubble types of galaxies (e.g. Shane \cite{Shane02}). More importantly, one of the aims of this study is to provide new data to help understand how the location of newborn stars is related to the galaxy morphology. So, it is imperative to select galaxies with similar morphologies in the different redshift bins. In Fig. \ref{Fig:HisTS} we present the distributions of morphological types for objects in the Local sample (left; taken from HyperLeda) and of S\'ersic index for objects in the High-z sample (right; Barden et al. \cite{Barden05}). The morphological type is related to the S\'ersic indexes (Barden et al. \cite{Barden05}, Shen et al. \cite{Shen03}, Ravindranath et al. \cite{Ravindranath04}), and under this premise we can compare the morphologies of the two samples with the available data. Ravindranath et al. (\cite{Ravindranath04}) explored the relation between the S\'ersic index and morphological type of galaxies when the effect of redshift on their observation is simulated. They retrieved the S\'ersic indexes for local galaxies artificially redshifted to z$\sim$1 (as they would be observed in the GOODS-S field). In Fig. 1 in their paper it can be seen how galaxies with 0.5$<n<$2, the range in which most of the objects in our High-z sample are found, have preferably morphological types, T, around 5, fairly similar to the most frequent morphological type in the Local sample. Thus, the morphological ranges are quite similar amongst low and intermediate redshift samples, and the most frequent morphological types are in the range 2$\lesssim$T$\lesssim$6, i.e. Sab-Scd.

\subsection{Preparation of the Images for Analysis}

Due to the heterogeneity of our data, we need to perform several operations on the images before the analysis is done. Here we describe these operations, first for the Local sample and then for the High-z sample.

The \GALEX-\NUV images from the \ANG are already sky subtracted, calibrated and registered to a World Coordinate System (WCS), and the objects of interest are fully included, as it is an atlas centred on targets, not on coordinates. In contrast, the same galaxies in the SDSS are in many cases not in the field centre, and may even be split amongst two or more ``fields'' (using the \SDSS nomenclature for the released individual images), depending on size and position, and the images are not sky subtracted (though \SDSS provides estimates of the background values). We thus have to produce our own sky-subtracted mosaics of the galaxies from the \SDSS images. The background is estimated from the median of counts in regions where there are no detected objects, in an iterative procedure, for more accuracy. The mosaics produced are WCS-registered to the \GALEX images, but with the angular scale of the \SDSS, 0.396''/pixel. 

One problem that arises is that in these images there may be contributions by stars from our own galaxy, and also field galaxies, whose presence is spurious to our aims. We remove these intruding objects in two steps. In a first step, we create a mask of stars from the USNO-B star catalogue (Monet et al. \cite{Monet03}). On the position of each star, and in a circular area with diameter equal to a certain factor,$F_{\star}$, times the full width at half maximum (\FWHM) of the \PSF in the \gs band, we have flagged the pixels as ``Not a Number'', NaN, thus being discarded from the computations. The aforementioned factor was fixed at $F_{\star}=$2 as a best solution, after some trial and error. This mask is visually inspected and manually edited to ensure that in a sensitive region on and around the object of interest, all stars, and only the stars, are effectively being masked. This is necessary as the USNO-B catalogue is not always complete in accounting for all detected stars at the depth of the images, and besides, some bright blue regions in the galaxies (zones with recent or ongoing intense star formation) are in cases wrongly classified as stars in the catalogue, and we do not want to blank out these zones, given the nature of our study. In a second step, we manually create a second mask, to blank out those regions of the images (i.e. set them as NaN, thus being ignored altogether) which are occupied by extended field galaxies or bright stars (whose bright wings of the PSF are not effectively masked in the first step). This masking does not prevent some galaxies from having to be discarded as not suitable for the study, as we mentioned above, when the impact of intrusive objects is too severe, or there are other problems with the quality of the images.

With the High-z sample the actions previous to the analysis are somewhat less involved. In first place, the \GOODS images are already registered to a common WCS, photometrically calibrated and sky subtracted. So we start dividing the images into ``stamps'' with a typical size of 5 arcseconds on a side, to speed up the computational tasks. Also here too there are spurious objects in the neighbourhood of the galaxies under study. But in this case, given the high galactic latitude location of the field, and given that it subtends a relatively small area on the sky, there are few stars of our galaxy in the image stamps of the objects. The presence of field galaxies is more important, in number, because of the depth of the images. In this case these intrusive objects are masked by blanking out those areas of the images which are classified by SExtractor as belonging to any object other than the detected target. We perform a visual inspection of each stamp after this masking to ensure that it looks reasonable, and discard those objects for which it does not work properly, as is the case when there is contact (apparent or physical) between the intruding objects and the target. This ``SExtractor masking'' procedure is also performed for the local objects, but in many cases it turns out to be inefficient in eliminating the stars, especially when they are superposed on the object of interest. That is why for objects in the nearby sample we need to apply, in addition, the more involved procedure described above.

\subsection{Differences in relevant observational parameters amongst the imaging datasets}

Another problem that arises when one aims at comparing the radial distribution of luminous flux at a given wavelength of galaxies at different redshifts, and observed through different instruments, is that, in general, the images used for this task trace those distributions in different ways, which may bias the study. The most important parameters which may differ amongst the images are: a) angular scale ($\gamma$, given in arcseconds/pixel); b) physical scale, defined as the proper length that the angular scale subtends at the distance of the object ($\Gamma$, given in kpc/pixel) c) PSF properties (most importantly its \FWHM) ; d) photometric depth ; and e) relative band-shifting due to differences in redshift. Variations in these parameters amongst images makes the direct comparison of the flux distributions as registered in them misleading, or at least, more difficult to interpret. In this work we take into account these issues when interpreting our results, as we explain in this subsection.

\subsubsection{Resolution}

The objects in the Local sample extend in distance between 4 and 60 Mpc. With the \GALEX angular scale ($\gamma$$=$1.5''), this means a physical scale $\Gamma$ varying in the range 0.03$\leq\Gamma\leq$0.44 kpc/pixel, while with the \SDSS pixel (0.396'') the same parameter ranges between 0.008 and 0.12 kpc/pixel. On the other hand, the small angular scale of the \HST-ACS images ($\gamma$ = 0.03''/pix after ``drizzling'', i.e. with high spatial resolution), translates, within the redshift range 0.5$\leq$$z$$\leq$1.1, into a physical scale 0.18$\leq\Gamma\leq$0.25 kpc/pixel. So, the \HST images in fact provide a better resolution per pixel than \GALEX for the farthest objects in the respective samples, though for the nearest galaxies, the \SDSS and \GALEX images provide significantly higher resolutions (roughly by an order of magnitude).

With regard to the PSF of the images, we assume it to be Gaussian, in all cases to a good approximation, with the only parameter necessary to describe it being the \FWHM. This parameter also varies over a wide range amongst the datasets. For \GALEX-\NUV images the \FWHM$ is \sim$5'', and this is the value we adopt for these images. For the \SDSS images this parameter is, in general, different for every object, because the images were taken under varying seeing conditions. A guide value in the \gs band would be $\sim$1.5''. For this dataset we use, in each case, the value of the \FWHM in the \emph{Field} image where the centre of the object lies, in every band, as reported in the corresponding \emph{tsField} file (again using the \SDSS nomenclature). For the High-z sample, we assume a \FWHM of 0.09'' for all bands, as the minor differences amongst them are not significant to our study.

As with the size of the pixel, there are also significant variations in the projected dimension of the \FWHM, measured in kpc, at the distance of the galaxies. For example, for an object at 60 Mpc the \GALEX \FWHM of 5'' subtends $\sim$1.45 kpc, while the 0.09'' \FWHM of the \GOODS images translates into a better resolution of $\Gamma \sim$0.7 kpc at $z$=1.

\begin{figure*}
\centering
\includegraphics[width=5.5cm]{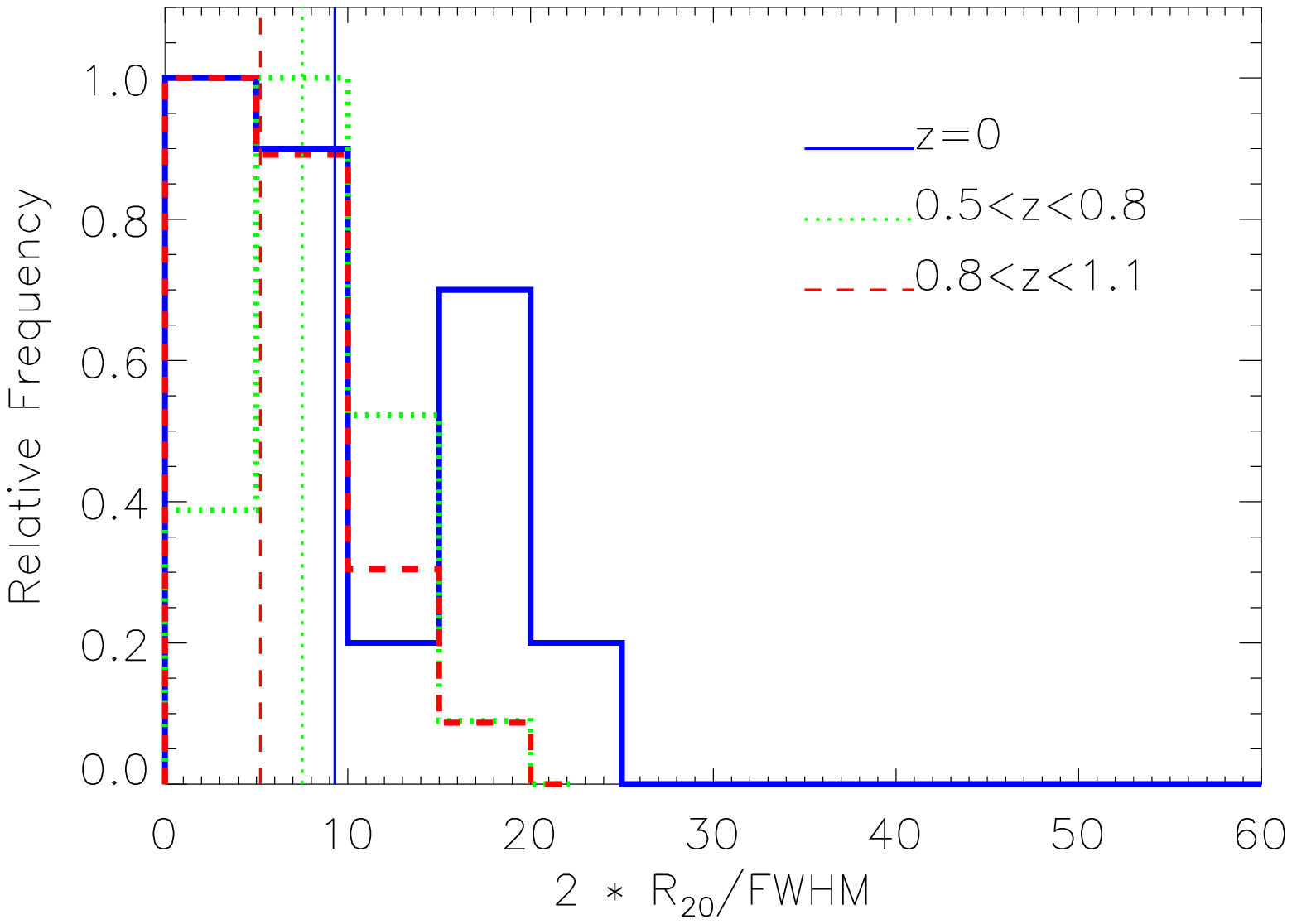}
\includegraphics[width=5.5cm]{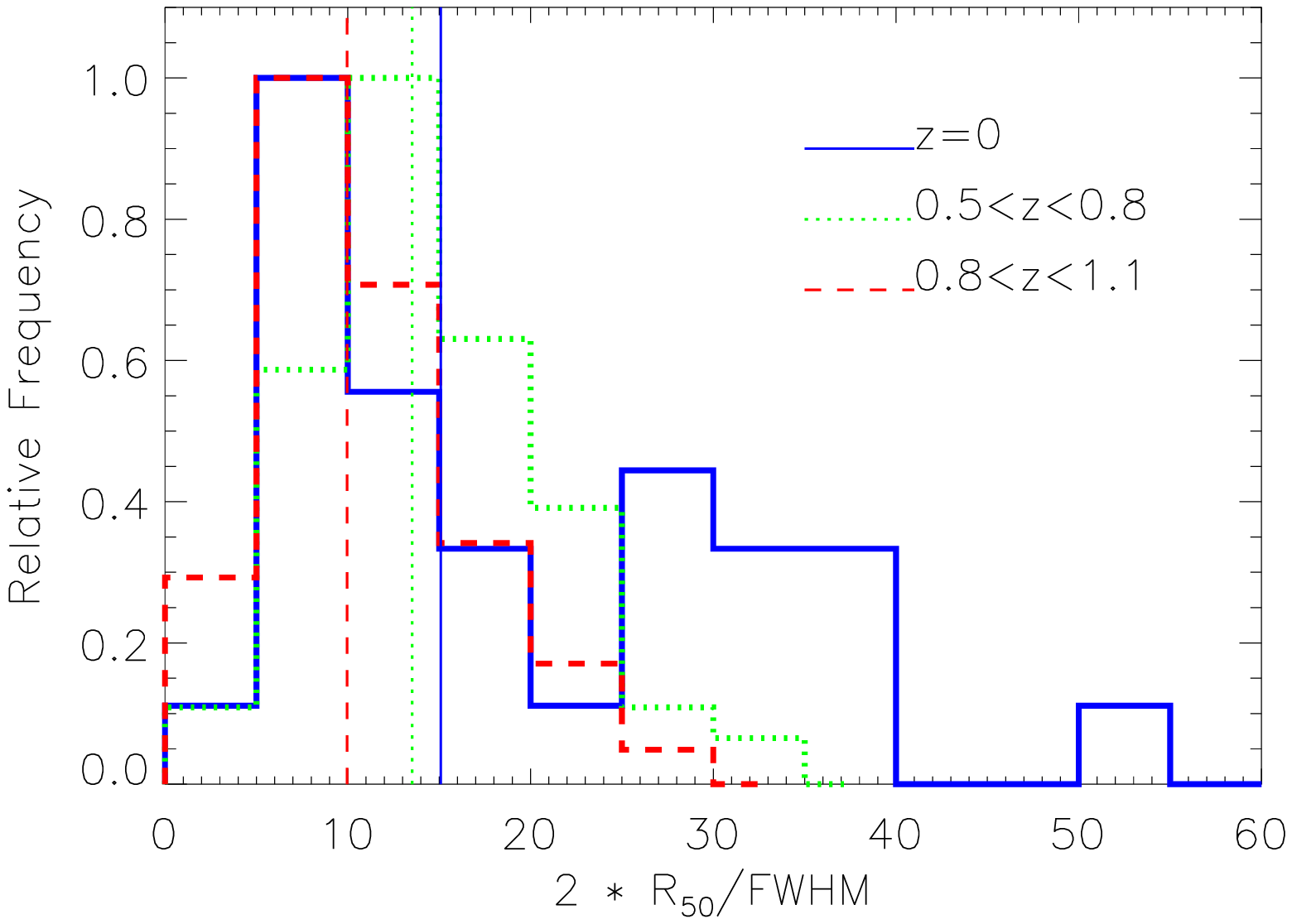}
\includegraphics[width=5.5cm]{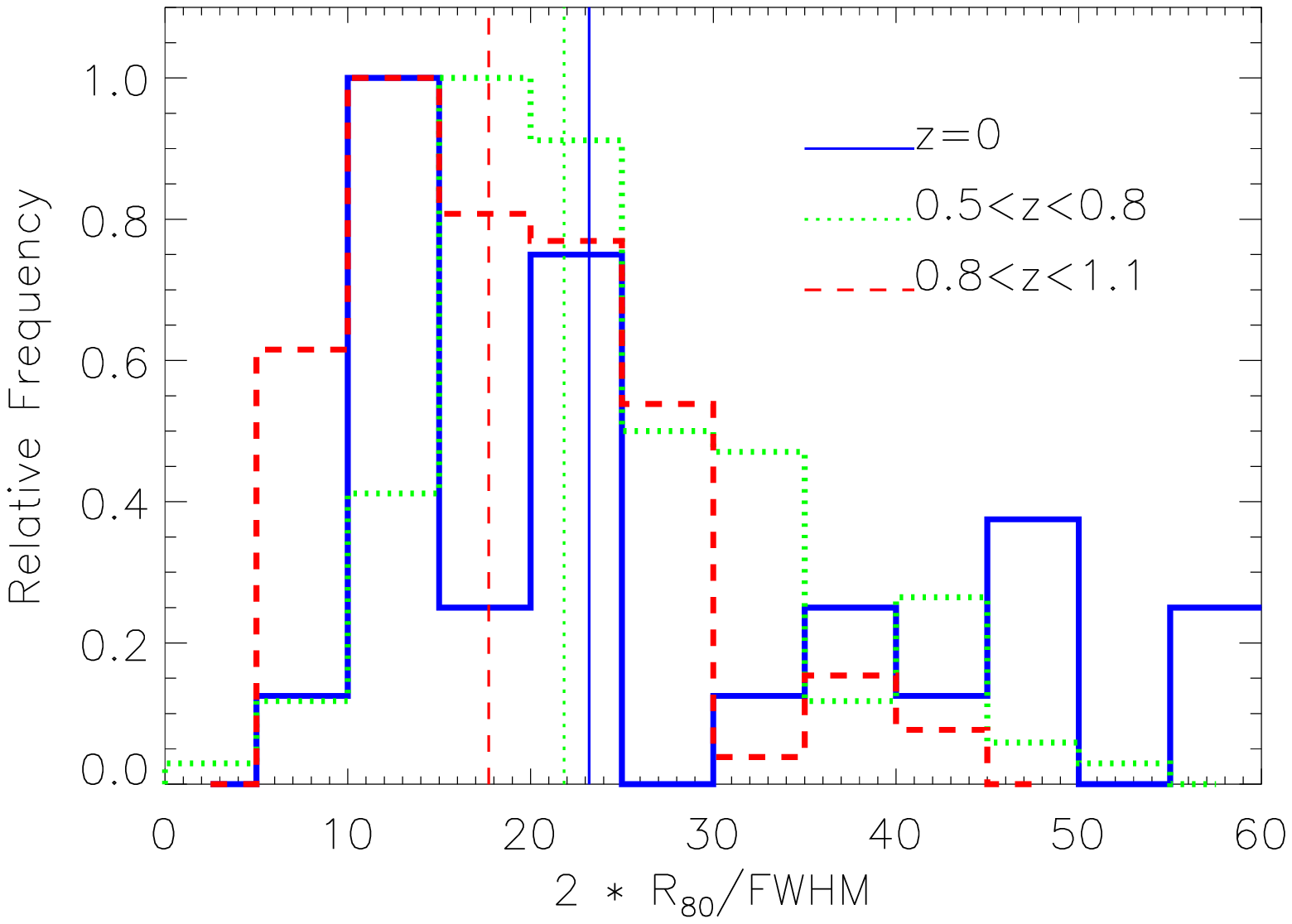}
\caption{Quantifying the resolution of our images: ratio of the diameter of the isophote which encloses a certain fraction of the total flux of the galaxy in \NUV (20\%, 50\% and 80\% from left to right) to the \FWHM of the \PSF, depending on the redshift range. The samples and subsamples are coded as: blue-continuous for the Local Sample, green-dotted for the mid-z subsample (\midz) and red-dashed for the far-z subsample (\farz). For objects of the mid-z and far-z subsamples, the FWHM is that in the \bb band. The vertical lines mark the median value of each distribution, and this applies to all other histograms shown in this article.}
\label{Fig:R2FWHM}
\end{figure*}

In this paper we present an analysis of effective radii (i.e. the major axis of the -assumed elliptical- isophote which encloses a 50\% of the total flux, \Rfifty), and of ``Concentration'', in whose expression appear \Rtwenty and \Reighty, which enclose 20\% and 80\% of the \NUV flux respectively. In order to show that we can measure the above quantities accurately, we compare these radii with the FWHMs. In Fig. \ref{Fig:R2FWHM} we present the ratio of these radii (referred to \NUV band) to the FWHM of the PSF in \NUV, for different redshift bins. We use this code to differentiate the samples: blue-continuous line for the Local sample ($z$$\sim$0), green-dotted for the mid-z subsample (\midz) and red-dashed for the far-z subsample (\farz), and this same code is used throughout this work in histograms of properties of these samples. In the left panel we see the aforementioned ratio for \Rtwenty, which would be the radius most affected by differences in spatial scale as the redshift changes. The worst cases are for galaxies in the mid-z and far-z subsamples, the minimum values of the ratio 2$\cdot$\Rtwenty/FWHM being 0.9 and 1.7 (for two objects at $z=$0.57 and $z=$1.07 respectively). The median values are between 5 and 9 (i.e. large enough to not be affected by the PSF), with the objects in the High-z sample having comparatively the worst resolutions. In the other two panels the median values of this ratio range between 10 and 15 for \Rfifty and between 17 and 23 for \Reighty. So, the dimensions which are of interest to us, these radii, are sufficiently larger than the FWHMs, though there is a non negligible difference between the effective resolution for the Local and High-z samples.

The figures just reported could raise doubts about whether our analysis is somehow biased due to differences in spatial resolution with redshift and datasets. To double check this issue we also study what happens when the images of the objects are degraded to the same spatial resolution. This common resolution is given by the lengths, in kpc, that the pixel and FWHM of the \GALEX images subtend at 60 Mpc, the maximum distance for galaxies in the Local sample. At that distance, the angular scale of 1.5''/pixel gives a physical scale of $\Gamma=$0.44 kpc, and the FWHM of 5'' subtends 1.45 kpc (we dub this magnitude \FWHMkpc). By means of interpolation and convolution, we produce images in rest-frame \B-band, \Brest, and rest-frame \NUV-band, \NUVrest, of all galaxies under study at that resolution (after rounding off to $\Gamma=$0.5 kpc, \FWHMkpc$=$1.5 kpc). How this is done is explained in the Appendix. In Fig. \ref{Fig:Proj} we can see examples of this transformation for two galaxies, at low redshift (Messier 95, D=12 Mpc ; above) and intermediate redshift (GEMSz033236.03m274423.8, z=0.95 ; below), presenting their ``original'' resolution (left), and degraded to the common resolution (\ResFIFTY ; right). Throughout this work we comment on how the use of images with a common resolution affects the reported results.

\subsubsection{Band-shifting}

Another key issue to take into account in our study is that as we observe objects in a range of redshifts, the observing bands trace different parts of the rest-frame SEDs of the objects. This must be accounted for if meaningful results on the distribution of the underlying stellar populations are to be obtained. We have decided on a strategy such that the selection of observing bands in which to extract the different measures depends on the redshift ``bin'' in which the object lies. This is a relevant issue for the High-z sample, but the observing bands in the Local sample must be taken into account as well to allow for comparison. As we describe further on, several parameters of the galaxies are measured in a band selected to trace the rest-frame \B-band, at all redshifts. Thus we refer to it as the \Brest band. In a broad sense, we use this band to characterize the distribution of ``older'' stars in the galaxies, to complement and contextualize the description of those newborn/very young stars, traced by the rest-frame \NUV images. For the Local sample, this \Brest band is the \gs band. In the High-z sample, it depends on the redshift bin. For the mid-z subsample (\midz) it is the \ii band,and for the far-z subsample (\farz) it is the \zz. In a similar way we also define a \NUVrest band to refer to the band that we use to trace the rest-frame \NUV flux in each redshift bin. For objects in the Local sample, this corresponds to the \GALEX-\NUV band itself, and for those in the High-z sample it is the ACS-\bb band.

\subsubsection{Photometric Depth}\label{Sec:PhotoDepth}

We also must discuss the differences in depth between the datasets, which are significant. To better understand these differences we will give some illustrative numbers. The 1$\sigma$/pix noise levels, and angular scales, $\gamma$, for the \GALEX, \SDSS and \GOODS data, in the \NUVrest band, are respectively 28 \magarcsq - 1.5'', 24 \magarcsq - 0.396'' and 25 \magarcsq - 0.03''. Then we assume characteristic distances for each sample: 30 Mpc for galaxies in the Local sample and $z=$0.8 for those in the High-z sample. If we compute how many pixels are needed on the given images, at the corresponding distances, to cover an area of 1 $kpc^2$ we get these figures: \GALEX, 21 pixels; \SDSS, 301; and \GOODS, 20. If we take the given noise levels, and take into account that when we average the intensities over N pixels the signal-to-noise ratio increases by the $\sqrt N$, then the corresponding 1$\sigma$/$kpc^2$ values would be 29.7 mag/$kpc^2$ (GALEX, Local), 27.1 mag/$kpc^2$ (SDSS, Local) and 26.6 mag/$kpc^2$ (GOODS, High-z). But at $z=$0.8 we have 2.6 magnitudes of cosmological dimming, and so, the actual value to consider for GOODS-High-z would be 24 mag/$kpc^2$. So, on equivalent areas, GALEX goes $\sim$10 times ``deeper'' than SDSS, and SDSS goes $\lesssim$20 times deeper than \GOODS (cosmological dimming included). In section \ref{Sec:discussion} we discuss how this may affect our results.

\begin{figure*}
\centering          
\includegraphics[width=3.5cm]{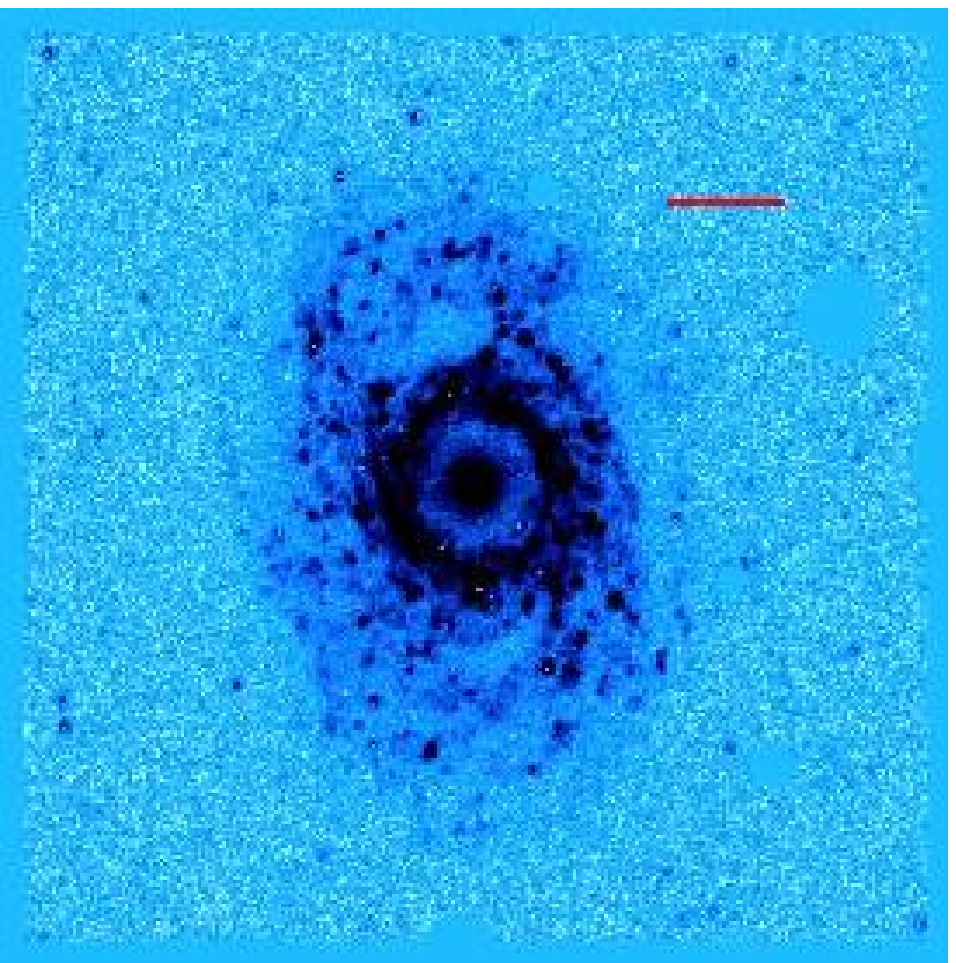} 
\includegraphics[width=3.5cm]{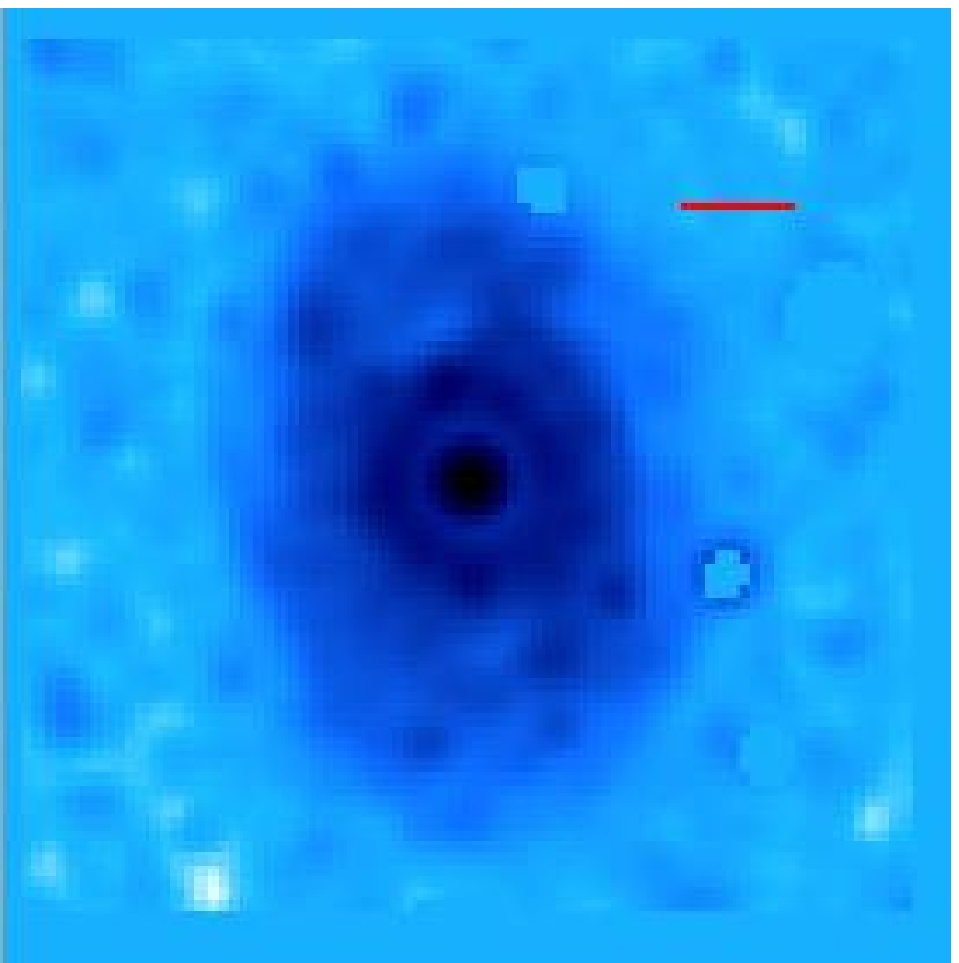}\\
\includegraphics[width=3.5cm]{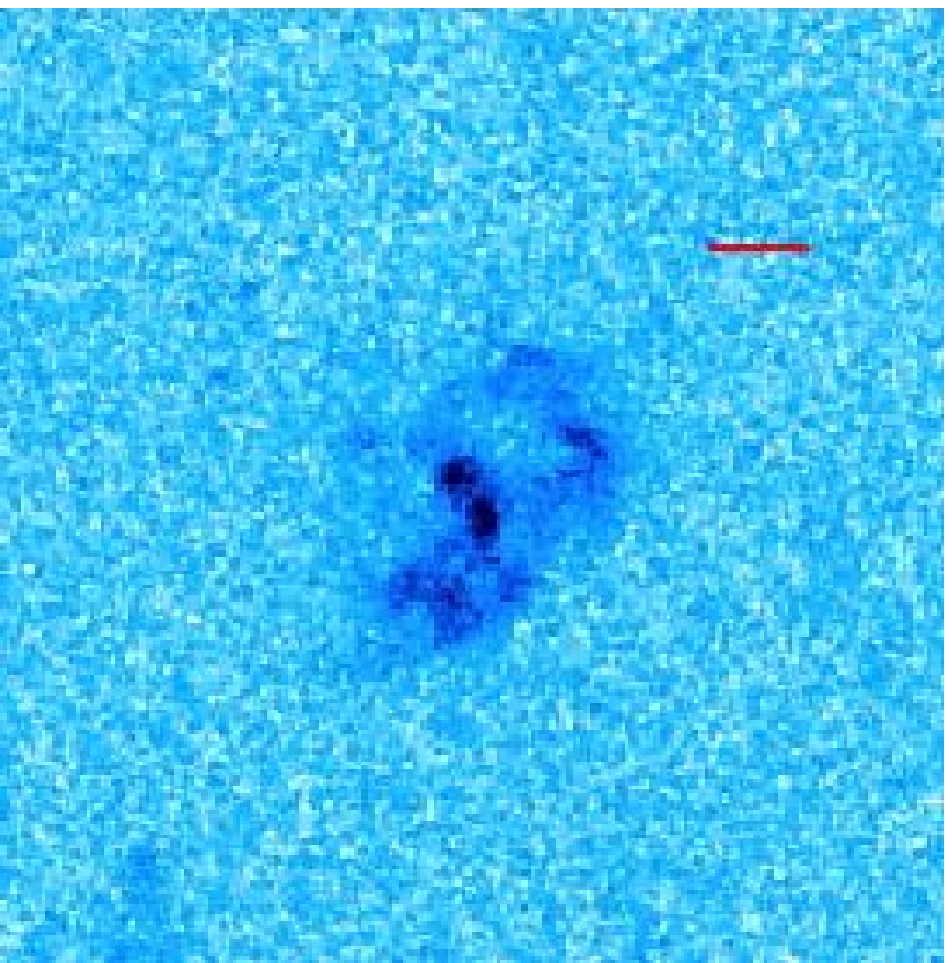} 
\includegraphics[width=3.5cm]{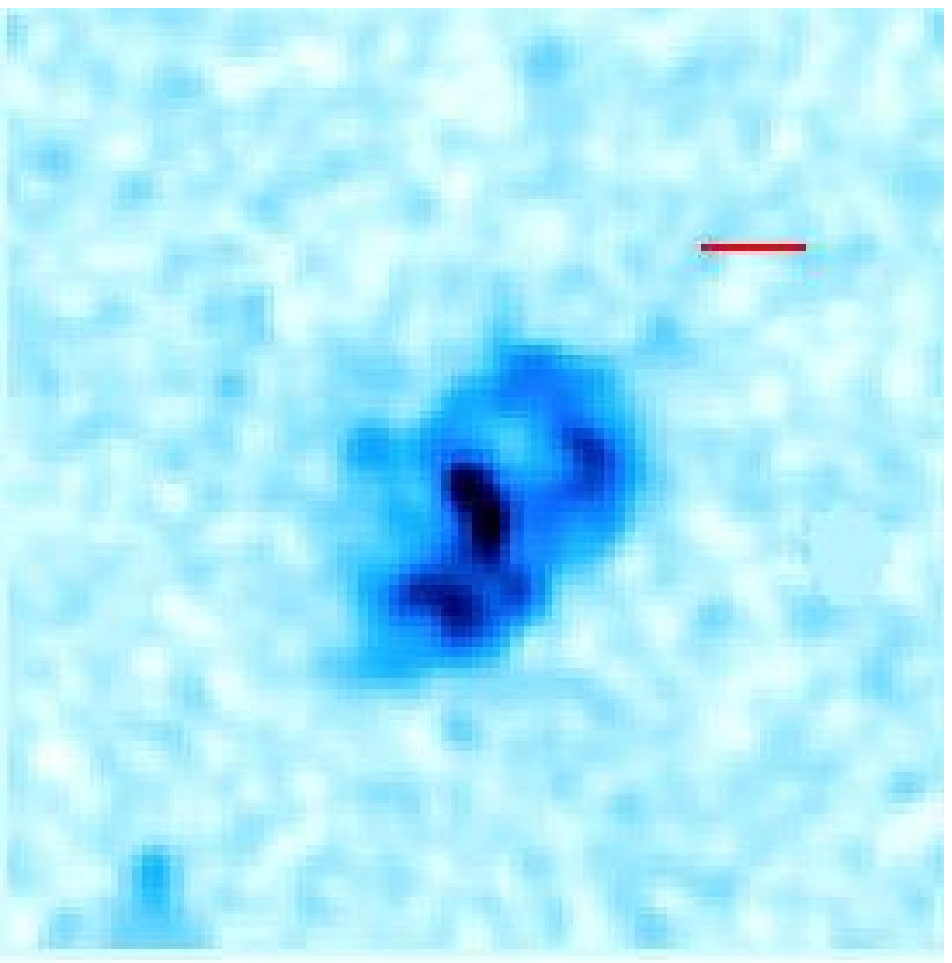} 
\caption{Examples of projections of images to a common resolution. The galaxy portrayed on the top line is Messier 95 ($D=$12 Mpc) in \GALEX-\NUV band, at its intrinsic resolution on the left ($\gamma=$1.5''/pixel), and with degraded resolution ($pixel=$0.5 kpc, $FWHM=$1.5 kpc), on the right. On the bottom line we show the analogous sequence for GEMSz033236.03m274423.8 ($z$=0.95), in band \GOODS-\bb. The red bar marks a length of 5kpc at the distance of the galaxies.}
\label{Fig:Proj}      
\end{figure*}

\section{Data Analysis \& Results}\label{Sec:analysis}

\subsection{Extraction of Radial Surface Brightness Profiles}

The main goal of this work is to trace the evolution of the radial distribution of SF in disc galaxies, using as proxy for this distribution that of the rest-frame \NUV flux. To accomplish this, we use the analysis of surface brightness (\mur) profiles in different bands. These \mur profiles were extracted using photometry on quasi-isophotal elliptical apertures. The intensities were estimated as the median of the flux in the area between elliptical apertures of increasing semi-major axis length (hereafter we term these lengths, loosely, as radii). The ellipticity and position angle of the apertures are fixed to those retrieved by SExtractor for the whole distribution of light of the object in the \Brest band. The pixels that go into this calculation are those in the \Brest isophotal area of the object (see below for an explanation of this). The centre of the apertures is also fixed. For the Local sample, the centre is that reported in $HyperLeda$ for the object, which after inspection seems adequate in all cases. For the objects in the High-z sample the process is more involved, because these objects show, in many cases, appreciable asymmetries. The procedure to pin-point the centre of the apertures is this: in a first iteration, the first moments (in ``x'' and ``y'') of the brightness distribution of flux of the object in the \zz band are employed. \zz is the reddest band available for the High-z sample, and thus traces best the distribution of old stellar populations, and thus, stellar mass. After visual inspection of the resulting profile and the image of the object, the centre was refined, if needed, with help of the task ``imexam'' from IRAF\footnote{http://iraf.noao.edu}, to match what was visually estimated, in each case, as the dynamical centre of the object. In a regular disc galaxy, as are those under study, this centre coincides with the central bulge or nucleus. 

The radius of the annular apertures was linearly increased in steps of 1 pixel (the corresponding angular scale varies depending on the dataset; see Table \ref{Tab:1} for reference), up to a radius which is 50\% larger than the radius of a circle with the isophotal area of the object in the \Brest band. The isophotal area is that covered by the set of connected pixels with intensities above the detection threshold (1$\sigma$/pixel) which constitute a detection (in our case: 24.7 \magarcsq in \gs, 25.2 \magarcsq in \ii, and 24.8 \magarcsq in \zz, for objects in the respective subsamples). For objects with a more or less regular morphology, as are those selected, the median intensities in these annulii are a good approximation to isophotal intensities. The error in the intensity, $\delta$I, is given by the $\sigma$ of the distribution of fluxes inside the annulus, divided by the square root of the corresponding pixel area. The intensities (``I'') are transformed to surface brightnesses ($\mu$), expressed in \magarcsq in the AB system, using the corresponding magnitude zero point and angular scale, $\gamma$, through the expression $\mu$ $=$ zero - 2.5$\cdot$log(I/$\gamma^{2}$). For the errors in $\mu$, $\delta\mu$, the formula used is $\delta\mu$ $=$ 2.5$\cdot$log( 1 + $\delta$I/I). We produce surface brightness profiles of the objects in all available bands, according to the scheme described. For this task we use several scripts written by us in Python language\footnote{http://www.python.org} which ``glue'' together different tasks of IRAF, SExtractor and DS9\footnote{http://heq-www.harvard.edu/RD/ds9/}.

The previously described intensity profiles are used in the analysis of stacked profiles (sect. \ref{Sec:Stacked}), and in the estimate of centre-to-border differences in surface brightness (sect. \ref{Sec:Concentration}). To measure effective radii, and other magnitudes which rely on the amount of flux contained in different radii (e.g. the concentration parameter, \C), we recur to the growth curves of the objects. These are computed by integrating the flux in the elliptical apertures previously described. The corresponding radii are the semi-major axes of the elliptical apertures.

\begin{figure*}
\centering          

\includegraphics[width=4cm]{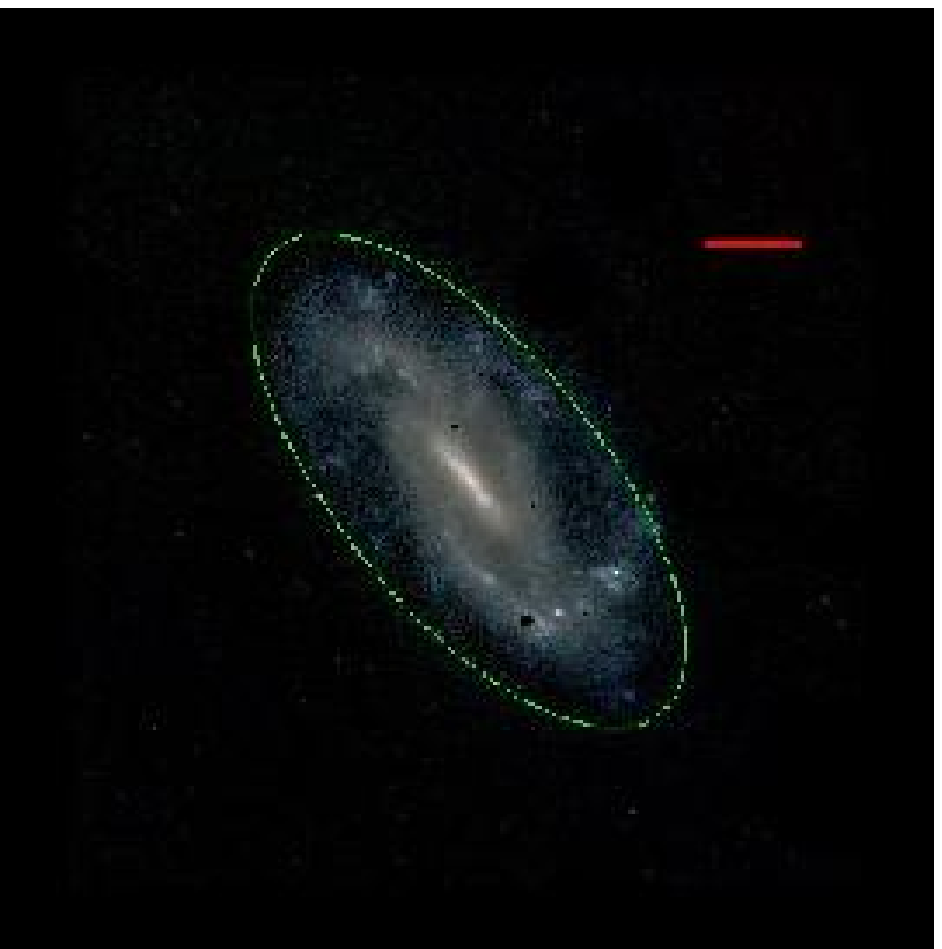} 
\includegraphics[width=4cm]{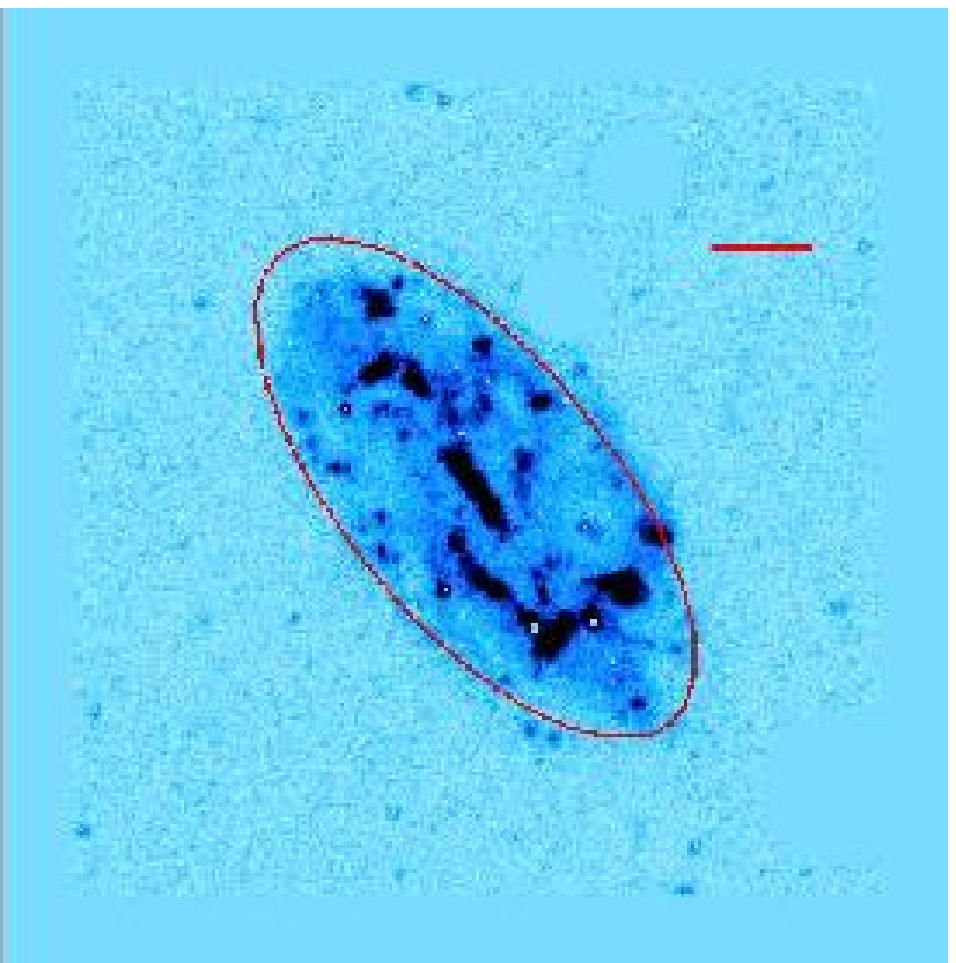} 
\includegraphics[width=6cm]{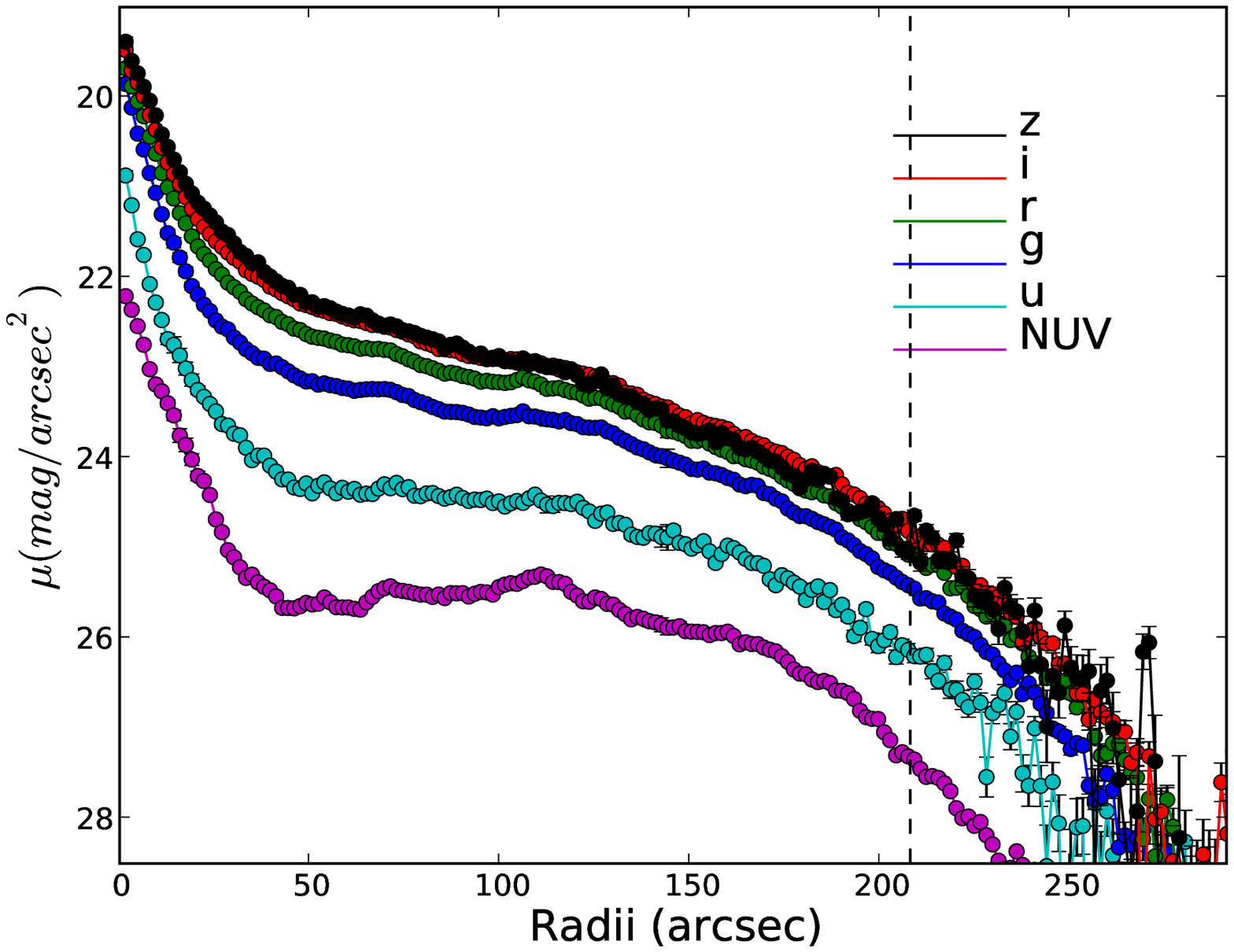} \\

\includegraphics[width=4cm]{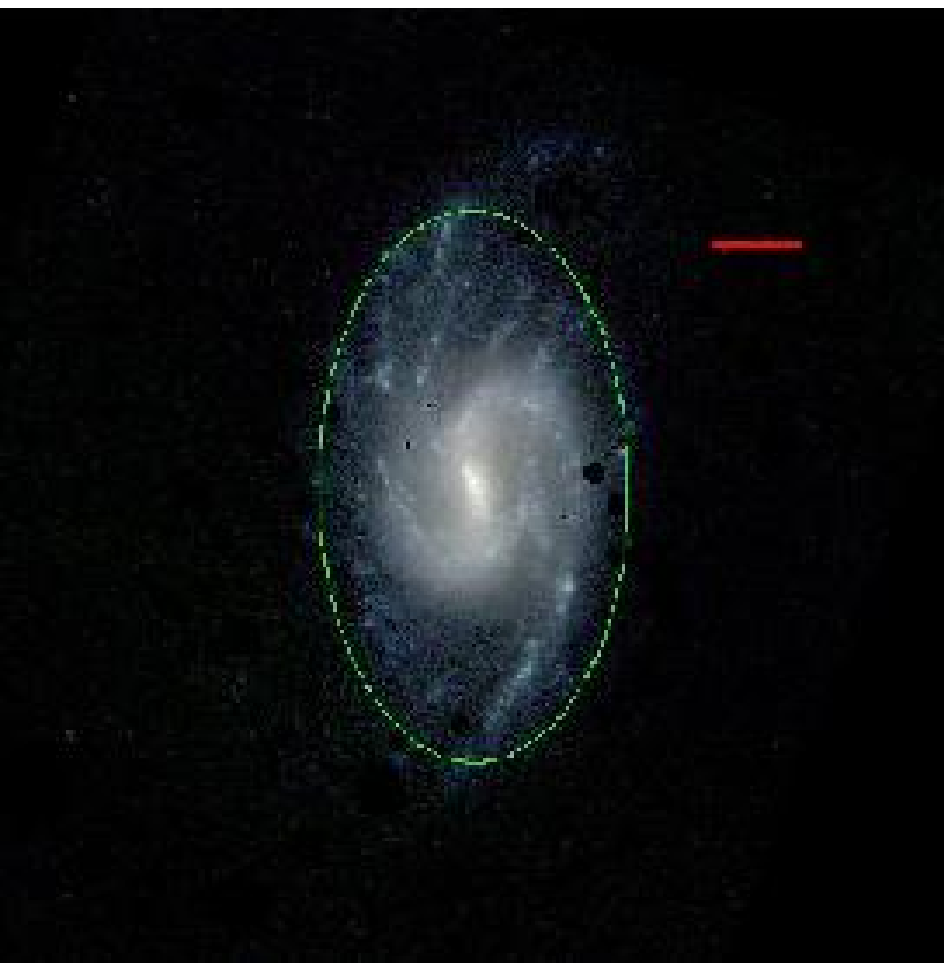} 
\includegraphics[width=4cm]{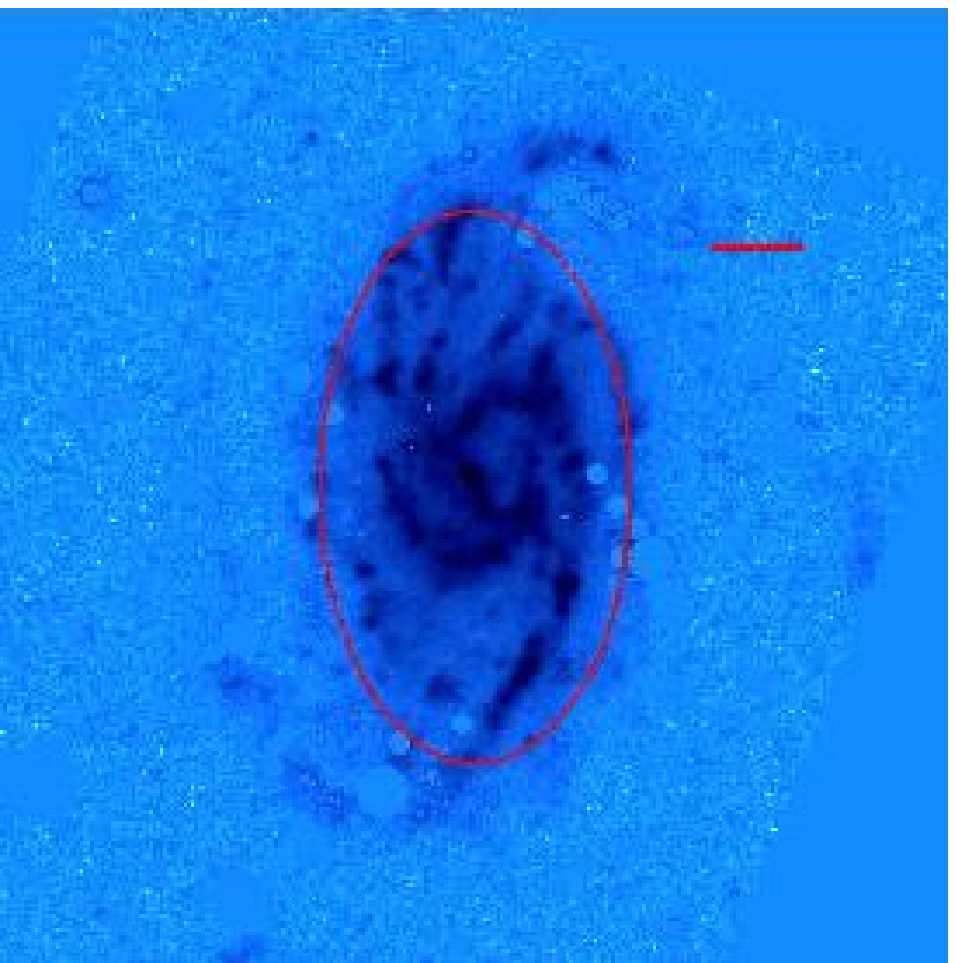} 
\includegraphics[width=6cm]{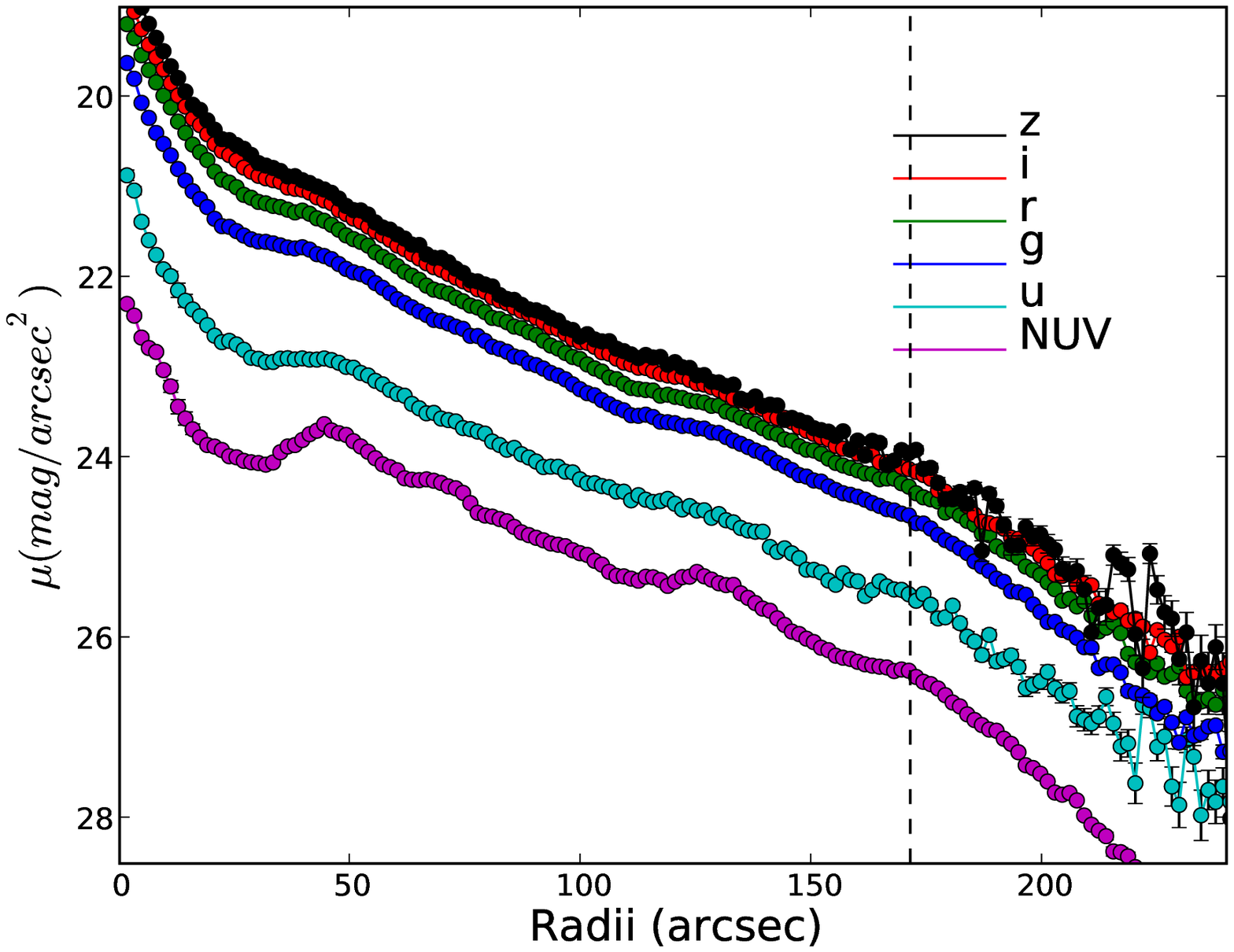}\\

\includegraphics[width=4cm]{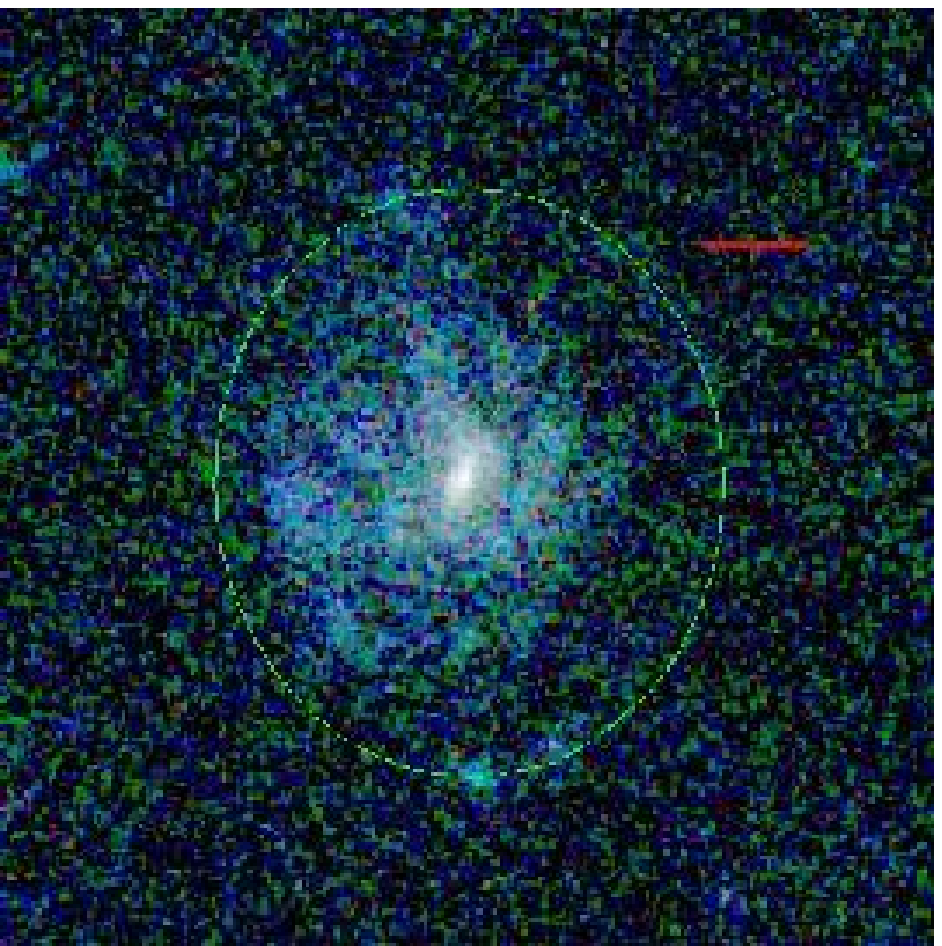} 
\includegraphics[width=4cm]{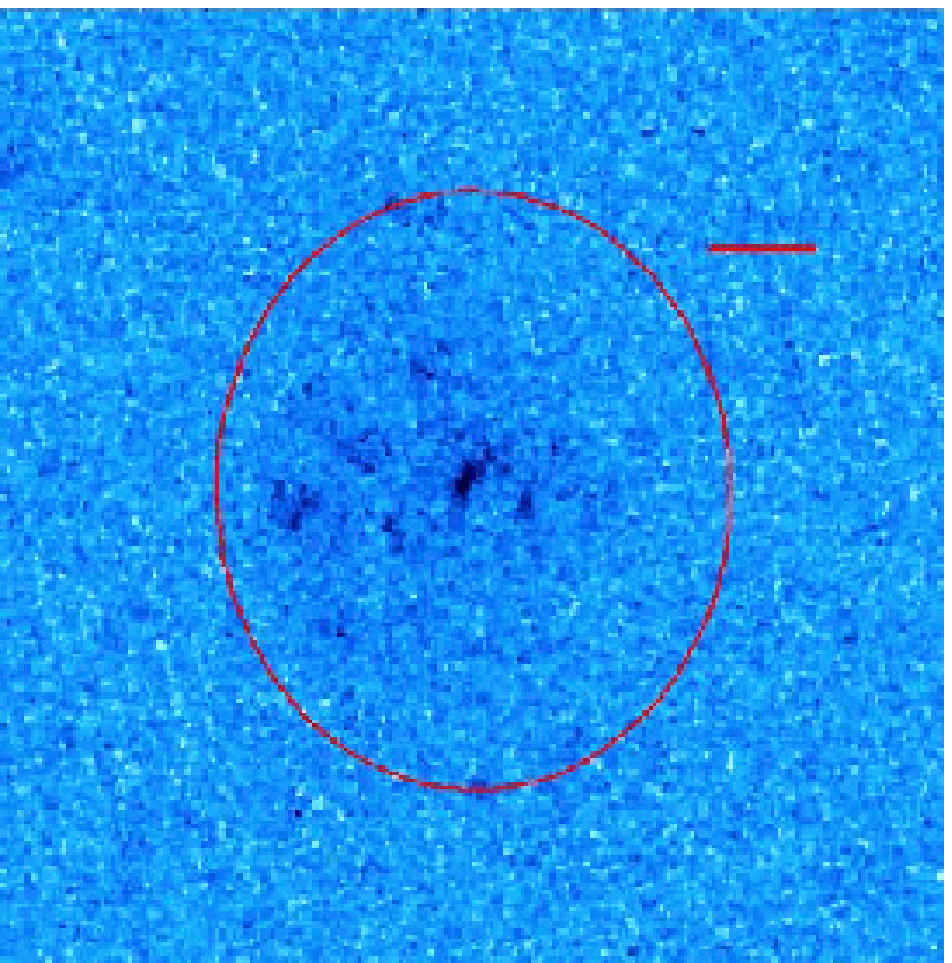}
\includegraphics[width=6cm]{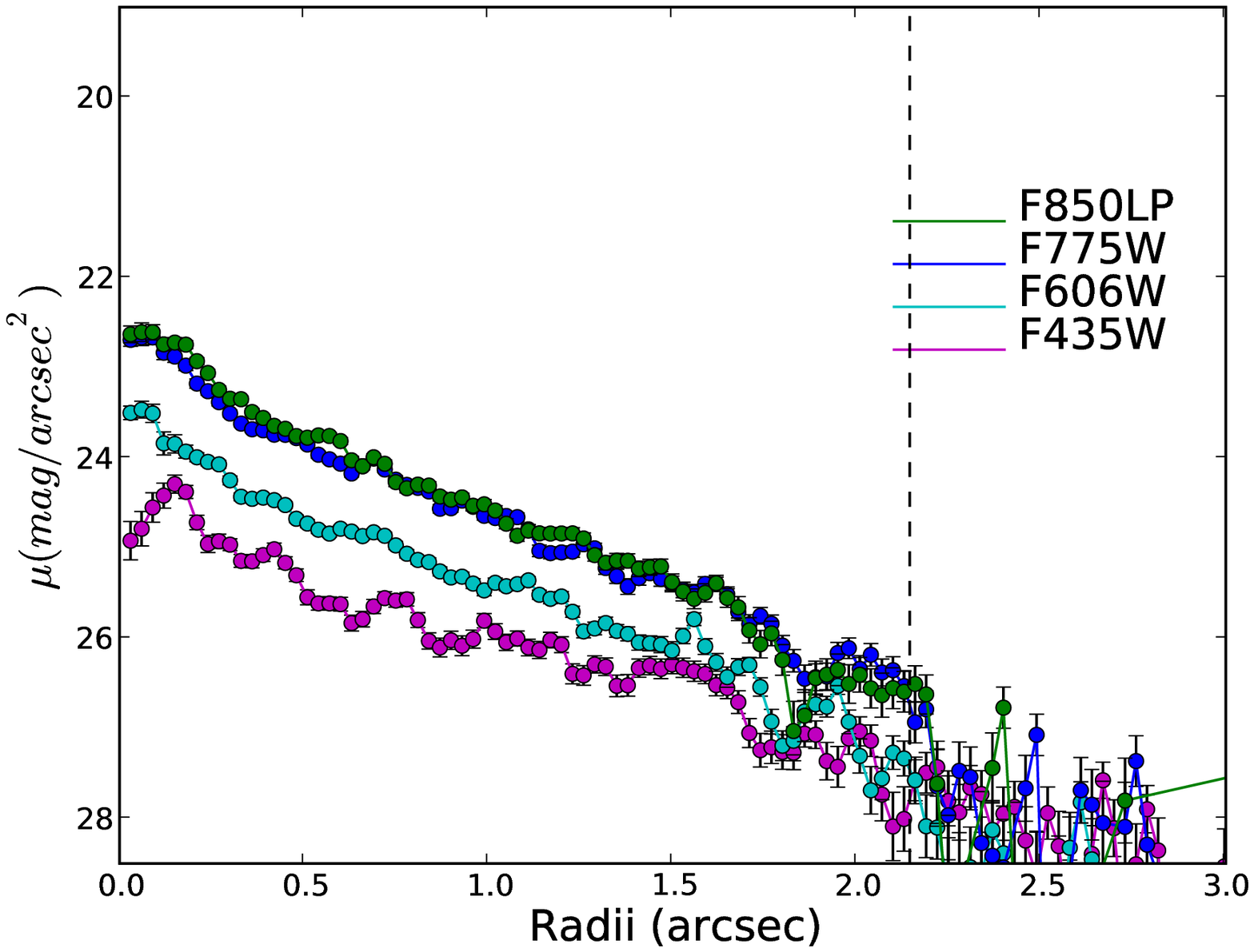}\\

\includegraphics[width=4cm]{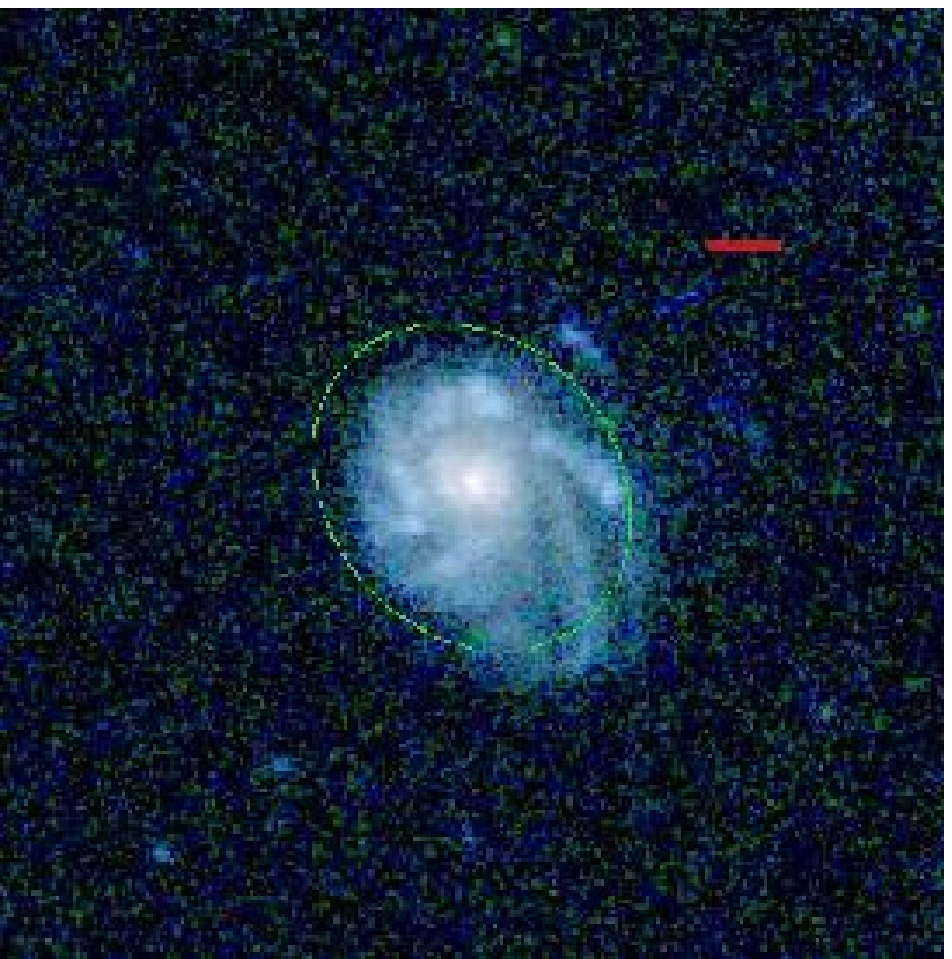} 
\includegraphics[width=4cm]{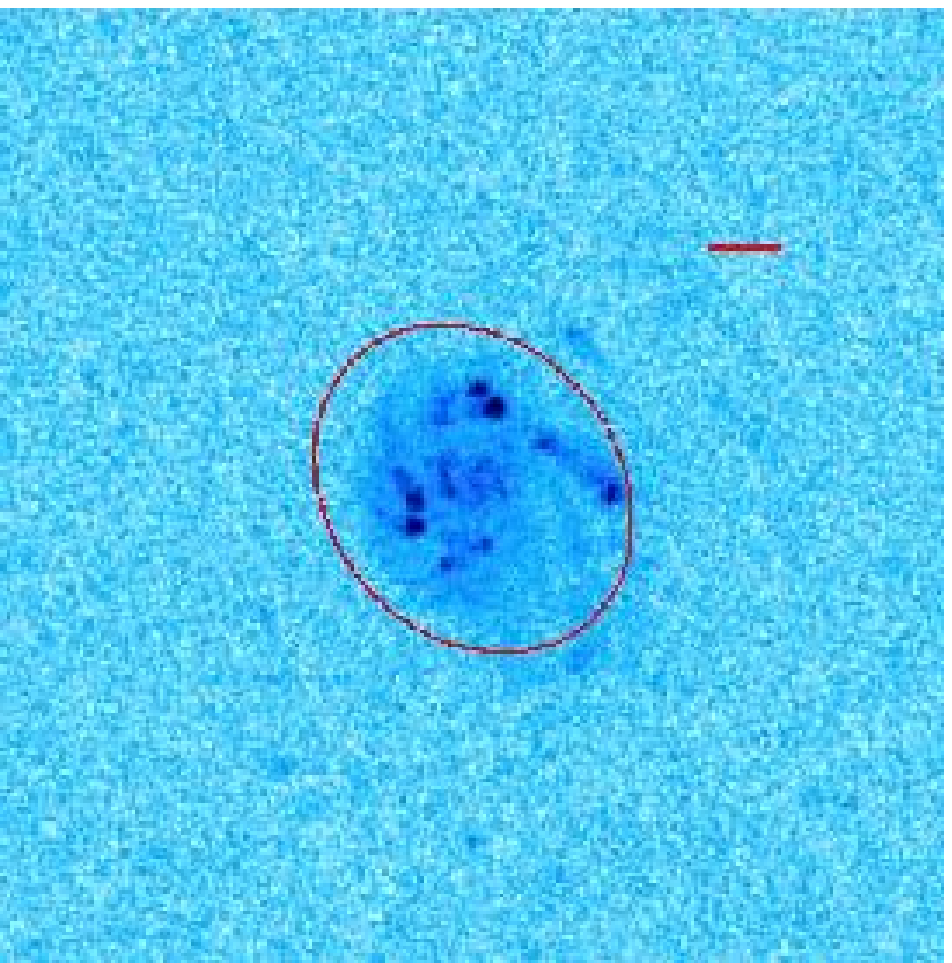} 
\includegraphics[width=6cm]{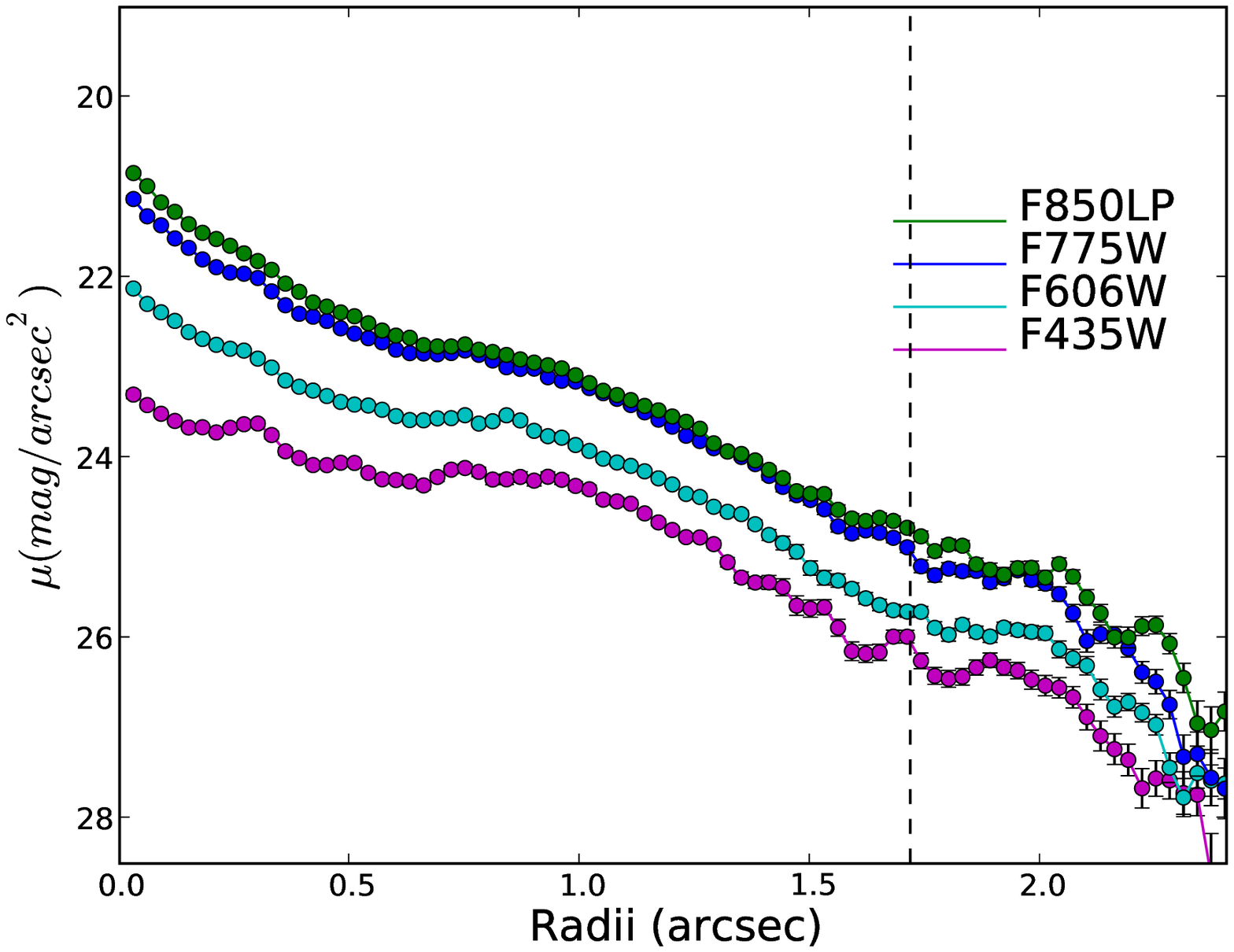} \\

\includegraphics[width=4cm]{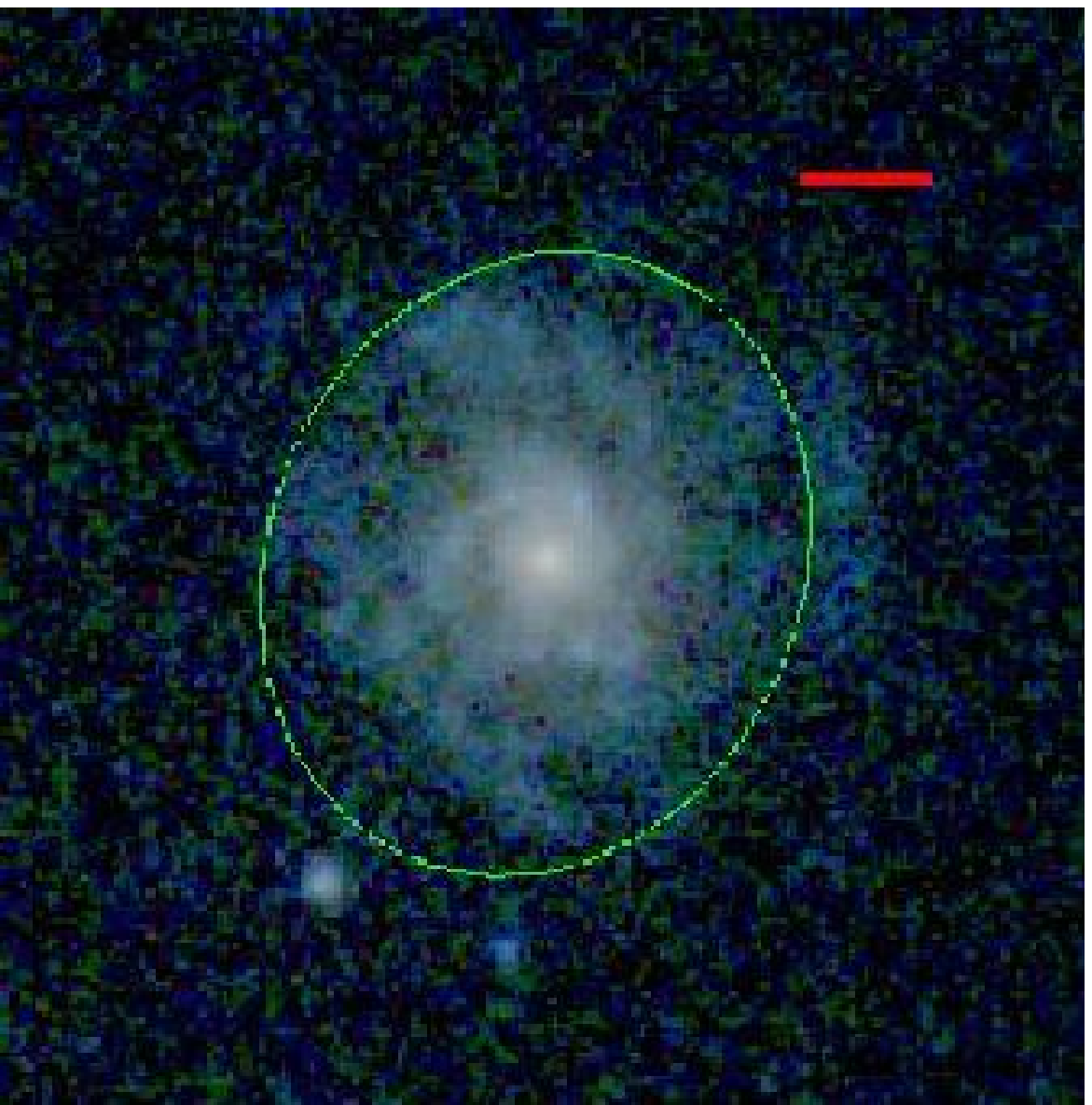} 
\includegraphics[width=4cm]{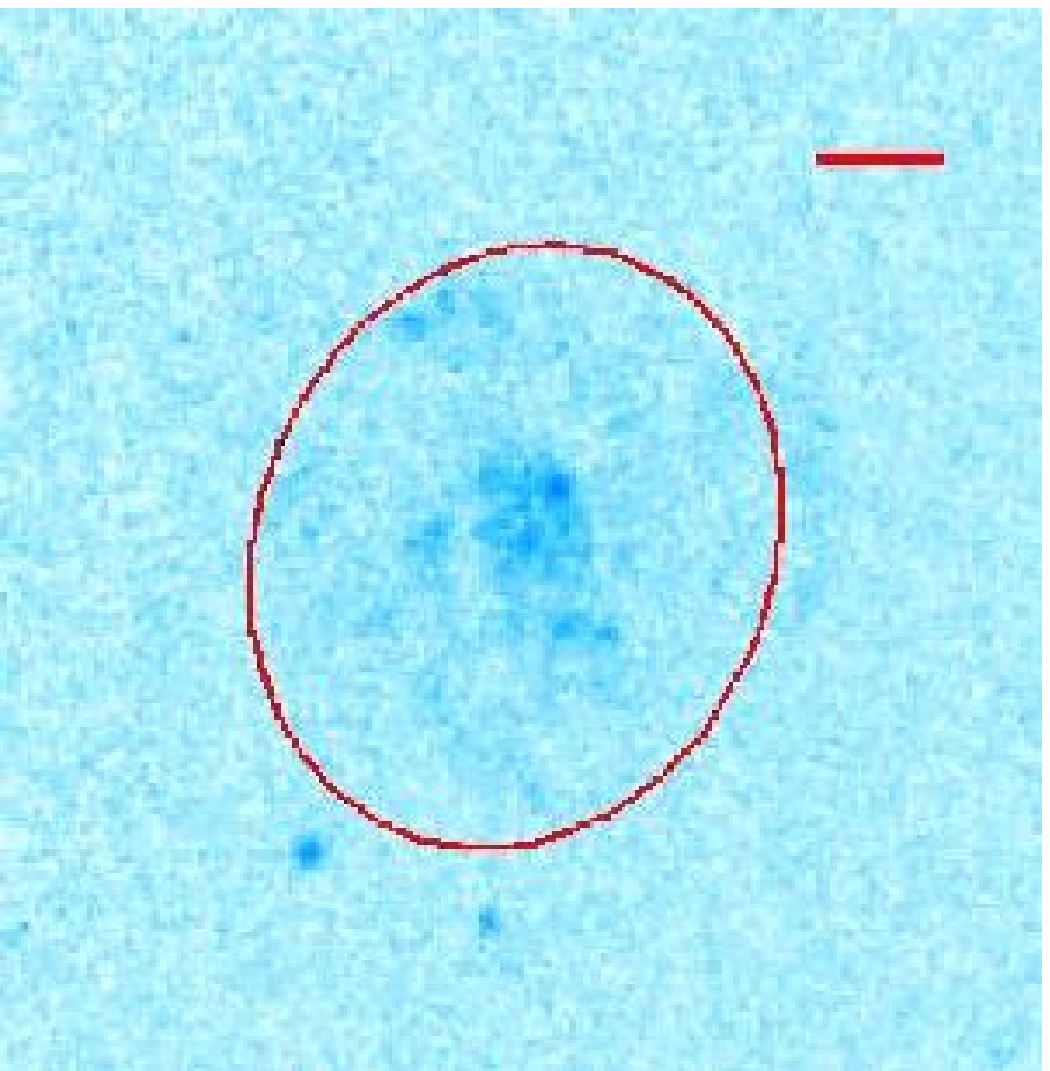} 
\includegraphics[width=6cm]{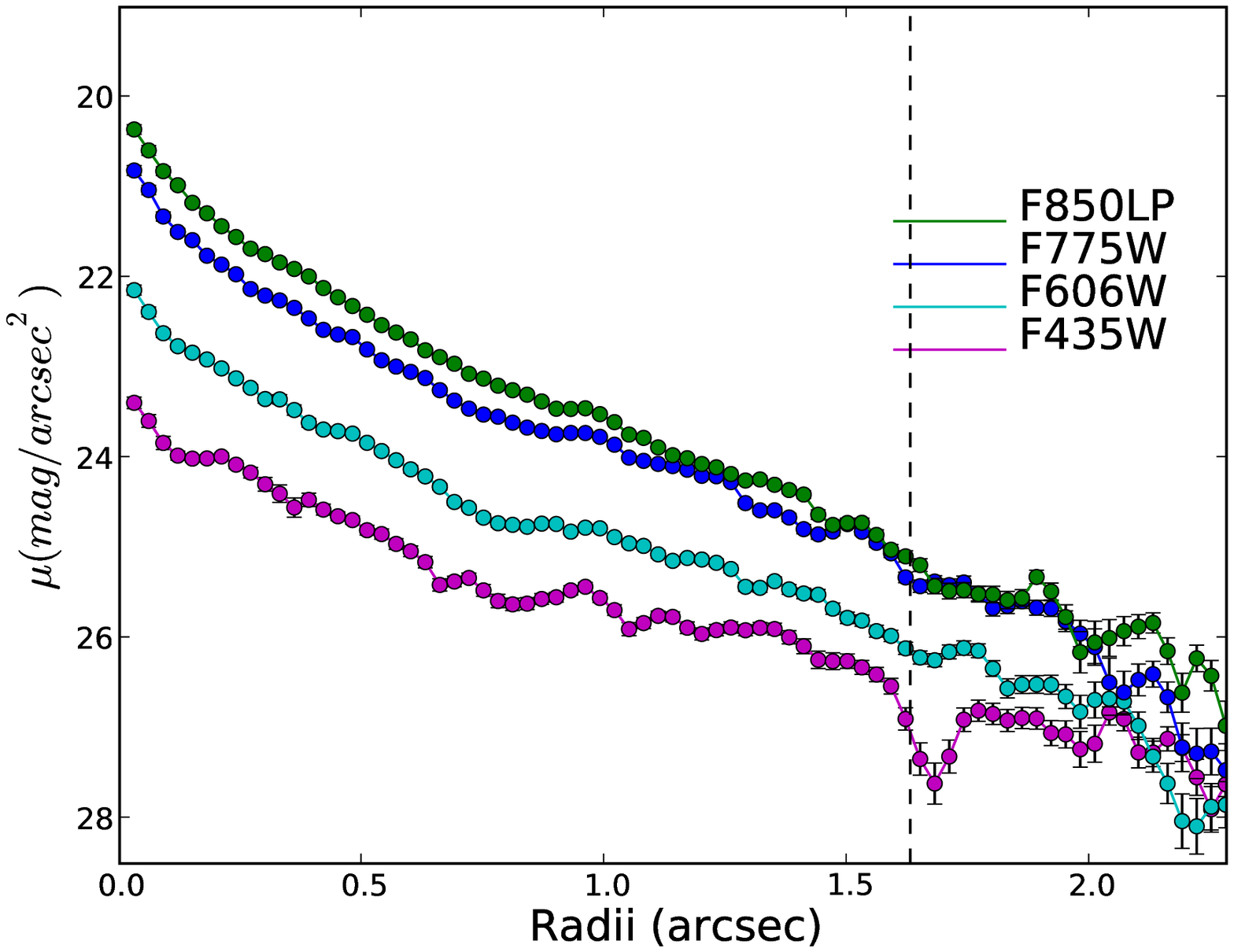} \\

\caption{Examples of images and surface-brightness profiles of objects under study. From above: NGC3319 ($D=$15 Mpc), NGC3359 ($D=$18 Mpc), GEMSz033251.15m274756.0 ($z=$0.54), GEMSz033237.54m274838.7 ($z=$0.67) and GEMSz033248.47m275416.0 ($z=$0.78). From left to right we show an rgb-composite, the image in \NUVrest and the observed surface brightness profiles in all available bands (cosmological dimming has not been accounted for in these figures). The bands used as the 'r' 'g' and 'b' channels are, respectively: \is, \rs and \gs for objects in the Local sample, and \zz, \ii and \vv for those in the High-z sample. The ellipse marks the Petrosian radius ($\eta$=0.2) in the \Brest band and the solid horizontal lines indicate the equivalent to 5 kpc at the distance of the object. In the panels with the \mur profiles, the vertical dashed line marks the position of the Petrosian radius.}
\label{Fig:Obj}
\end{figure*}

In Fig. \ref{Fig:Obj} we show several examples of objects in the Local and High-z samples, as seen in rest-frame visible bands (as ``rgb composites'', on the left), in rest-frame \NUV (centre), and also present their surface brightness profiles in all available bands (left). In the profiles we see clear truncations, also termed Type II profiles (Freeman \cite{Freeman70}, Erwin et al. \cite{Erwin05}), in NGC3319 ($D=$15 Mpc, first from above) and NGC3359 ($D=$18 Mpc, second). One object shows a Type II profile (GEMSz033237.54m274838.7, $z=$0.67, fourth), though this ``downbending'' seems not to be a genuine disc truncation, as it happens well inside the disc. Another case also shows hints of truncation (GEMSz033251.15m274756.0, $z=$0.54, third), though not so clearly, as it happens in a region of the profile with low signal-to-noise ratio. The fifth object, GEMSz033248.47m275416.0 ($z=$0.78), shows a single exponential profile (Type I). As we can see, the \NUV images show in all cases a patchy appearance, with varying degrees of similarity to the visible images. In the local objects it is easier to recognize structures such as spiral arms in the \NUVrest images, and also in GEMSz033237.54m274838.7, though in this case the spiral arms show some interesting asymmetry, perhaps the signature of a past interaction. The other two galaxies, taken from the High-z sample, show less defined morphologies, though the presence of a stellar disc is clear in all cases.

As we have already stated, we use the \NUV surface-brightness profiles and growth curves as proxies for the distribution of the SF in galaxies. Nonetheless, the dust extinction plays a significant role in the shaping of these profiles that cannot be ignored, as the extinction curves usually adopt a power-law profile in which there is higher absorption at shorter wavelengths ($A(\lambda)\propto \lambda^{-1}$ to a first order approximation). A detailed treatment of this issue would require us to produce radial extinction curves for every object, which is by no means a straightforward task. To be done reliably  it requires more data than we have (it would be desirable at least to extend the spectral coverage into the rest-frame near IR). Present state-of-the-art instrumentation (i.e. the absence of high-resolution IR facilities) prevents us from tackling the problem in such depth. So, in the next subsections we present results which do not include extinction corrections, and so our picture of the radial distribution of the SF and its evolution since $z\sim$1 should be taken with caution. 

Nonetheless, as a first and crude approach to solving this problem, in Sect. \ref{Sec:Dust} we explore how the results might be affected by a radially resolved correction on dust extinction, following the work by Boissier et al. (\cite{Boissier04}). At the very least that exercise serves to visualize that the effect of dust cannot be ignored in this kind of studies, and also as an indication of the sense in which more reliable corrections would alter the presented results.

\subsection{B-band Luminosities and Stellar Masses of Galaxies.}

As we stated in section \ref{Sec:data}, we have estimates of the \B-band luminosities, \mb, and stellar masses, \mstar, of the galaxies in the High-z sample from the COMBO-17 dataset. We also need similar estimates for the Local sample, so as to compare, amongst redshift bins, the relation between different measures of the radial distribution of \NUV-flux and these physical properties. With this aim we have, first, performed aperture photometry of the galaxies in the Local sample, to build their SEDs in \NUV and the 5 \SDSS bands. Then, using the HyperZ code (Bolzonella et al. \cite{Bolzonella00}), we fit these photometric data to model SEDs to obtain estimates of the desired properties of the galaxies. We use the stellar population synthesis models implemented in HyperZ, which are based on synthetic spectra from the GISSEL98 (``Galaxy Isochrone Synthesis Spectral Evolution Library'') spectral evolution library of Bruzual \& Charlot \cite{BruzualCharlot93}. The Initial Mass Function (IMF) used is that of Miller \& Scalo \cite{MillerScalo79}, with lower and upper cut-offs in stellar mass at $M_1 = 0.1$ \msun and $M_2 = 125$ \msun. With respect to the Star Formation History (SFH) we have opted for Single Stellar Populations (SSP) of different ages (from 1 Myr to the age of the universe at the redshift of the object). We assume a solar metallicity ($Z=Z_{\odot} \sim 0.02$) and the SEDs are attenuated, using a free parameter, $A_v$, according to the law of reddening by inner dust absorption given in Calzetti et al. \cite{Calzetti00} for starburst galaxies. We restrict this parameter to be within 0$<$Av$<$2 mag. The reddening produced by dust in our own galaxy is also introduced in terms of the excess in the $B-V$ colour for each galaxy ($E(B-V)$; estimates on this parameter are taken from HyperLeda), given as input. The redshift, which is normally the parameter needed when using HyperZ, is fixed, and given by the corresponding distance according to HyperLeda.

\begin{figure*}
\centering
\includegraphics[width=8cm]{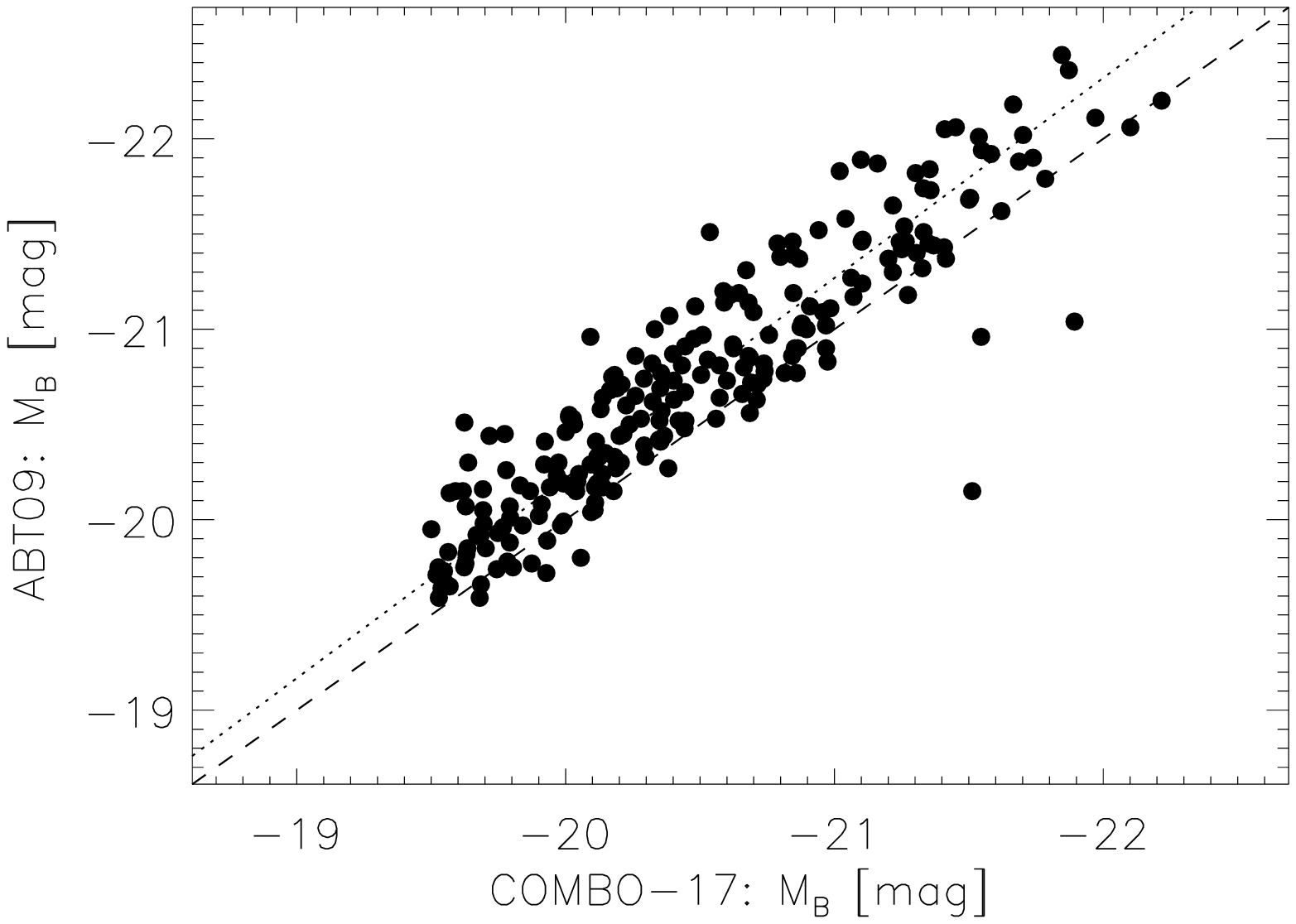} 
\includegraphics[width=8cm]{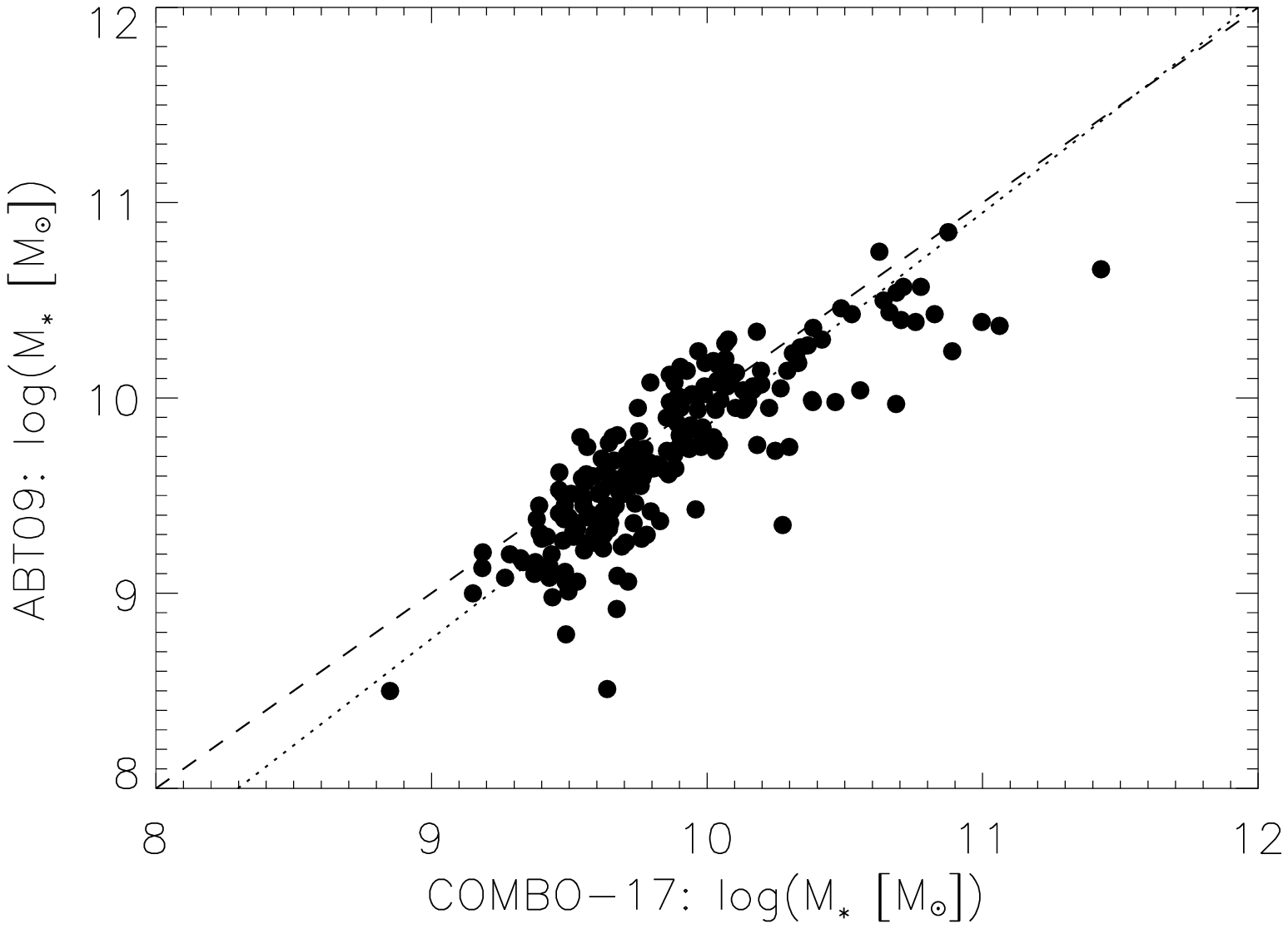}
\caption{Left: Comparison our estimates of \B-band luminosities (\mb) of galaxies in the High-z sample, obtained with HyperZ, to those from COMBO-17. Right: Analogous to the left panel but for stellar masses (\mstar). For both panels, the dashed lines represent a one-to-one relation, and the dotted, linear fits to the distributions of points.}
\label{Fig:LMcompare}
\end{figure*}

The reliability of part of the analysis presented in this work depends on how consistent are our previous estimates of stellar masses and luminosities of galaxies with those from COMBO-17. In order to check this, we proceed as follows. We have measured the \mb and \mstar of the High-z galaxies with HyperZ, in an analogous way to that described above for the Local sample. The only differences in the input data are the photometric values (here we use the 4 available \GOODS bands), and the reddening from the Milky Way, which is fixed to be E(B-V)$=$0.007 mag (Schlegel et al. \cite{Schlegel98}). Then we compare the estimates of \mb and \mstar thus produced with those from COMBO-17, and the results are shown in Fig. \ref{Fig:LMcompare}. The \mb estimates (left) show good agreement, our values being 0.2 dex brighter than those from COMBO-17, with relatively little dispersion ($\sigma$=0.3 mag). The \mstar estimates have a somewhat larger dispersion ($\sigma$=0.2 dex), and there is some bias, our mass estimates being lower than those from COMBO-17 by 0.14 dex ($\sim$30\% less massive) (at \mstar = $10^{10}$ \msun). The main contributors to this bias are, presumably, three factors: a) the photometry; b) the choice of IMF; and c) the SFH of the model SEDs. The photometry from COMBO-17 was obtained from ground observations (MPG/ESO 2.2m telescope on La Silla), and applying aperture corrections depending on morphological parameters to retrieve the fluxes, while we have just integrated the fluxes in the, elliptical, quasi-isophotal aperture with semi-major axis equal to the rest-frame \B-band Petrosian radius of the object. Second, we use the IMF from Miller \& Scalo (\cite{MillerScalo79}) while COMBO-17 uses that from Kroupa et al. (\cite{Kroupa93}). We have tried other SFHs, apart from single bursts, such as a constant SFR or an exponentially decaying SFR, but the best results, in terms of scatter and bias when comparing with COMBO-17, were those obtained with the SSP models, which are thus our choice.

To conclude this point, and with aim of minimizing the possibility of having any bias when comparing results which depend on the stellar masses or luminosities of galaxies at different redshifts, throughout this work we use our own ``HyperZ'' estimates of these quantities at all redshifts, rather than the values from COMBO-17 available for the High-z sample. This includes the results shown in Fig \ref{Fig:HisLM} on the distributions of stellar masses and luminosities for the different redshift samples, and already commented in section \ref{Sec:data}.

\subsection{Effective Radii (\Reff)}\label{Sec:Reff}

\begin{figure}
\centering
\includegraphics[width=8cm]{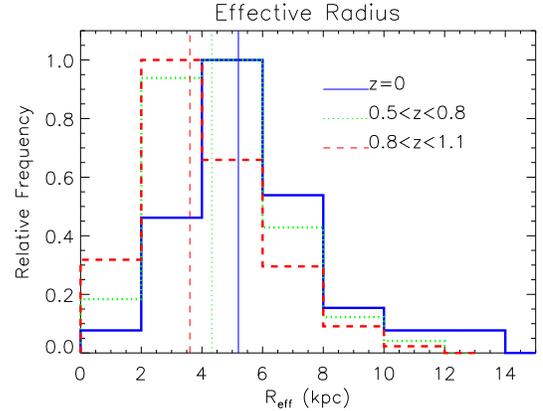}
   \caption{Effective radii of the rest-frame \NUV flux distributions.}
      \label{Fig:HisReff}
\end{figure}

The most basic characterization of the radial distribution of the Near UV flux is the effective radius. This is defined as the radius which encloses half the total flux of an object, given a certain geometry of apertures (i.e. their centre, ellipticity and polar angle, assuming they are elliptical). We use the same apertures employed to retrieve the radial surface brightness profiles described in section \ref{Sec:data}. In Fig \ref{Fig:HisReff} we show the distributions of effective radii in \NUVrest for galaxies in different reshift bins. The same combinations of line style and colour that were used in Fig. \ref{Fig:R2FWHM} apply here to code the different redshift subsamples, and this code also applies to the vertical lines which mark the median values of the corresponding distributions. We see that there are some differences amongst the distributions of effective radii: for the Local sample the median value is \Reff$=$5.2$\pm$0.4 kpc, 4.3$\pm$0.2 for the mid-z subsample (\midz), and 3.6$\pm$0.2 kpc for the far-z subsample (\farz). We recall here that no selection criteria were applied to the Local or High-z samples to put a limit on the maximum size of the galaxies, and so this difference, though not very significant given the error bars, is not due to any obvious selection effect. According to this result, there is an increase by 44$\pm$20\% in the \Reff(\NUV) between z$\sim$1 and z$\sim$0. If a Kolmogorov-Smirnov test is applied, the probability that the Local and mid-z samples of \Reff(\NUV) were extracted from the same distribution is rejected at 99.5\% confidence level, while for the Local and far-z samples, this rejection is at a level $>$99.9\%. If the same exercise is performed on images degraded to a common resolution (\ResFIFTY), the values are not significantly altered.

However, we do not claim that these figures are, on their own, proof of radial growth of the star-forming disc (or more precisely, growth in the distribution of \NUV flux), because the galaxies in these samples do not share exactly the same ranges of luminosity and stellar mass, as seen in section \ref{Sec:data}. We thus present more reliable tests on this matter in next subsections.

At this point it is convenient to explain why we select a maximum distance, \Dmax, of 60 Mpc in the selection of the Local sample, a limit that provides the largest possible sample while still having a negligible bias in size. To reach this conclusion we tested how the limit in distance affected the median value of \Reff(\Brest), \medReffB, in the Local sample. The median value of \medReffB with \Dmax$=$60 Mpc is 4.3$\pm$0.4 kpc (31 objects), whereas if that limit distance is 80 Mpc the sample is increased to 52 objects, at the cost of an increase by $\sim$35\% in \medReffB. To give a reference for considering the retrieved value of \medReffB, we refer to Shen et al. (\cite{Shen03}), who estimated median sizes in a complete sample of galaxies, using different bands and estimators, in the local universe. The median \rs-band luminosity in the Local sample is -20.95 mag. With this luminosity the disc-like objects in Shen et al. (\cite{Shen03}) have \Reff(\rs)$=$3.7 kpc. Shen et al. base their results on the extraction of radial profiles using circular apertures, i.e. they obtain circularized radii ($r_c$). In contrast, our results are obtained taking as radii the semi-major axes of elliptical apertures ($r_e$) which fit the overall geometry of the objects. As an approximation, $r_c = r_e * q^{1/2}$, where $q$ is the axial ratio of the elliptical apertures. In the Local sample, the median value $\tilde q=$0.74 (corresponding to an inclination of 42$\deg$), and so, the value by Shen et al. would correspond to $\sim$4.3 kpc if they had used elliptical apertures. This result is in good agreement with ours, given that the effective radii measured in \rs and \gs are very similar (e.g. de Jong \cite{deJong96}).

\begin{figure}
\centering
\includegraphics[width=8cm]{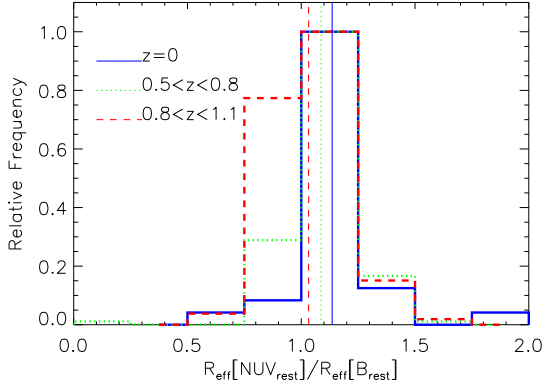}
   \caption{Ratio of the effective radii in rest-frame \NUV, \NUVrest, and rest-frame \B, \Brest bands.}
      \label{Fig:HisReffNUV2B}
\end{figure}

Another interesting test is to compare the spatial extension in the \NUV to that measured in the \B-band, which traces older stellar populations. In Fig. \ref{Fig:HisReffNUV2B} we present the distributions of the ratio of the \Reff in \NUVrest to that in \Brest for galaxies in the redshift bins considered. Also in this case there is some difference between the low and intermediate redshift samples. Galaxies in the High-z sample show median values of this ratio around unity (1.09$\pm$0.01 and 1.03$\pm$0.02 for the mid-z and far-z subsamples respectively), while for those in the Local sample the median value is at 1.14$\pm$0.04. This means that local and higher redshift galaxies are slightly more extended in \NUVrest than in \Brest, but at $z\sim$1 the difference is less. The null hypothesis with respect to the Local and far-z samples can be rejected at a level of confidence $>$99.9\% (K-S test).

A possible explanation of our findings could be the emergence of the bulge/pseudo-bulge in the galaxies as cosmic time evolves. In fact, all other structures in a galaxy being equal, a brighter bulge or pseudo-bulge would make the effective radius of the galaxy ``shrink''. This would be more evident in \B than in \NUV, as the populations in these structures are relatively old.  We need to assume that there is not a parallel increase in the SFR in central regions that may cancel out this difference between bands. Galaxies of earlier morphological types have, by definition, more prominent bulges. It could be argued then, that perhaps the apparent evolution in the ratio \Reff(\NUV)/\Reff(\B) could be due to a difference in the distribution of morphological types, in which there would be a larger fraction of earlier types in the Local sample with regard to the High-z sample, and this could lead to a larger median ratio of effective radii between both bands in the former. In consequence, we have tested the ratio \Reff(\NUVrest)/\Reff(\Brest) for local galaxies with 4$<T<$10 (Sc to Sm), and the resulting median value is 1.20$\pm$0.08, an even larger ratio.

So, it seems that this increase in the ratio of effective radii is related to a progressive outwards migration of the SF relative to the distribution of older stars in the discs. We have also compared the aforementioned ratios when we use equal resolution images (\ResFIFTY), and the result stands, though the differences decrease a little, with median values of the ratio \Reff(\NUVrest)/\Reff(\Brest) of 1.13$\pm$0.03 (z$\sim$0), 1.06$\pm$0.01 (\midz) and 1.02$\pm$0.01 (\farz). Again, a K-S test rejects the hypothesis that the Local and far-z samples come from the same parent distribution with high probability (confidence level $>$99.9\%). In this way, it seems that the ratio \Reff(\NUVrest)/\Reff(\Brest) has slightly increased since z$\sim$1, by $\sim$10\%.

\subsection{Median Intensity/Colour profiles}\label{Sec:Stacked}

\begin{figure*}
\centering          

\includegraphics[width=7cm]{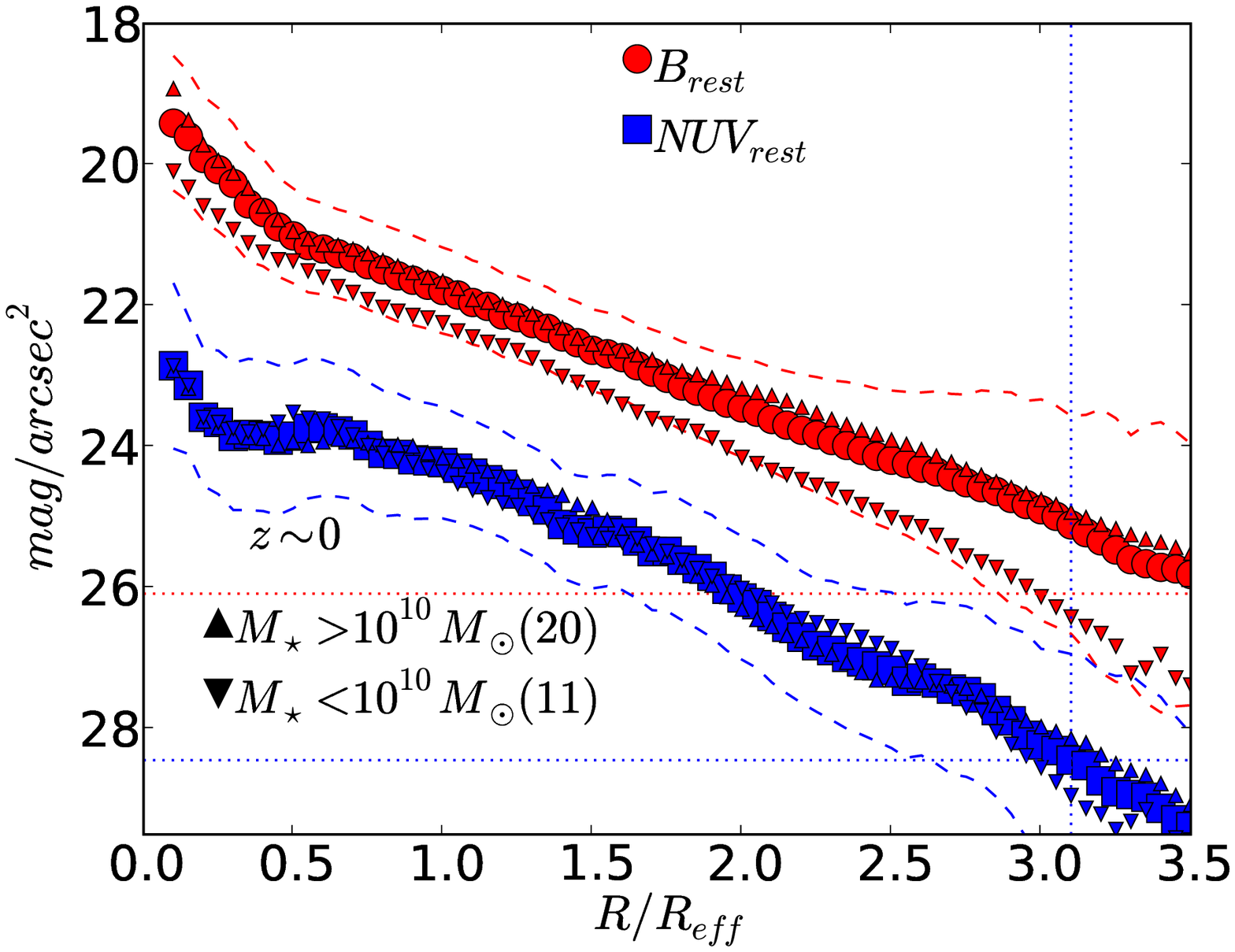} 
\includegraphics[width=7cm]{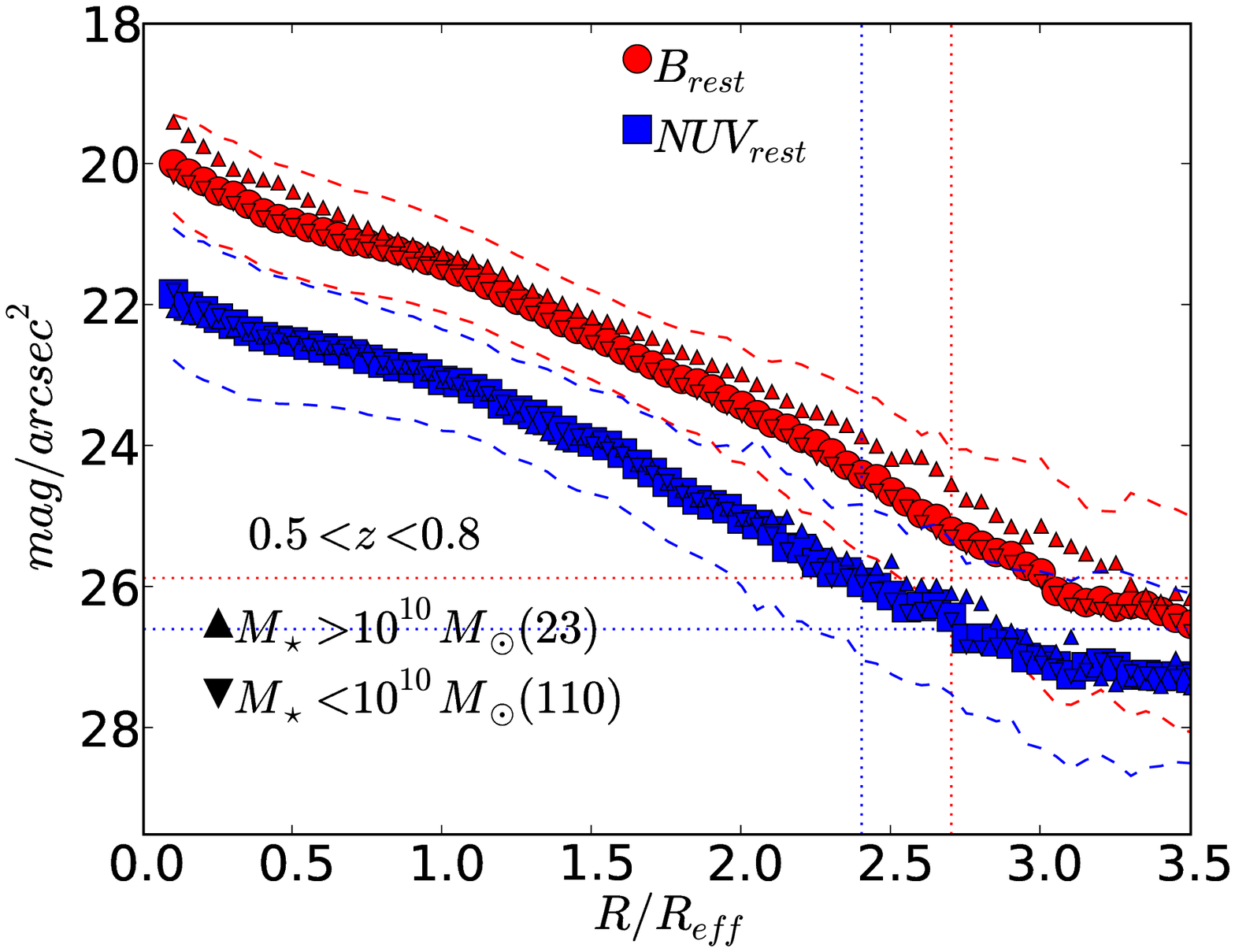}\\
\includegraphics[width=7cm]{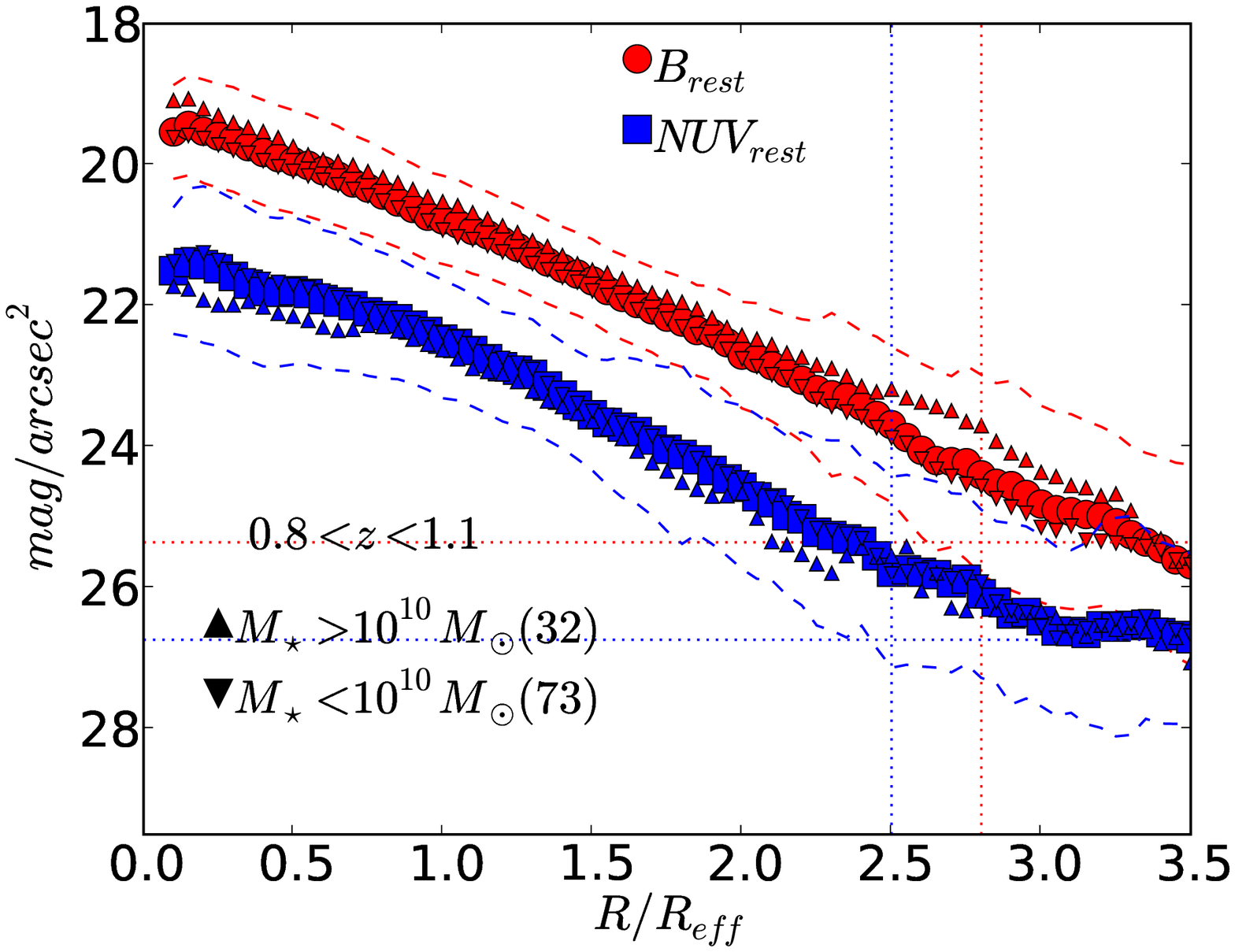}
\includegraphics[width=7cm]{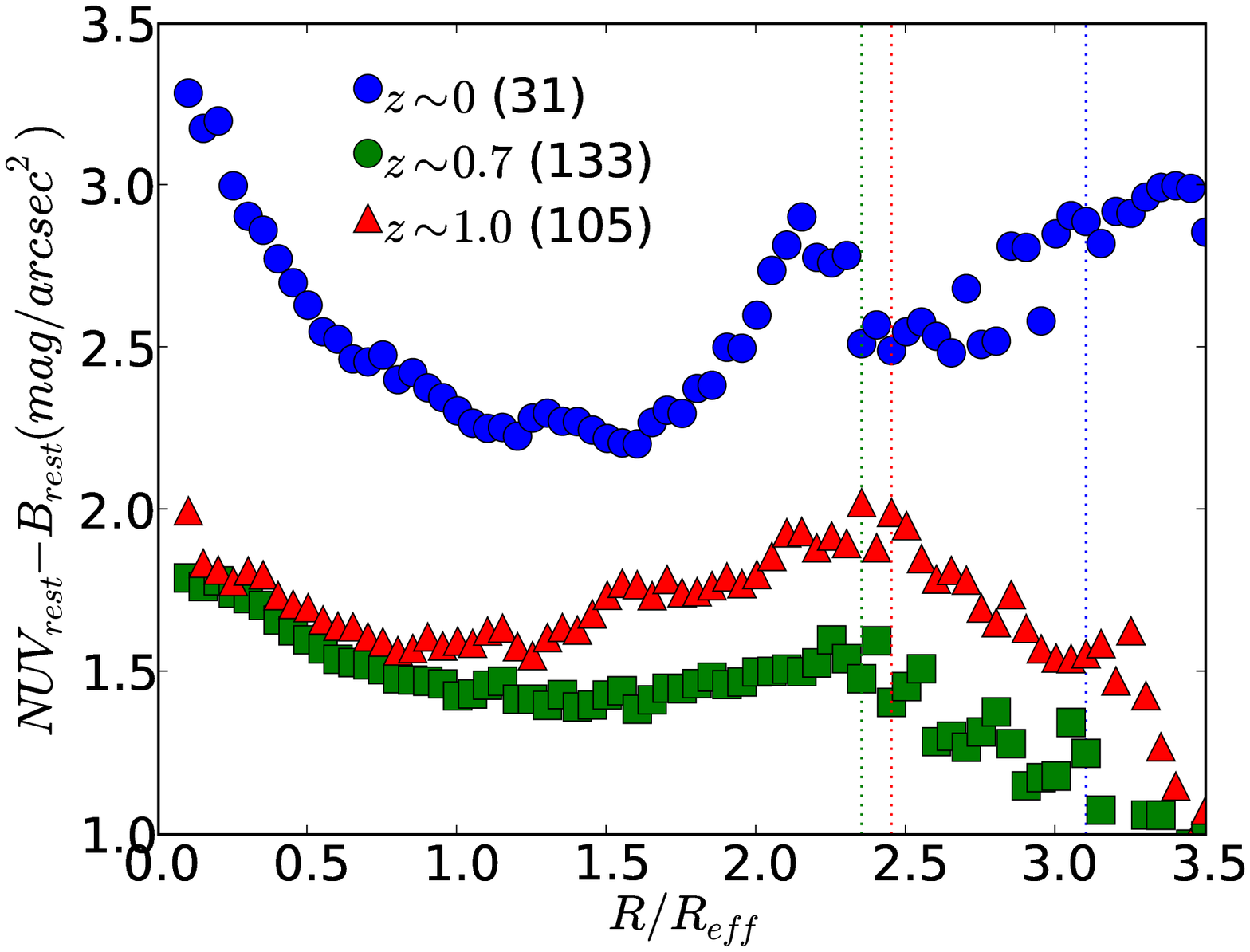} \\

\caption{{\bf Median surface brightness (\mur) profiles (cosmological dimming corrected) of the galaxies in different redshift bins: $z\sim$0 (upper left),  \midz (upper right), and \farz (lower left), in \Brest (red dots) and \NUVrest (blue squares). The vertical dotted blue and red lines mark the radii up to which the \NUVrest and \Brest profiles are reliable, given the uncertainty in the estimate of the background, $\sigma_{back}$ (i.e. the last radius with $\delta\mu=[\mu(I-\sigma_{back})-\mu(I+\sigma_{back})]<0.2 mag$, where $\sigma_{back}$ is the standard deviation of the background fluctuations in the typical annulus used to estimate it). The horizontal dotted lines are the corresponding surface-brightness levels at those fiducial radii. In the lower right panel are the corresponding median colour profiles (\NUVrest-\Brest). There again, the vertical dotted lines in each color give the fiducial radii for the correponding color profiles ($[\delta^2\mu_B+\delta^2\mu_{NUV}]^{\frac{1}{2}}<0.2$mag). In all cases the radial scale is relative to the corresponding effective radii, \Reff, of the objects in the \Brest-band. In the \mur profiles, the dashed lines give the standard deviation of the values of $\mu$ at each radius. The upward pointing triangles give the mean profile for galaxies with \mstar$>10^{10}$\msun, while the downward pointing triangles are the equivalent for galaxies with \mstar$<10^{10}$\msun.}}
\label{Fig:MedProfs}
\end{figure*}

In order to see with more clarity what the differences between the aforementioned effective radii in \Brest and \NUVrest mean, we also explore the median surface brightness profiles of the galaxies in our sample. In Fig. \ref{Fig:MedProfs} we present median surface brightness profiles (\mur) in \NUVrest (blue squares) and \Brest (red dots) for the galaxies in the Local sample (upper left) and at intermediate redshifts (upper right, \midz and lower left, \farz). In the lower right panel are shown the \NUVrest-\Brest median colour profiles, for every redshift bin. In all panels, the radial coordinates have been scaled by the effective radius in \Brest for each galaxy before taking the median. The dashed lines in the \mur profiles give the dispersion (standard deviation). For the colour profiles, median values of the dispersion are 1.0 mag ($z\sim$0), 0.8 mag ($z\sim$0.7) and 1.0 mag ($z\sim$1).

In Table \ref{Tab:1}, 7th column, we list the surface brightness level of the 1$\sigma$ fluctuations of the background, estimated in an annular aperture used with this aim in the production of intensity profiles and growth curves. These levels are fainter than the values of $\sigma_{arcsec^2}$, for local objects, and brighter in the case of galaxies in the High-z sample, as the area of this annulus (and thus the number of pixels considered) is respectively larger, and smaller, than 1 $arcsec^2$. The background subtraction is a significant contributor to error in the profiles. We want therefore to estimate up to which radius the individual surface-brightness, and colour, profiles are reliable, taking into account this uncertainty in the background estimate. We are conservative in this regard, and following Pohlen \& Trujillo (\cite{Pohlen06}, see section 3.3 in their article), we adopt the following strategy to find these fiducial radii and surface-brightness levels. First, the corresponding 1$\sigma_{back}$ fluctuations, converted to intensities, are added and subtracted  to the median intensity profile in each band, giving the profiles $\mu(I+\sigma_{back},r)$ and $\mu(I-\sigma_{back},r)$, respectively. Then, the fiducial radius is the largest for which the difference between these profiles, $\delta\mu=| \mu(I+\sigma_{back},r) - \mu(I-\sigma_{back},r)| < 0.2$ magnitudes. This is computed using the observed profiles, before cosmological dimming, or any other corrections are applied to the profiles (see below). In Fig. \ref{Fig:MedProfs} we show the corresponding fiducial radii and surface-brightness levels as vertical and horizontal dotted lines, in the same colour on the graph as the corresponding intensity profile. In \Brest band, the fiducial radii/surface-brightness levels are at 3.7\Reff(\Brest)/26.1 \magarcsq, and 2.7/25.9 and 2.8/25.4 for the Local, mid-z and far-z sample, respectively. In \NUVrest the corresponding values are 3.1/28.5, 2.4/26.6 and 2.5/26.7. So, the fiducial surface brightness levels are $\sim$2.5 \magarcsq brighter than the 1$\sigma_{back}$ fluctuations.

For the colour profiles a similar approach is followed. Given the profiles of $\delta\mu(r,band)$ in each band (\NUVrest and \Brest), we obtain the fiducial radius as the largest for which $(\delta^2\mu(r,NUV_{rest})+\delta^2\mu(r,B_{rest}))^{\frac{1}{2}}< 0.2$ mag. These radii are indicated as vertical dotted lines in the lower right panel of Fig. \ref{Fig:MedProfs}, again in the same colour on the graph as the corresponding colour profile. The fiducial radii are at 3.1, 2.3 and 2.5 \Reff(\Brest), for the Local, mid-z and far-z samples.

The profiles of the High-z sample shown in Fig. \ref{Fig:MedProfs} have been corrected for cosmological dimming, and a simplistic k-correction has been applied assuming a ``flat'' SED ($f_{\nu}$ = constant) and no inner dust absorption (Av$=$0 mag). These corrections are applied after the analysis of the reliability of the profiles, described above, is performed. This is the reason why the lines of fiducial radius and surface-brightness do not cross at a point of the intensity profile for the mid-z and far-z subsamples, but at points 0.66 and 0.96 magnitudes fainter, respectively. Given these assumptions, the median value of surface brightness, $\mu$, in \NUVrest, at \R$=$\Reff, for galaxies at z$\sim$0 is roughly 1/5 of the value at z$\sim$1 ($\Delta\mu$=1.8 mag). This is related to the significant decrease in the SFR the discs have suffered in the last $\sim$8 Gyr. In the \B band this evolution is less significant, with $\Delta\mu$=1.0 mag; i.e. the surface brightness of discs (at \R$=$\Reff) in the \B-band nowadays is $\sim$40\% of the value at z$\sim$1.

Fig. \ref{Fig:MedProfs} helps us to understand Fig. \ref{Fig:HisReffNUV2B}. The \Brest and \NUVrest profiles at intermediate redshift fall off exponentially, falling roughly ``parallel'' at all radii, because of the intense star-forming activity, which is largely dominating the appearance of the galaxies not only in \NUVrest, but also in \Brest. At z$\sim$0 the profiles also fall off, to a good approximation, with exponential rates, but the \Brest profile has a slightly steeper slope than the \NUVrest in the inner parts of these median profiles. This difference may contribute to make the ratio \Reff(\NUVrest)/\Reff(\Brest) at $z$$\sim$0 larger than the value at $z$$\sim$1. This central excess in brightness of the \Brest profile at $z\sim$0 could be the signature of older stellar populations being ``piled up'' in evolving bulges or pseudo-bulges. At $z\sim$0, in \NUV, there is also an excess in the central region over the disc profile. The central emission in \Brest band at z$\sim$0 is above the corresponding value at $z\sim$0.7 ($\Delta\mu \sim$ 0.6 mag), but only slightly above the corresponding value at $z\sim$1 ($\Delta\mu \sim$ 0.1 mag). In the \NUVrest band there is not a clear difference amongst redshift bins in this respect, though in the highest redshift bin (\farz) the central emission is the brightest. It must also be noted that the \Brest and \NUVrest profiles at $z\sim$0.7 (upper right panel) show slight upbendings at small radii, similar to, but far less significant than those at $z\sim$0, which we thus interpret as a hint of progressive evolution of pseudo-bulges in the explored range of redshifts. If we use images with equal resolution the shapes of the resulting median profiles do not differ significantly (the contribution of the bulges are more spread in radius, as expected), and also fit into the given interpretation.

Also in Fig. \ref{Fig:MedProfs} we show the median profiles when galaxies are segregated by their stellar masses: galaxies with \mstar$>10^{10}$\msun are represented by upward pointing triangles, and those with \mstar$<10^{10}$\msun by downward pointing triangles. It can be seen that the shapes of the \mur profiles when galaxies are segregated by mass are very similar to those of the whole populations, though there are minor differences may be seen. In \Brest, more massive galaxies have somewhat higher levels of surface brightness than less massive. In \NUVrest quite the opposite happens, median profiles of less massive galaxies, both in the Local and High-z samples, being somewhat brighter than those of more massive galaxies. This suggest a somewhat higher efficiency in SFR per unit area in less massive galaxies, a result that fits within the downsizing scenario.

Another interesting feature may be found in the \NUV profiles shown in Figure \ref{Fig:MedProfs}. Both for local and intermediate-redshift galaxies, the median profiles in this band are slightly ``downbending'', with the ``elbow'' of the profile located at \R$\sim$\Reff, something that is not seen in the \B-band profiles. Here we recall that the intensity profiles of stellar discs may be classified, in general, in three types: Type I, in which the intensity decays exponentially with a certain scale radius, $h$ (the slope of the \mur profile, $m$, is inversely proportional to this scale radius); type II, in which the profile declines exponentially with a certain $h$ out to a break point, past which a smaller value of $h$ pertains, i.e. the descending slope of the profile becomes steeper; and type III, in which there is also an exponential decay in the intensity and again a change in the slope of the profile at a certain radius, but in this case the profile becomes shallower outside it (see, e.g. Erwin et al. \cite{Erwin08} for a review on related phenomenology). Stellar disc truncations are a subtype of type II profiles in which the break radius is located at what seems to be the ``edge'' of the stellar disc, beyond the extent of the spiral arms.

The interpretation of the downbending feature in this figure is not straightforward, since the profiles shown here are the average of profiles which have not been segregated according to type, and thus are mixed, supposedly in proportions similar to those found in field galaxies at z$\sim$0 and at z$\sim$1: $\sim$60-50\% of types II, $\sim$30-40\% of types I, and $\sim$10\% of types III (see e.g. Pohlen \& Trujillo \cite{Pohlen06}, Azzollini et al. \cite{ATB08b}). In any case, it seems that the \NUV profiles show, commonly, a change in their slopes at some point in the discs, past which the drop in intensity becomes steeper.

Recently, it has been shown that truncated discs usually show radial colour profiles (\us-\gs) with a minimum value at the position of the break (i.e. it is a bluest point), and they become gradually redder inwards and outwards from this point (Azzollini et al. \cite{ATB08a}, Bakos et al. \cite{Bakos08}). So, the colour profile has a characteristic ``blue valley'' shape. Even when the galaxies have not been segregated by the type of their intensity profiles (i.e. in types I, II, or III) the median colour profiles \NUV-\B presented in Fig. \ref{Fig:MedProfs} (lower right) show a similar feature, most easily recognizable in the median colour profiles for galaxies at $z\sim$0 (blue) and $z\sim$1 (red). This feature is clearly recognizable also when the FWHMs of the PSFs of the images are equal in both bands, as is the case when a uniform resolution (\ResFIFTY) is simulated.

The local colour profile (blue squares) is clearly redder than those at $z\sim$0.7 (green squares) and $z\sim$1 (red triangles), by $\sim$0.7 mag, and the ``valley'' shape is more prominent, with a difference in colour $\Delta$C$\sim$1.1 mag between the minimum and maximum (at the centre of the galaxies). In contrast, galaxies in the High-z sample show a more moderate minimum to maximum difference in colour of $\Delta C \sim$0.4 mag. We see that the minimum in colour takes place in the range 1-1.5\Reff. From these colour profiles we see how the central part is getting progressively redder in relation to the colour at the break. We interpret this as further evidence of secular bulge growth as cosmic time progresses.

One question to be answered in the light of these results is whether the reported progressive rise of the inner parts of the median profiles, both in \Brest and \NUVrest bands, is due to a mere selection effect of the samples at different redshifts. As we said before, objects in the Local sample were selected to have morphological types in the range $0<T<10$, and those in the High-z sample, to have S\'ersic index $n<$2.5. We consider this as the best approach to have a sample of disc galaxies not biased towards a particular Hubble morphological type, as already explained in Section \ref{Sec:data}. By definition, earlier type disc galaxies have more prominent bulges, and so, if these were over-represented in the Local sample, with regard to the High-z sample, this would cause a trend similar to that observed in the shape of the inner profiles with redshift. So, to settle this controversy it would be desirable to select the objects according to exactly the same criterion, for example the S\'ersic index, at all redshifts. Being short of that type of data, we have performed another test, to at least discard evidence for a bias within our limitations. We produced median intensity profiles for galaxies in the Local and High-z samples, dividing each of them into two more subsamples, termed as ``earlier'' and ``later'', according to their morphological types. In the Local sample, the  ``earlier'' subsample is composed of objects with $0<T<4$ (S0a to Sbc, 12 galaxies), and ``later'' of those within 4$\leq T<$10 (Sc to Sm, 19 galaxies). In the mid-z and far-z subsamples the ``earlier'' range corresponds to $1.25\leq n<2.5$ (89 galaxies), and the ``later'' to $0<n<1.25$ (149 galaxies). As would be expected, the median \NUVrest and \Brest profiles of the ``earlier'' subsamples show the trend to have brighter central parts with regard to the discs at lower redshift, as happens for the whole samples, yet amplified (i.e. with a larger contrast in brightness between the central part and the disc). Nonetheless the same effect, though somewhat less pronounced, is also present for the ``later'' subsamples. So, this effect does not seem to be due to a minor fraction of galaxies with prominent bulges in the Local sample, but is a common condition affecting, though in varying degree, most of the galaxies classified as disc-type, according to the prescribed criteria.

There is also the important effect that variable dust extinction along radius may have on the shape of the \Brest and particularly \NUVrest, which so far has not been considered. We postpone the consideration of those effects until Sec. \ref{Sec:Dust}, where the effects of dust are treated with some detail, within the limitations of this work.

\subsection{\Reff - \B-Luminosity Relation}

A set of tests of particular interest for the study of disc growth is to compare the effective radii of these discs, as measured in different wavelengths, to the luminosities and stellar masses of the galaxies. Here we focus on the \Reff as measured in the \NUVrest, and so on the relation between the size of the star-forming disc and the stellar content of the galaxies, as a function of redshift/time.

\begin{figure}
\centering
\includegraphics[width=8cm]{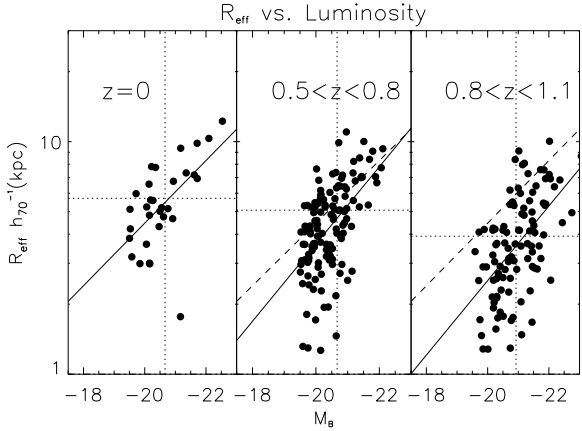}
   \caption{Effective radius of the rest-frame NUV flux distribution against the \B-band luminosity (\mb), for galaxies in different redshift bins: z$=$0 (Local sample, right panel), \midz (mid-z subsample, middle) and \farz (far-z subsample, right). The dotted vertical and horizontal lines mark the centroid of the distributions (median values), and the continuous line is the one that best fits the distributions. The dashed line in the middle and right panels gives the fit to the Local sample (left).}
      \label{Fig:ReffvsLum}
\end{figure}

In a first test, we study the relation between the \Reff in \NUVrest and \B-band luminosity (\mb) of galaxies in different redshift bins (see Fig. \ref{Fig:ReffvsLum} for results). In the three panels we show \mb against \Reff in \NUVrest, for galaxies in the Local sample (z$\sim$0), mid-z subsample (\midz) and far-z subsample (\farz) respectively. The first thing to note is that there is significant dispersion in this relation at all redshifts. The vertical and horizontal dotted lines mark the median values of the distributions for galaxies in the range -20$<$\mb$<$-22 mag. The objects have \Brest luminosities which gather around \mb$\sim$-21 mag at all redshifts. The observed distributions have been fitted to a line with a least squares deviation fit, in an iterative ``bootstrap'' method, to get the most probable values for the slope and the y-intercept ordinate and their errors. These best-fit lines are shown as continuous in each panel, while the dashed lines in the second and third panels correspond to the best-fit line for the Local sample relation, as reference. From visual inspection it seems clear that, at a given luminosity, \Reff is smaller as redshift increases.

\begin{figure}
\centering
\includegraphics[width=8cm]{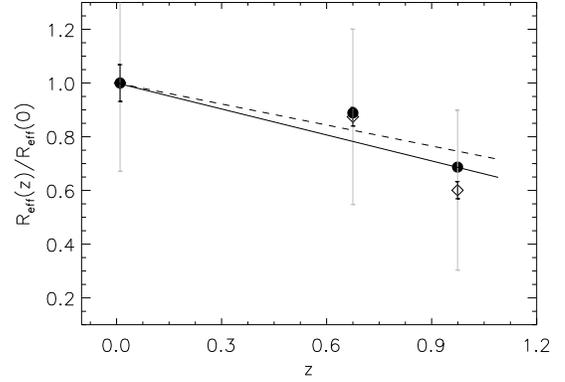}
\caption{Evolution of the relation \Reff(\NUVrest) - \mb. The empty diamonds give the value of \Reff at \mb=-21 mag from the best fit lines in Fig. \ref{Fig:ReffvsLum}, whereas the black dots are the corresponding median values of \Reff. All measures have been scaled by the best fit value at z$\sim$0. The line that best fits the diamonds (continuous) gives an evolution in \Reff(\NUVrest) at the given luminosity by a factor 1.48$\pm$0.09 between z$=$1 and z$=$0. The best fit line to the circles (median values of \Reff, dashed line) represents an increase by 1.35$\pm$0.07 in the same redshift interval. The long-gray error bars give the standard-deviation ($\sigma$) in \Reff($z$)/\Reff($z=$0, while the short-black give the associated error ($\sigma/\sqrt(N)$, where $N$ is the number of galaxies in the subsample).}
\label{Fig:EvoReffvsLum}
\end{figure}

In order to compare the radial extents of the SF of the galaxies in a meaningful way, we can look at the values of the best-fit lines for the given relation, \Reff(\NUVrest) - \mb, at a fixed value of luminosity, and plotted against redshift. This is presented in Fig. \ref{Fig:EvoReffvsLum}. In it we show the best-fit \Reff value of the aforementioned relation at \mb=-21 mag as a function of redshift, in units of the corresponding value at z$\sim$0 (Local sample). The values taken from the best-fit lines are presented by empty diamonds, while the filled circles correspond to median values. The long gray error bars show the standard deviations $\sigma$, while the short black bars give the corresponding errors ($\sigma/\sqrt{N}$, where $N$ is the number of galaxies in the subsample). We see how the median values and the best-fit values are in fair agreement. The best-fit points have been fitted to a line (continuous) which is forced to pass through the value at z$=$0. From this fit we obtain an increase by a factor 1.48$\pm$0.09 in \Reff(\NUVrest) at \mb=-21 mag between $z=$1 and $z=$0. This analysis on the \Brest images yields a very similar growth factor (1.42$\pm$0.07). These numbers are not significantly altered when the images are degraded to the common resolution (\ResFIFTY). If the median values are used, with the original resolution, the growth factor in \NUVrest is somewhat lesser (1.35$\pm$0.07), and the same applies to \Brest, with a factor 1.26$\pm$0.05, though these are still compatible with the best-fit values given the error bar. Again, very similar values are obtained if equal resolution images are used (\ResFIFTY). For comparison, Trujillo \& Aguerri (\cite{Trujillo04}) found that galaxies with $M_V <$ -18.5 show an increase in their $V$-band \Reff by a factor 1.4$\pm$0.2 since $z\sim$0.7. Barden et al. \cite{Barden05} reported an increase in the same parameter by a factor 1.6$\pm$0.1 since $z\sim$1 (for galaxies with $M_V<$-20 mag).

\subsection{\Reff - Stellar Mass Relation}

In the time interval between z$\sim$1 and z$\sim$0, disc galaxies have undergone a significant, overall decrease in \B-band luminosity (e.g. Rudnick et al. \cite{Rudnick03}, Barden et al. \cite{Barden05}), which is parallel to, and a consequence of the also reported decline in global SFR described in the Introduction. This situation adds some difficulty to the interpretation of figures such as \ref{Fig:ReffvsLum} and \ref{Fig:EvoReffvsLum}. This is so because the observed evolution, with galaxies of a given luminosity having larger \Reff's at lower redshifts, has different interpretations depending on the relative contribution from changes in both parameters, \Reff and \mb. For example, in Fig. \ref{Fig:ReffvsLum} we can make the best-fit line for the far-z sample better match the Local sample either by moving it upwards, as would happen if galaxies grew in \Reff but not in \mb, or shifting it leftwards, which would imply a pure decrease in \mb with time, but no change in the \Reff of galaxies.

\begin{figure}
\centering
\includegraphics[width=8cm]{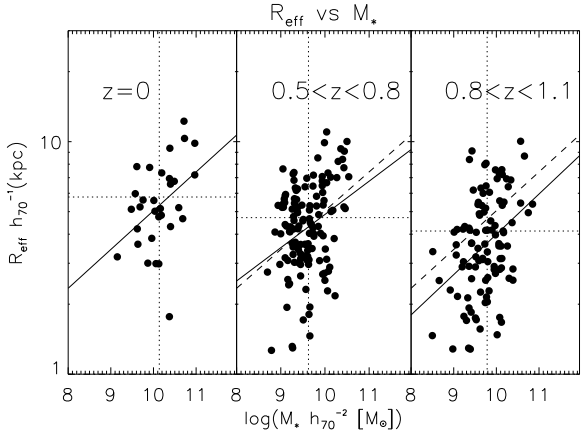}
   \caption{Effective radius of the rest-frame NUV flux distribution against the stellar mass (\mstar) for galaxies in the three redshift bins considered. This figure is analogous, in scheme and meaning of line styles, to Fig. \ref{Fig:ReffvsLum}.}
      \label{Fig:ReffvsMass}
\end{figure}

To overcome the posed problem, we explore the relation between \Reff and stellar mass, \mstar. Working in terms of stellar mass is useful because it removes the evolution that is simply due to the ageing of the stellar populations. Here we recall that we want to test for evolution in the spatial distribution of SF in disc galaxies in relation to the process of stellar mass build-up, and so it makes sense to compare the radial extent of the \NUVrest emission of galaxies with the same mass, at different redshifts. In Fig. \ref{Fig:ReffvsMass} we present the results of a test of this kind. There we can see the \Reff (kpc) in \NUVrest against the \mstar (\msun), for galaxies in different redshift bins. Here we also see a significant dispersion in the relation, as in Fig. \ref{Fig:ReffvsLum}. The dotted vertical and horizontal lines give median values in the represented parameters for galaxies within $10^9<$\mstar$<10^{11}$ \msun.

To the reported dispersion in the relations (\Reff-\mb and \Reff-\mstar), in addition to their intrinsic origin, contribute the uncertainties in the distances to the objects, those in \mstar and \mb, and also, of some importance, the patchy distribution of \NUV flux, dominated by conspicuous star forming regions. 

\begin{figure}
\centering
\includegraphics[width=8cm]{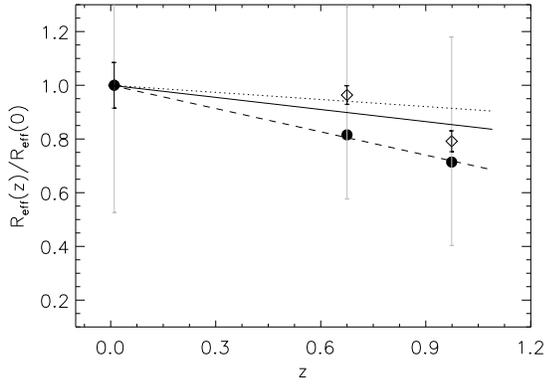}
   \caption{Evolution of the relation \Reff(\NUVrest) - \mstar (see Fig. \ref{Fig:ReffvsMass}). From the linear fits to this relation at \mstar$=10^{10}$ \msun (empty diamonds, continuous line), an increase in \Reff by 1.18$\pm$0.07 between $z=$1 and $z=$0 is deduced. If, instead, the median values of \Reff are used (filled dots, dashed line), the corresponding factor is 1.41$\pm$0.09. The dotted line shows the evolution for the \Reff(\Brest) - \mstar relation (1.10$\pm$0.05), from best-fit values. The long-gray error bars give the standard-deviation ($\sigma$) in \Reff($z$)/\Reff($z=$0), while the short-black give the associated error ($\sigma/\sqrt(N)$, where $N$ is the number of galaxies in the subsample.}
      \label{Fig:EvoReffvsMass}
\end{figure}

To test for evolution in the \Reff-\mstar relation, here we choose to compare the best-fit values at \mstar $= 10^{10}$ \msun, and the result can be seen in Fig. \ref{Fig:EvoReffvsMass}. As in Fig. \ref{Fig:EvoReffvsLum} here we present the ratio \Reff($z$)/\Reff($z$$=$0) (best-fit values: empty diamonds) against redshift. From a linear fit (continuous line), we deduce an increase by a factor 1.18$\pm$0.06 in \Reff(\NUVrest) at \mstar $=$ $10^{10}$ \msun since $z=$1; i.e., a moderate change ($\lesssim$20\%) in that quantity in the given period. If we measure the evolution in this relation by the median values of \Reff (filled dots, dashed line), the change is more significant (1.4$\pm$0.1), but this must be taken with caution, as the samples have slightly different median values of stellar mass, and \Reff correlates with this parameter.

Also, in Fig. \ref{Fig:EvoReffvsMass} we show, as a dotted line, the evolution in the relation between the effective radius in \Brest and stellar mass. The corresponding evolution is an increase by a factor 1.10$\pm$0.05 between $z=$1 and $z=$0, i.e. slightly less evolution than in \NUV band. For comparison, Barden et al. (\cite{Barden05}) reported practically no evolution in \Reff in rest-frame $V$ band for a fixed stellar mass between $z\sim$1 and $z\sim$0, in agreement with our result within $\sim2\sigma$. These figures also remain virtually unchanged when images of a same resolution (\ResFIFTY) are used.

\subsection{Radial concentration and central emission of \NUV-Flux}\label{Sec:Concentration}

\begin{figure}
\centering
\includegraphics[width=8cm]{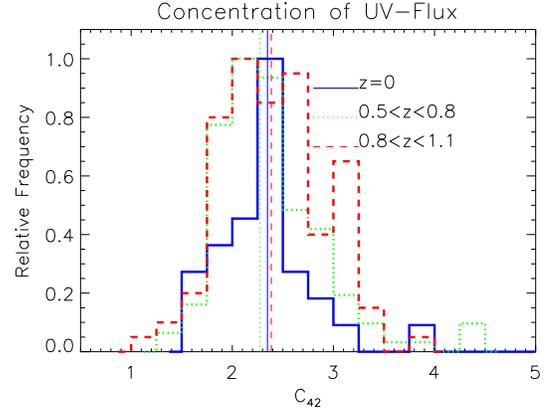}
   \caption{Concentration of \NUV flux for galaxies in different redshift bins. \C is defined as $5 \cdot log(R_{80}/R_{20})$, where \Reighty and \Rtwenty are the radii which enclose 80\% and 20\% of the total flux of the object respectively. For the three redshift bins the median value of \C is around 2.3-2.4}
      \label{Fig:HisConc}
\end{figure}

So far we have presented results on the size of the galaxies as seen in \NUVrest, and parameterized by the effective radii. In a step forward to characterize the SF distribution we have explored the radial concentration of the \NUV emission in disc galaxies. This gives a measure of how relevant is the SF in the central regions of the galaxies in relation to that in the rest of the discs. In Fig. \ref{Fig:HisConc} we show the distributions of the concentration parameter \C in \NUVrest band for samples in different redshift bins. The \C parameter (Kent \cite{Kent85}) is defined as $5\cdot log(R_{80}/R_{20})$, where \Reighty and \Rtwenty are the radii which enclose 80\% and 20\% of the total flux of the object respectively. We find that the median values of \C are around 2.3-2.4, and show no evolution with redshift within the error bars (Local: 2.35$\pm$0.13, \midz: 2.28$\pm$0.05, \farz: 2.39$\pm$0.05). The minor differences amongst redshift bins remain minor when we analyse images at equal resolution, and are again not significant, given the error bars.

But, what does this \C parameter really mean? It is larger for those distributions of light which are spread over a larger area but at the same time have a larger fraction of the total flux accumulated near its centre. It may be better understood with some examples: for a disc with a flat surface brightness profile, \C$=$1.51, for an exponential disc it is \C$=$ 2.75, and for a de Vaucouleurs profile (a S\'ersic with n$=$4), it is \C=5.16 (Kent \cite{Kent85}). So, the \NUV profiles are between flat and exponential, being more similar to the latter, and their concentration does not seem to evolve in this period, at least by this parameter.

\begin{figure}
\centering
\includegraphics[width=8cm]{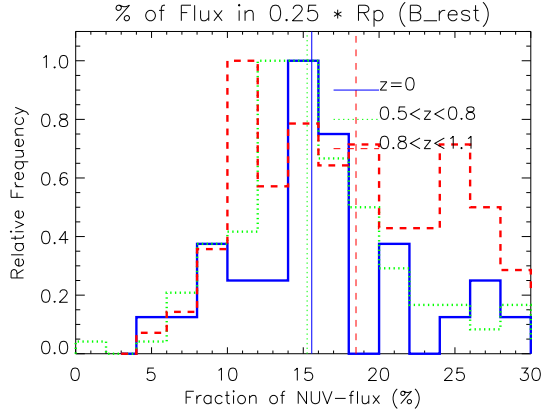}
   \caption{Fraction of \NUV flux contained within $0.25 R_P$ ($R_P$ is the Petrosian radius in \Brest), for different redshift bins.}
\label{Fig:HisPetroFracNUV}
\end{figure}

The previous measure of \NUV concentration depends only on the radial distribution of the emission in that band. But we also want to test how the SF is taking place in relation to the older stellar populations observed in the rest-frame \B-band. In a first test, we have compared the fraction of the total \NUV flux contained in an aperture with radius $R=$\Rp/4, (\Rp is the Petrosian radius, measured in the \Brest band), for galaxies in the three redshift bins under consideration, and the results can be seen in Fig. \ref{Fig:HisPetroFracNUV}. The median values of the fractions are 15.6$\pm$1.3\% for the Local sample, 15.3$\pm$0.7\% for the mid-z subsample (\midz) and 18.5$\pm$0.8\% for the far-z subsample (\farz). If all of the \NUV flux were contained in the Petrosian radius, and it were distributed uniformly in the area defined by that radius, the resulting fraction would be 6.3\%, and so the results reflect a significant enhancement of the \NUV emission in the central parts of the discs, which was expected from visual inspection of the images (also see Fig. \ref{Fig:MedProfs}). Also noteworthy is that the given measure of central concentration of \NUV flux remains virtually unchanged between z$\sim$0 and z$\sim$1, given the error bars. A K-S test shows that the null hypothesis can not be rejected with a confidence larger than $\sim$98.5\% in the best case, for the Local and far-z samples. The fractions obtained with images of uniform resolution (\ResFIFTY) are not significantly different from the given.

We have also performed this test with a smaller fraction of the Petrosian radius, \Rp, to perform this test, \Rp/10, and no significant difference amongst redshift bins is found either. For z$\sim$0 the fraction is 3.6$\pm$0.9\%; for \midz, 3.3$\pm$0.3\%, and for \farz, 3.6$\pm$0.2\%. At equal resolutions (\ResFIFTY), the given fractions do not show any clear trend with redshift either: 3.1$\pm$0.5\% at z$\sim$0 versus 2.6$\pm$0.1\% for \midz and 3.0$\pm$0.2 for \farz.

\begin{figure}
\centering
\includegraphics[width=8cm]{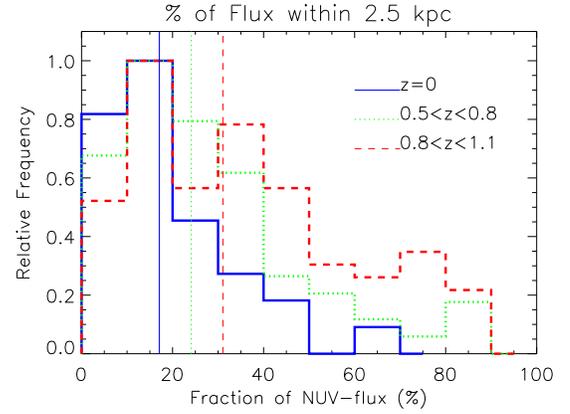}
   \caption{Fraction of \NUV flux within R$<$2.5 kpc, for different redshift bins.}
\label{Fig:HisRkpcFracNUV}
\end{figure}

The preceding measure of the fraction of \NUV flux contained in the ``central'' part of the galaxies is referred to the size of the galaxies in the \B band, through the Petrosian radius. The mean values of \Rp are $\sim$11 kpc for the Local sample, and $\sim$8 kpc for the High-z sample. So, given that these differences exist, it is also interesting to define that ``central'' region without referring to any measure of the size of the objects whatsoever. To accomplish this, we also present results on the fraction of \NUV flux contained within R$<$2.5 kpc, and this may be seen in Fig. \ref{Fig:HisRkpcFracNUV}. Galaxies at z$\sim$1 have a fraction of flux within \R$<$2.5 kpc which is a factor $\lesssim$2 larger than that at z$\sim$0 (17.1$\pm$2.5\% for z$\sim$0, 24.1$\pm$1.8\% for \midz and 31.1$\pm$2.2\% for \farz). Applying a K-S test, the null hypothesis for the local and $z\sim1$ samples can be discarded with a confidence $>$99.9\%. These fractions are practically unaltered at equal resolution, and what we see here is that galaxies in the past had a larger fraction of \NUV flux within a fixed radius ($R<$2.5 kpc), than nowadays. The difference is analogous, though reduced, if instead of $R<$2.5 kpc we use $R<$1 kpc, with a fraction of $\sim$4\% at $z\sim$0 and $\sim$6\% at $z\sim$1. But, given the error bars, the differences are not significant in that case.

\begin{figure}
\centering
\includegraphics[width=8cm]{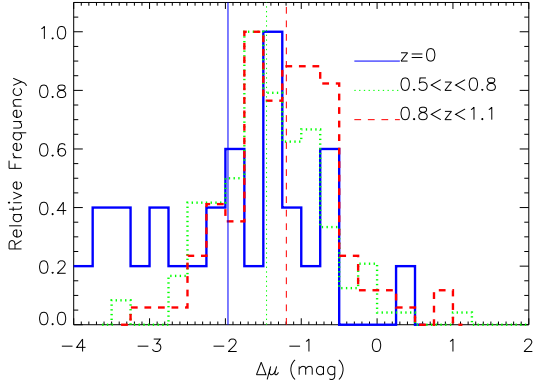}
   \caption{Difference in rest-frame \NUV surface brightness, $\Delta\mu$, between $R=$0 and $R=$\Reff ($\Delta\mu = \mu(R=0) - \mu(R=R_{eff})$), for different redshift bins. Median values of $\Delta\mu$ are -2.0$\pm$0.2 mag ($z\sim$0), -1.5$\pm$0.1 mag (\midz) and -1.2$\pm$0.1 mag (\farz).}
\label{Fig:HisMusNUV}
\end{figure}

Another interesting question, and also related to the previous one, is how intense is the \NUV emission in the central parts of the galaxies. We have chosen to give this measure in relation to the brightness at the effective radius, \Reff. In Fig. \ref{Fig:HisMusNUV} we present the difference in surface brightness $\Delta\mu$, measured in rest-frame \NUV, for the explored range of redshifts, and defined as $\Delta\mu=\mu(R=0)-\mu(R=R_{eff})$. There we can see that this difference is more significant today than in the past, with median values of -2.0$\pm$0.2 ($z\sim$0), -1.5$\pm$0.1 (\midz) and -1.2$\pm$0.1 \magarcsq (\farz). This is equivalent to an increase of brightness in the centre of the galaxy relative to the level at \Reff, from a factor of $\sim$3 at $z\sim$1 to $\sim$6 at $z\sim$0. In the \B-band these differences are even more significant, with values of $\Delta\mu$ of -3.1$\pm$0.2 ($z\sim$0), -1.7$\pm$0.1 (\midz) and -1.5$\pm$0.1 \magarcsq (\farz), with an increase in relative brightness from a factor of $\sim$4 at $z\sim$1 to $\lesssim$20 at $z\sim$0. For both bands, the hypothesis that the local and $z\sim$1 samples are a realisation of the same parent distribution can be discarded with probabilities higher than 99.9\%.

If images of a same resolution (\ResFIFTY) are used, the sense of these differences holds, though the values of $\Delta\mu$ are somewhat lessened, and more similar to each other, as the central emission is smeared over a larger area when the images are degraded by convolution. This smearing effect is more prominent on images in the Local sample, as the differences in spatial scale ($\Gamma$, \FWHMkpc) are larger within it. Under equal resolution, the values of $\Delta\mu$ in \NUVrest still show evolution ($\Delta\mu(NUV_{rest})=$-1.17$\pm$0.12 at $z\sim$0, -1.11$\pm$0.05 at $z\sim$0.7 and -1.05$\pm$0.05 \magarcsq at $z\sim$1), although is less dramatic, and not significant enough under a K-S test (confidence level at 81.4\%). In contrast, in \Brest they still show significant evolution, the samples being drawn from different distributions with a confidence higher than 99.9\% ($\Delta\mu(B)=$-1.85$\pm$0.08 at $z\sim$0 versus -1.21$\pm$0.03 at $z\sim$1).

These results suggest that the slopes of the profiles, at least in the \B band, have become steeper over time, or being more precise, that the emission in the centres of the galaxies has greater contrast with respect to that in the discs. In fact, from the inspection of Fig. \ref{Fig:MedProfs} (left and central panels) it seems that the latter interpretation seems more adequate. There it can be seen that the slopes of the averaged profiles in \B do not seem to vary significantly between z$\sim$1 and z$\sim$0, but there is an abrupt increase in the emission towards the inner parts of the profiles for galaxies at z$\sim$0, in contrast to the profiles at larger redshifts, which are flatter at smaller radii.

The preceding results may seem, prima facie, and to some extent, contradictory. First we have that the concentration (\C) of the \NUV emission does not seem to evolve in this range of redshifts. The same may be said about the fraction of flux inside a certain fraction of the Petrosian radius, \Rp (i.e. \Rp/4), for which if there is any evolution at all it is minor. We have also seen that the difference in brightness between the central parts of the galaxies and the disc (at \Reff), is somewhat larger nowadays than in the past. But, and here is the apparent contradiction, the fraction of flux inside a fixed radius (e.g. \R$<$2.5 kpc) is larger at $z\sim$1 than at $z\sim$0 (this also applies, though not so significantly, if the central region is defined as that within \R$<$1 kpc). The key to solve this ``paradox'' is that the first two measures of concentration are referred to the distribution of flux itself (in rest-frame \NUV or \B), and thus they are not affected by the fact that the objects are somewhat more compact (smaller) in the past, which the last measure shows as evident. Or as an alternative way to explain this, it seems that the centre of the galaxies has become slightly brighter in \NUV in relation to the discs with time, but at higher redshift, the overall \NUV emission, which had a somewhat flatter profile, was also more compact (see also Fig. \ref{Fig:HisReff} and comments on this).

Now it is useful to look again at Fig. \ref{Fig:MedProfs}, to better understand these results. There we can see how the averaged \NUV profiles (blue squares) for the Local sample show enhanced central emission in comparison with the profiles for the High-z sample. The radial scale is relative to the effective radii, and so, in that figure, the differences in compactness (i.e. size) of the objects are not evident. It is also important to note that the \B profiles (red circles) also show enhanced central emission in the Local sample. This seems to be related to the larger contribution of central bulges or pseudo-bulges to the profiles. So, it is doubtful that the slight increase in central \NUV emission can be attributed to an enhancement of nuclear SF with respect to past times. Most probably this central \NUV emission is due to intermediate age stars (with ages $>$100 Myr), at least in part. But a detailed analysis of the spectral energy distributions and stellar populations, in the central regions and in the discs, is necessary to solve this uncertainty. It must also be noted that the central excess of \NUV brightness is roughly at the same surface brightness level at z$\sim$1 and at z$\sim$0, and in \B the central emission at z$\sim$0 is only slightly above the level at z$\sim$1. At the same time, the surface brightness of the discs has decreased significantly in both bands. So, it could be that the bulges/pseudo-bulges have ``outshone'' the discs over time, but, alternatively, it could also be that, as the discs have gradually faded, because of declining star-forming activity, the bulges have been left ``exposed''. At this point it is not possible to settle this dilemma. Nonetheless, we present some evidence in section \ref{Sec:SigmaProfs} indicating that the second interpretation may be closer to reality.

\subsection{Stellar Disc Truncations}

\begin{figure}
\centering
\includegraphics[width=8cm]{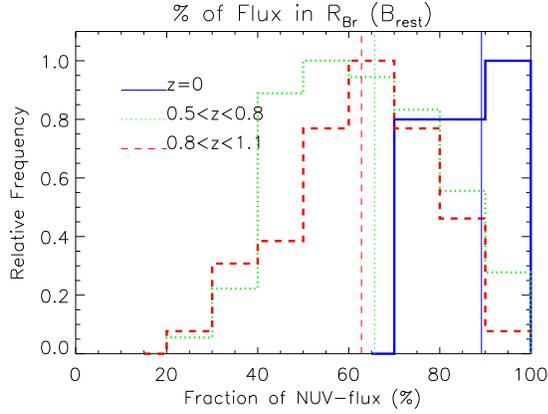}
   \caption{Fraction of \NUV flux within the ``break'' radius (measured in \Brest) for galaxies in the Local and High-z samples. The classification of profiles and estimates of \Rbr for objects in the High-z sample is taken from Azzollini et al. \cite{ATB08b}.}
      \label{Fig:HisRTrFracNUV}
\end{figure}

The galaxies in the High-z sample are taken from the survey on disc truncations presented in Azzollini et al. (\cite{ATB08b}). Thus, we have estimates of the ``break'' radii, \Rbr, for those galaxies common to both samples which have truncated discs (137 galaxies). In the Local sample we find 13 truncations, and we have estimated their break radii. In both ranges of redshift, the band used to characterize the profiles is the rest-frame \B-band. It is interesting to investigate the relation between the extension of the \NUV emission ($\sim$SF), in relation to \Rbr. This is because \Rbr is thought to be related to a threshold in the SF of the galaxies. We present results on this in Fig. \ref{Fig:HisRTrFracNUV}. In it we can see the fraction of \NUV emission contained in an aperture with radius equal to \Rbr, for galaxies in the Local sample, and mid-z (\midz) and far-z (\farz) subsamples. The distributions within the High-z sample ($z\sim$1) are very similar, with median values of 66$\pm$2\% (\midz) and 63$\pm$2\% (\farz). In the Local sample the fraction is significantly higher, with a median of 89$\pm$2\%. The histograms of the mid-z and far-z subsamples reflect significant scatter, and it is interesting that the distributions extend downwards to $\sim$20\%, implying that there are discs in which most part of the \NUV flux outside the break radius, contrary to what happens for the histogram at $z\sim$0. Here, again, using images with equal resolution does not have significant consequences for the results. 

So, it seems that local truncations include most of the \NUV flux, and hence, presumably also the star-forming activity. This fits into a scheme for the formation of truncations in which the suppression of SF outside the break plays a major role. Nonetheless, the fraction of \NUV flux within \Rbr at $z\sim$1, though high, seems to challenge this picture, if we take the emission in that band as a tracer of on-going SF. But several things must be said before reaching any conclusion. First, it is known that the profile classifications are somewhat less reliable at higher redshfit due to poorer signal-to-noise ratios. For example, in ATB08b, and from simulations of artificial galaxies, it was estimated that $\sim$30\% Type I galaxies of magnitude \zz$\sim$23 are wrongly classified as type II. Second, the \NUV flux is not a perfect tracer of SF, as has been said, and stars with ages of a few $\times10^2$ Myr also contribute to the emission at those wavelengths. So, stars recently formed inside the break which have migrated outwards, as in the models by \Roskar et al. (\cite{Roskar08a}, \cite{Roskar08b}), could also contribute to the \NUV emission outside the break. Nonetheless, it could also be possible that there might be somewhat more SF outside the breaks at higher redshift than in the local universe perhaps as a consequence of a more important merger activity in past times, and this could explain the reported difference, or part of it.

\subsection{The Attenuating Effect of Dust in the \B and \NUV bands}\label{Sec:Dust}

In the preceding sections we have presented results on the radial distribution of \NUV flux, analysing several parameters that characterize these distributions and also median surface brightness and colour profiles. We are ultimately interested in the spatial distribution of the SF, but as we have said, the effect of dust may play a significant role in the shaping of the \NUV profiles, and so hinders an interpretation of the given results in terms of SF. Here we show some tests we have performed in order to estimate the effects of dust attenuation on the results.

In Boissier et al. (\cite{Boissier04}), an analysis of the radial FIR to FUV ratio in a sample of 6 local disc galaxies was used to provide a recipe for dust attenuation in the UV (at 2000 \AA) with radius, as a function of the \B-band luminosity of the galaxy, and its inclination (see equations 5, 6 and 10 in their paper). Assuming the dependence of attenuation on wavelength ($k_{\lambda}$) given by Calzetti et al. (\cite{Calzetti00}), we can get an estimate of the extinction as a function of radius in the bands explored (\NUVrest and \Brest), $A_{band}(r)$. The resulting extinction profile significantly obscures the central parts of the galaxies ($A_{NUV}(0)\sim$2.5 mag), while it becomes a minor effect beyond $R\sim$2\Reff ($A_{NUV}(R>2 R_{eff})<$0.4 mag). 

\begin{figure*}
\centering          

\includegraphics[width=7cm]{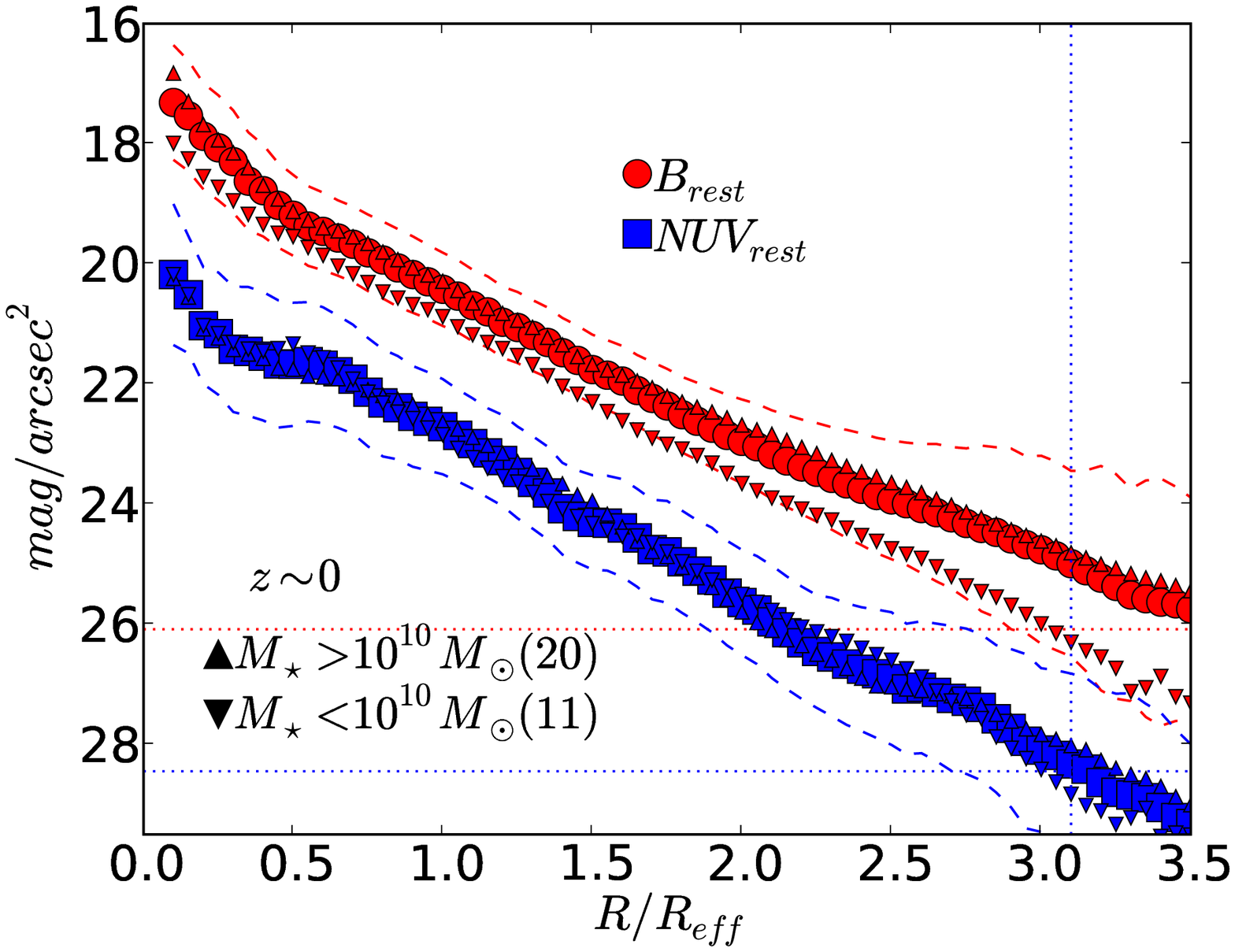} 
\includegraphics[width=7cm]{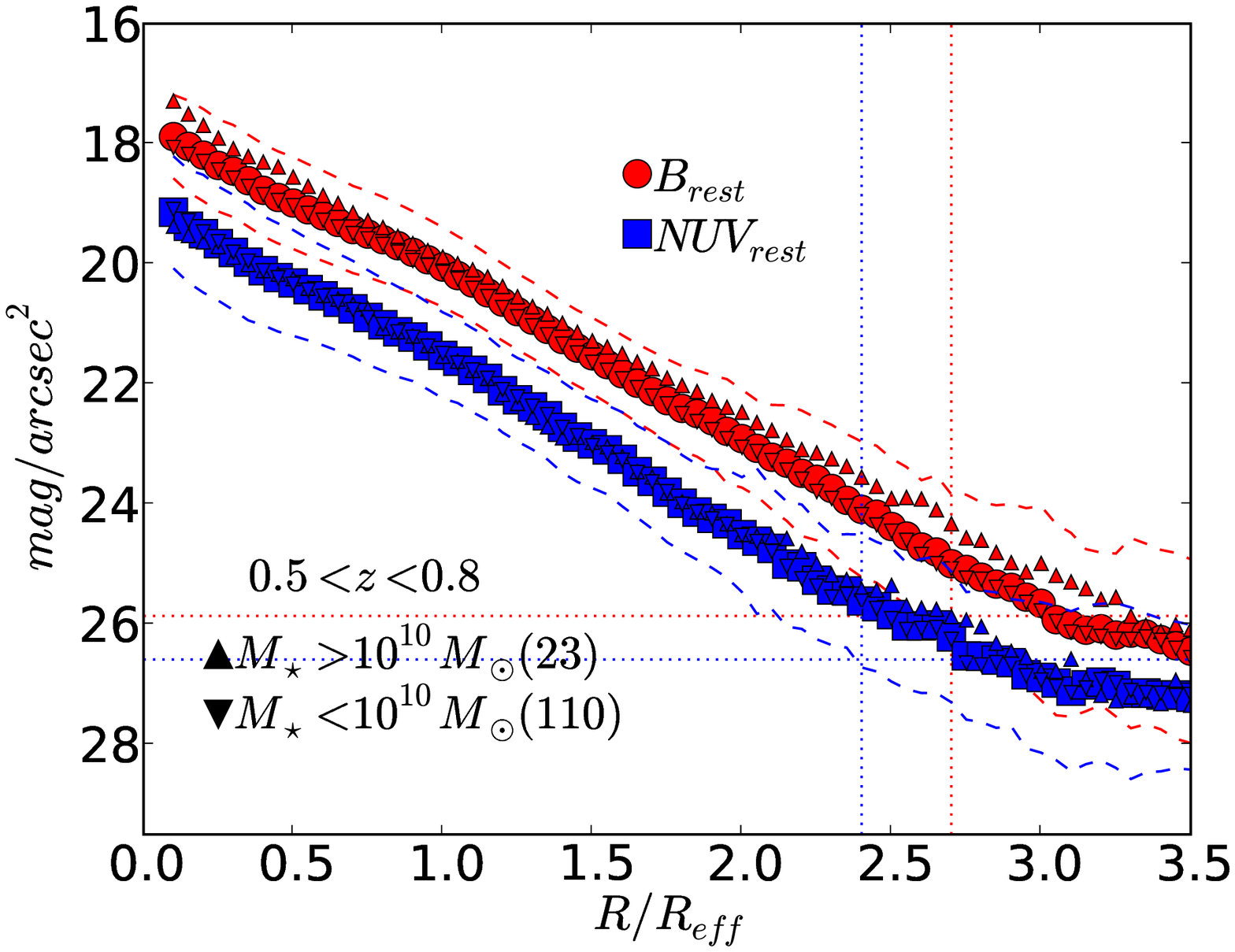}\\
\includegraphics[width=7cm]{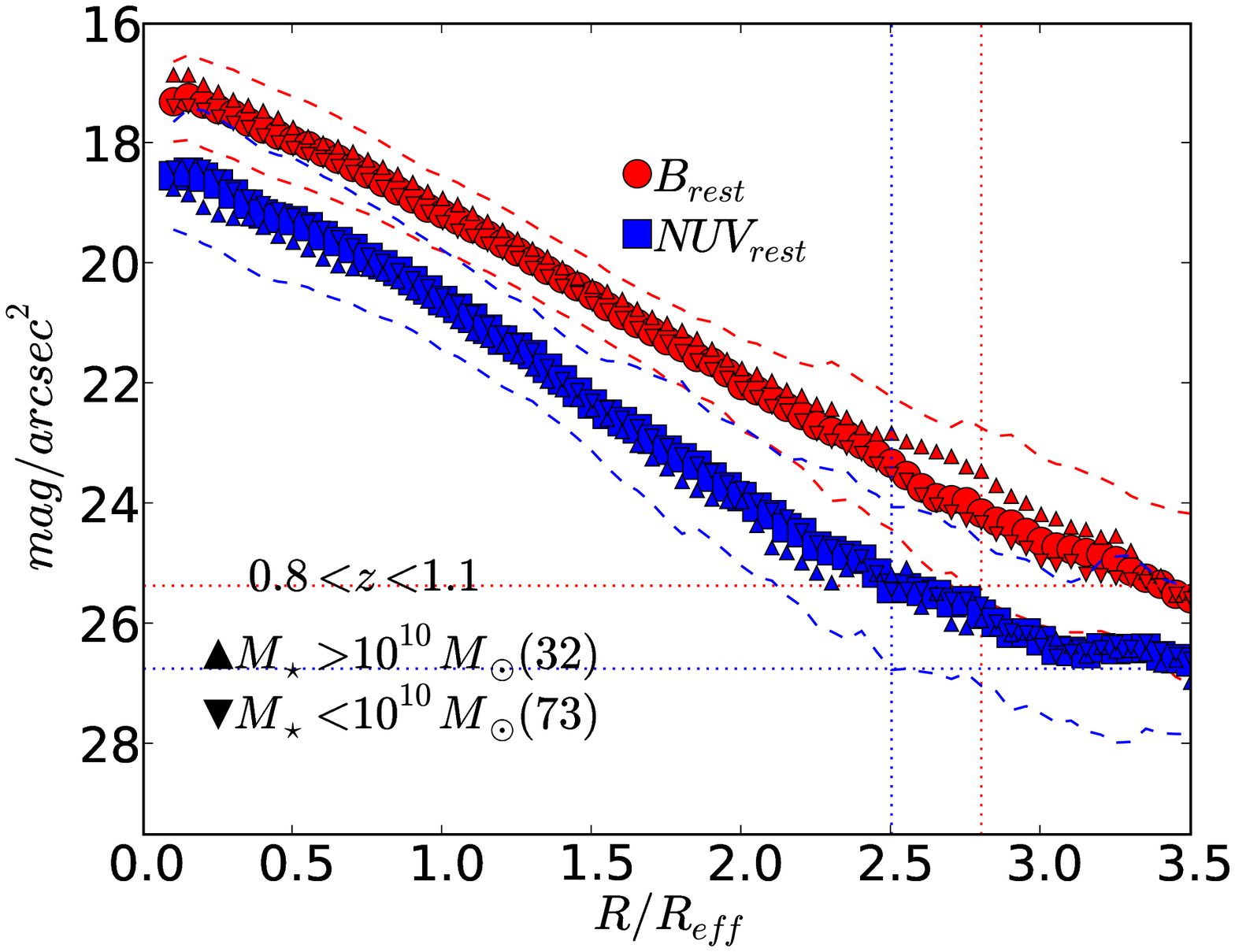}
\includegraphics[width=7cm]{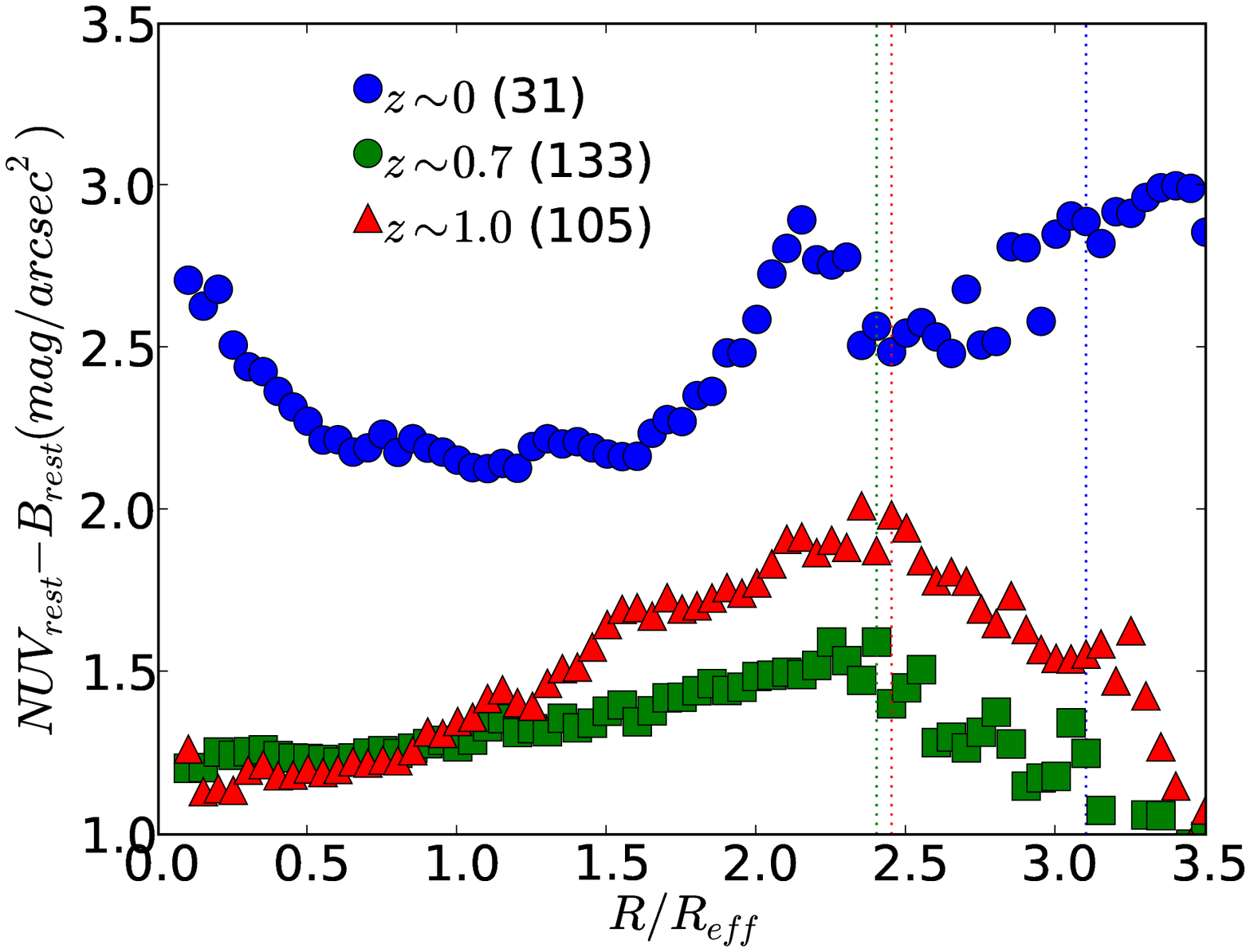} \\

\caption{This figure is analog to Fig. \ref{Fig:MedProfs} in all but the radial curve of extinction provided by Boissier et al. (2004) has been substracted from the median profiles. This also applies to the colour profile. The reliability of these profiles, particularly at intermediate $z$ should be considered with caution. See the text for details.}
\label{Fig:MedProfsND}
\end{figure*}

In Fig. \ref{Fig:MedProfsND} we show the median surface-brightness and colour profiles of Fig. \ref{Fig:MedProfs}, after allowing for the dust extinction in the \NUVrest and \Brest bands. We have used median values of inclination and absolute \B-band magnitude for each sample to compute the extinction profiles and these have been subtracted from the stacked profiles. We can appreciate how the \mur profiles become steeper in the inner parts after this correction. The previously noted hints of profile downbending in the \NUVrest profiles are washed out in the Local sample, and decreased significantly for the higher redshift samples. The colour profile for the Local sample still has a valley-like appearance, though for the mid-z and far-z sample this shape has disappeared, and instead we get rather flat colour profiles (out to the fiducial radius), with a positive slope (i.e. the median profiles become redder to the outer parts of the galaxies). The colour profile of the Local sample is now not so red in the central parts, as it was before the correction. 

Can we trust these extinction corrected profiles? Before interpreting these figures, it must be emphasized, though, that the extinction curve by Boisser et al. was derived from the analysis of a very small sample of local galaxies. So the reliability of the applied correction is even less at intermediate redshifts than it is in for the Local sample, as there might be evolutionary changes in the radial distribution of dust attenuation in this time interval. This is probably the case as as the SF distribution is quite different at intermediate redsdhifts than currently, and consequently the dust distribution is also likely to be different. Nonetheless, this result indicates that in the progressive reddening of the inner parts of the profile since $z\sim$1, the ageing of the stellar populations must play a major role.

Also in Fig. \ref{Fig:MedProfsND}, we can see how the rise of the central parts of the \mur profiles at lower redshift, in comparison to the profiles at higher redshift bins, still holds. We have not modelled any possible evolution in the radial distribution of dust extinction, and so this difference was not expected to be altered after this correction. Only a detailed analysis of the dust extinction profile in each galaxy, from the modelling of the SEDs of the galaxies, could provide a more accurate picture.

But we have also explored the net effect of dust on the effective radius and other parameters of the radial profiles commented in preceding sections. With this aim, we have taken the median profiles in \Brest and \NUVrest for the Local sample, subtracted the corresponding radial curves of extinction (Boissier et al. \cite{Boissier04}), and measured the parameters in both sets of profiles, those corrected and not corrected from extinction, for comparison.

The effective radii seem to be significantly affected by dust attenuation, and after dust correction, these radii decrease, in \Brest and \NUVrest, to 61\% and 64\% of the non-corrected values, respectively. This is a direct consequence of the assumed radial extinction curve, which implies higher absorption in the central parts of the galaxies. However, the differential effect that dust has on \Reff in both bands could only account for part ($\lesssim$6\%) of the measured difference between the effective radii, by which the \Reff in \NUVrest is $\sim$10\% larger than in \Brest.

The concentrations (\C) are also affected by the dust correction. The dust corrected value in \NUVrest, is \C$=$2.52, versus 2.25 without correction. The new value is  closer to the value for an exponential profile, but still below it (\C$=$2.75). Perhaps more interesting are the results with regard to the fraction of flux contained in a given fraction of the Petrosian radius or in a fixed radius. On one hand, the former is practically unaffected, the fraction of \NUV flux contained within \Rp/4 being 19.7\% and 21.6\% before and after the dust correction. This could be anticipated, as the extinction curves are very similar in shape between these two bands, as made evident by the little differential effect they have on the effective radii, commented above. On the other hand, the fraction of \NUV flux within a fixed radius (2.5 kpc) does show a significant difference, being 20.6\% before the correction, and 41.1\% after it. This is so because the profile is particularly attenuated in the inner parts of the galaxies.

The difference in surface brightness between the centre and \Reff is also affected by the dust correction. Without dust correction, $\Delta\mu(NUV)=$-1.5 \magarcsq, and after correction it becomes -1.7 \magarcsq (increasing). In contrast, in \B band, it goes from -2.4 to -2.2 \magarcsq (decreasing). This difference in the sense of the variation is also due to a larger dust absorption in \NUV than in \B (the reader must bear in mind that the definition of $\Delta\mu$(band) contains \Reff(band), which is also affected by the dust correction).

We have gone one step further, and performed the same analysis described here, this is, measuring the parameters studied before and after dust correction, but using the median \NUV and \B profiles at $z\sim$1 (far-z sample). Combining these results with those obtained using the $z\sim$0 profiles, it is possible to estimate the contribution that dust absorption makes to the observed variation of these parameters in this time interval. These net effects are: 1) \Reff in \B increases by 17\% since $z\sim$1 due solely to the effect of dust. This value is similar to the evolution found (an increase by 10\% at a fixed mass), and so perhaps the reported evolution could be mostly due to dust. 2) In contrast, dust does not seem to have a net effect on the evolution of \Reff(\NUV), and thus the reported evolution (an increase by $\sim$20\%), seems to reflect a change in the distribution of young stars in the galaxies. 3) As a consequence of 1) and 2), without the effect of dust, the ratio \Reff(\NUVrest)/\Reff(\Brest) would have increased by $\sim$25\% since $z\sim$1, above the reported increase by 10\%. 4) The Concentration, \C, in \NUVrest seems to decrease by 0.21 points since $z\sim$1. Taking this into account, \C would have increased slightly since $z\sim$1, were it not for the dust effect (the direct measure did not show a clear evolution of this parameter). 5) The dust makes the fraction of flux inside the central 2.5 kpc decrease by 12\%. The reported evolution is a decrease by $\sim$55\%, and so the evolution of the fraction of SF would be somewhat lower, closer to 40\%. 6) The fraction of flux inside \Rp/4 increases by 17\% due to dust, and so, as we measure no evolution in this parameter, this suggests that the fraction of SF inside that radius has actually decreased, since $z\sim$1, by that relative amount. 7) The net effect of dust on the difference in brightness between centre and \Reff is not big in either band. The variations attributable to dust would be -0.08 \magarcsq and +0.12 \magarcsq in \Brest and \NUVrest since $z\sim$1. Accounting for these trends, the real evolution of $\Delta\mu$ in \Brest and \NUVrest would be, respectively, slightly less and slightly more significant than those found.

After these tests, we  conclude that the effect of dust absorption on the profiles is significant, in absolute terms, for most of the studied parameters, especially for the effective radii, the exception being the fraction of \NUV flux inside a fraction of \Rp. Nonetheless, being short of a prescription for the evolution in the radial distribution of dust, we cannot estimate in a reliable way the effect of the dust absorption on the evolution of the explored parameters (e.g. \Reff(\NUV)). Assuming that the extinction law for the local galaxies (from Boissier et al. \cite{Boissier04}) also applies at $z\sim$1, which is at the very least dubious, it seems that the observed increase of \Reff(\NUV)/\Reff(\Brest) with time is a lower bound to the underlying evolution, once dust absorption is accounted for. On the other hand, the progressive decrease in the fraction of SF within the central 2.5 kpc would not be so large as direct measurement suggests, though still significant. Finally, there are two evolutionary trends whose reliability is reinforced after these tests. First, the central parts of the galaxies are reddenning largely due to ageing of the stellar populations (becoming brighter in \B with respect to \NUV). And second, the effective radius of the SF seems to be increasing with time. Instead of the last assertion, perhaps the reader would have expected the more straightforward ``the SF is migrating outwards''. But before stating that this is what is actually going on, we should read the next subsection.

\subsubsection{SFR Radial Profiles and their Integral in Time}\label{Sec:SigmaProfs}

Based on the \Brest and \NUVrest profiles presented in Figs. \ref{Fig:MedProfs} and \ref{Fig:MedProfsND}, we have also explored the stellar mass density profiles of our samples of galaxies. In particular, here we compare the estimated present profile of surface mass density with the accumulated mass profile, obtained from the time integral of the SFR radial profiles since $z\sim$1. We rely on the chosen dust correction (Boissier et al. \cite{Boissier04}) to estimate the SFR profiles from the \NUVrest surface brightness profiles, but it must be taken into account that this correction was derived from a reduced sample of local objects and we apply it to objects expanding in a wide range of redshifts. Furthermore, we sum the SFR over time to produce a profile of accumulated stellar mass, and compare it with the current stellar mass profile. But the \NUVrest profiles at intermediate redshifts used do not exactly correspond to the suspected progenitors of the galaxies in the Local sample. We recall that the Local sample has a median value of stellar mass  log(\msun)$=$10.1, while in the High-z sample it is log(\msun)$=$9.6. This difference is above the reported evolution in stellar mass between $z\sim$1 and $z\sim$0 ($\lesssim$50\%). So we warn against basing strong conclusions on this test. Nonetheless, the results are interesting enough to deserve mention.

The local \Brest profile provides the best proxy, available to us, for the radial distribution of mass, and we have converted it to a surface mass density profile ($\Sigma^{M_{\star}}(r)$, in units of $M_{\sun} pc^{-2}$), using Eq. (1) in Bakos et al. (\cite{Bakos08}). The mass to luminosity ratio in \gs band, $(M/L)_{g'}$, is estimated using results in Bell et al. (\cite{Bell03}, see Table 7), and assuming a Salpeter (\cite{Salpeter55}) IMF. We use median rest-frame \gs and \rs  profiles, without correcting for dust, to estimate the surface density mass profile. We proceed in this way because the recipe for $(M/L)_{g'}$ given in Bell et al. does not account for such extinction corrections.

With regard to the profiles of surface SFR density, i.e. $\Sigma^{SFR}(r)$, given in units of $M_{\sun} yr^{-1} pc^{-2}$, we use the \NUVrest profiles within each redshift bin, after dust correction. These are transformed to the sought-after profiles via Eq. (1) in Kennicutt (\cite{Kennicutt98}) (the Salpeter IMF is also assumed for this expression). These surface mass density and surface SFR density profiles are re-scaled into kpc, by means of the median values of \Reff(\Brest) (in kpc) within each redshift bin (see Section 3.3). Then the surface SFR density profiles are linearly interpolated and integrated in time to produce profiles of the stellar mass produced at each radius (in kpc) since $z\sim$1 (accumulated mass). 

\begin{figure}
\centering
\includegraphics[width=6cm]{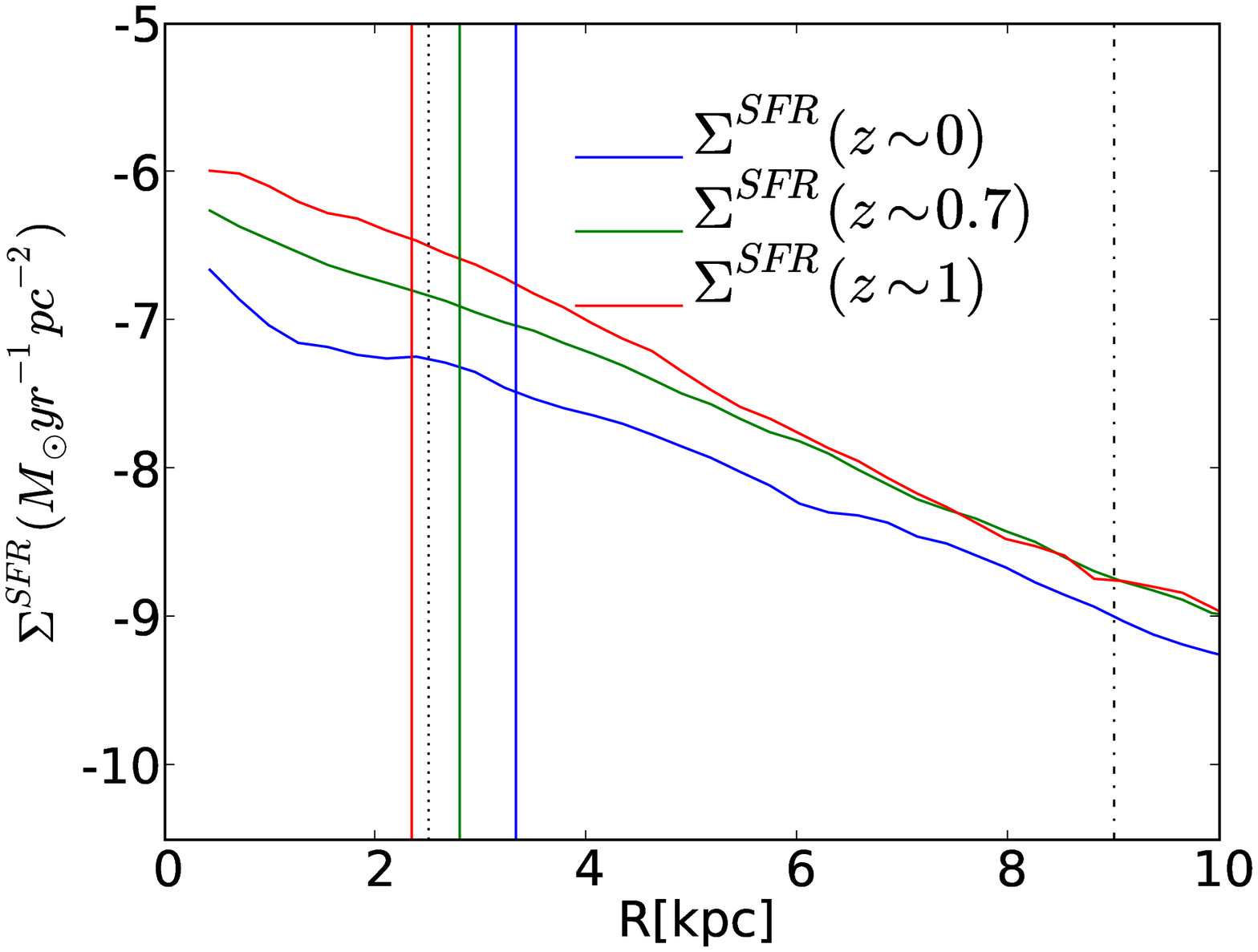}
\includegraphics[width=6cm]{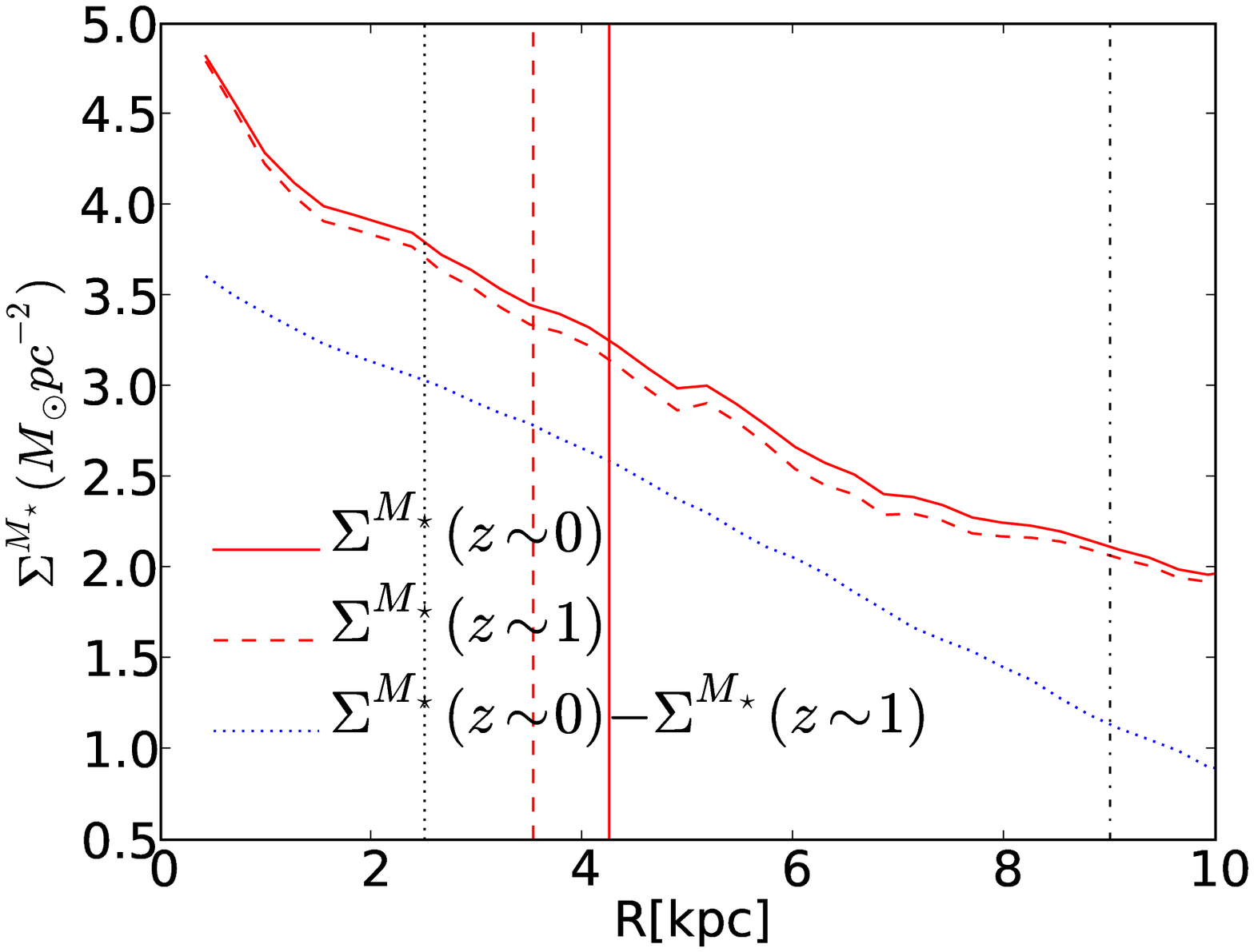}
\includegraphics[width=6cm]{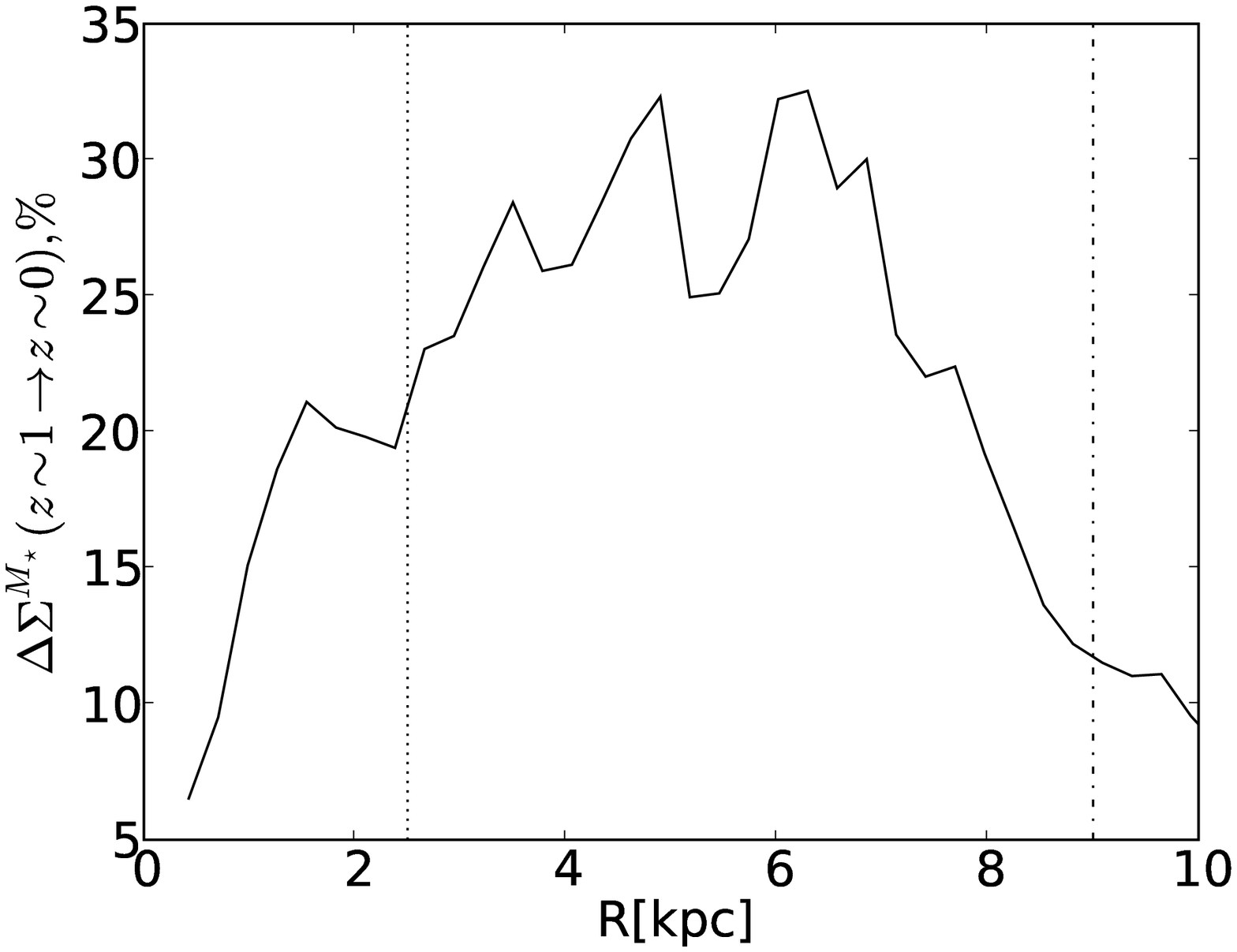}
   \caption{Upper: median SFR surface density profiles at different redshifts, derived from the dust-corrected \NUVrest profiles. The vertical lines are at the the corresponding median values of \Reff(\NUVrest), corrected from the effect of dust. In all panels, the vertical black dotted line marks a 2.5 kpc radius, and the vertical dot-dashed line marks the fiducial radius for the \NUVrest profile at $z\sim$1, which sets the limit for the reliability of the test, because of errors in the estimate of the background. Central: median stellar mass surface density profile at $z\sim$0, derived from the dust-corrected \Brest profile in Fig. \ref{Fig:MedProfsND} (continuous red curve). The blue dotted curve is the integral of the stellar mass born at each radius since $z\sim$1, from the linear interpolation and time integral of the SFR curves in the upper panel. The red dashed line is the result of subtracting this ``accumulated'' mass profile from the present stellar mass profile, i.e. a backwards extrapolation of the present day mass profile to approximate that at $z\sim$1. Lower: Relative increase in stellar mass since $z\sim$1 as a function of radius, from the ratio of the current stellar mass density profile to the backwards extrapolated profile at $z\sim$1. See the details in the text (Section \ref{Sec:SigmaProfs}).}
      \label{Fig:SigmaProfs}
\end{figure}

In Fig. \ref{Fig:SigmaProfs}, upper panel, we show the radial profiles of SFR surface density for the three redshift samples. We can see that the surface density of SFR has decreased since $z\sim$1, and that these decrease has been somewhat greater in the inner parts of the galaxies (within the inner $\sim$2.5 kpc). And it is remarkable that the profiles are roughly parallel beyond \R$\sim$4 kpc. These trends would cause the effective radii of \NUV to increase with time, as we had reported. The median effective radii, in kpc, of the dust-corrected \NUVrest emission can be seen as vertical lines in the same colours as the corresponding curves of SFR density. Also note that the innermost part of the local profile has an upward bend which is probably due to the contribution from not so young stars (with ages around $\sim10^9$ yr), and not an indication of recent SF. This central ``peak'' in \NUVrest (from which the SFR profile is derived) is responsible for the increase in the value of $\Delta\mu$ since $z\sim$1 that we have also reported.

The blue dotted curve in the central panel gives the amount of stellar mass (density) produced at each radius since $z\sim$1. Also in the same panel, and shown as a continuous red curve is the present stellar mass density, derived from the stacked \Brest profile ($z\sim$0), as explained above. If the accumulated stellar mass is subtracted from this profile we obtain an approximation to the mass profile at $z\sim$1, shown as the dashed red line. The dashed and continuous vertical lines show the effective radii of the \Brest profiles at $z\sim$0 and $z\sim$1. The backwards extrapolated mass profile also shows hints of a pseudo-bulge, as the local one does, and so this suggests two scenarios. First, it could be that this feature was already largely in place at $z\sim$1, but was outshone by the emission from young stars. This emission from young stars contributes significantly to the emission in \Brest at $z\sim$1, and so, this pseudo-bulge structure would not be recognizable in the $z\sim$1 profile in this band. This, however, would need confirmation from high-resolution observations in redder wavelengths at that redshift. Alternatively, it could also be that this pseudo-bulge structure we see in the local \Brest profile were the by-product of some mechanism of stellar mass redistribution within the galaxies in this time interval.

It is also noteworthy that the maximum variation in stellar mass is not at the centre. This is easier to appreciate in the lower panel, where we show the relative increase in mass since $z\sim$1, according to this test. The mean increase between \R$=0$ and \R$=$8 kpc is $\sim$120\%. But the maximum in the relative increase reaches $\sim$300\% at \R$\sim$1.5-2 kpc, possibly indicating a progressive dearth of the SF activity in the central part of the galaxies. It could be that a higher efficiency of the SF in the central parts (perhaps due to a higher gas density), had exhausted the gas reservoir more quickly in that area. This would be at a greater rate than that at larger radii, where a decline of SF with time also seems to happen, though not so marked. This indicates that either the parameters that regulate the efficiency of SF vary with radius, or that the gas replenishment is more effective in the outer parts of galaxies. An alternative, though admittedly less probable explanation, is the nuclear-activity-induced quenching of the SF.

Interesting as this may be, we need to present the caveats affecting this result. First and foremost, it is based on rather inaccurate dust extinction corrections which neglect possible evolution in the radial distribution of dust. Second, the local \Brest profile is not the most reliable proxy for the stellar mass density. Similarly, the \NUVrest profiles are also ``contaminated'' to some extent by the contribution from not-so-young stars. We must also note that the stacked profiles are for collections of galaxies which have similar stellar masses, luminosities and morphological types, but within a range, and they are not selected to represent the possible evolutionary path of the population of disc galaxies (i.e. trace their mass evolution). And last, but not least, we have ignored the effect of stellar mass redistribution within galaxies, and amongst galaxies (through mergers). These mechanisms most probably would make the true surface mass density profile at $z\sim$1 differ from the given approximation, which is based solely on the SFR curves. So, the main interest of this exercise is to test whether the SF activity in the inner parts of the discs could be, alone, responsible for the pseudo-bulge-like structure seen in the local \Brest median profile. And the answer seems to be ``no''. This test suggests that some other mechanisms must be collaborating in this phenomenon.

We have also checked that the profiles shown in Fig. \ref{Fig:SigmaProfs} are basically unaffected when equal resolution images are used. Besides, the fiducial radii in \Brest and \NUVrest are beyond the shown radial ranges, and the limit is imposed by the \NUVrest profile at $z\sim$1 (marked as a dash-dotted black vertical line in the panels in Fig. \ref{Fig:SigmaProfs}).

\section{Discussion}\label{Sec:discussion}

Barden et al. (\cite{Barden05}) found that the \B-band effective radii of stellar discs, at a given stellar mass (\mstar$\gtrsim10^{10}$\msun), do not evolve between z$\sim$1 and z$\sim$0 (although this result seems to be mass dependent, see Trujillo et al. \cite{Trujillo07}). Given that galaxies have increased their stellar mass content in this period (by $\sim$30\% according to Rudnick et al. \cite{Rudnick03}), these findings imply that individual disc galaxies have grown in size (\Reff), so as to keep their surface mass densities roughly constant. The absence of evolution in stellar mass density is at odds with some previous theoretical expectations. Mo et al. \cite{Mo98} proposed a model in which the size of a disc of fixed mass is to scale with time in proportion to the virial radius of the dark matter halo. In this model, this is the prescription for the evolution in effective radius, \Reff, at a fixed mass:

\begin{equation}
R_{eff}(z) = R_{eff}(0) \left[\frac{H(z)}{H(0)}\right]^{-2/3} \label{Eq:1}
\end{equation}

where \Reff(0) is the effective radius at $z=$0 and $H(z$) is the Hubble parameter as a function of redshift. According to this, stellar discs at $z\sim$1 would be a factor 2 denser than they are nowadays, in contradiction with the cited observational evidence.

When the expected disc size-stellar mass distribution is computed, including the evolution in the internal structure of dark matter halos and the adiabatic contraction of the dark matter by the self-gravity of the collapsing baryons, the predicted evolution in the mean size at fixed stellar mass since $z\sim$1 is about 15-20\% (Somerville et al. \cite{Somerville08}), in reasonable agreement with the aforementioned observational constraints.

Our results on the evolution of the effective radii of galaxies in the \B-band agree fairly well with the results by Barden et al. \cite{Barden05}, and so, they are also consistent with the model by Somerville et al. \cite{Somerville08}. But the effective radius in \NUV seems to grow somewhat more than in \B-band in this time interval (relative to \Brest, by $\sim$10\%). We have seen that at least part of this difference is due to a change in the shape of the profiles in both bands. It seems that the profiles have changed from ``flat'' in the central parts of the galaxies at z$\sim$1, to more ``cuspy'' at z$\sim$0, an evolution which is stronger in \B-band. We attribute this change in shape to the interplay of two processes: a) a progressive piling up of stellar mass in the central parts, i.e. the signature of growing bulges and pseudo-bulges in discs, and b) to the ageing of stellar discs, due to a decreased star-forming activity per unit area, leading to a progressive dimming of the discs in the \NUV and \B bands. From our results the relative contributions of these mechanisms to the final observed evolution of the shape of the profiles cannot really be disentangled. The bulges/pseudo-bulges, which are easily seen in the z$\sim$0 profiles, could already have been in place at z$\sim$1, and not have evolved significantly in this interval, but could be outshone by bright, actively star-forming discs at z$\sim$1. The fact that central brightness in the mean \NUV and \B profiles has not evolved significantly over this range of redshifts, as seen in Fig. \ref{Fig:MedProfs}, while the brightness of the discs has significantly decreased, argues in favor of this interpretation. Furthermore, when we try to reproduce the mass density profile at $z\sim$1 from the present curve ($\sim$\Brest profile), and the time integral of the SFR profiles ($\sim$\NUVrest profiles), though applying less than fiducial dust corrections, leads to a past mass profile where this pseudo-bulge feature is already present. On the other hand, the strong evolution in colour in the central parts (\R$\sim$0), being much redder nowadays (even after dust-corrections), favours the interpretation that there might have been substantial secular evolution of bulges/pseudo bulges in the last 8 Gyr.

The concentration (\C) of the \NUV profiles does not seem to vary significantly in this time interval (0$\lesssim z \lesssim$ 1), and the values of this parameter indicate that globally the profiles are close to exponential, but somewhat ``flatter''. The fraction of flux enclosed within a fraction of the Petrosian radius (measured in \B) seems also not to evolve. But the fraction of flux within a fixed radius, given in kpc (e.g. \R$<$2.5 kpc) shows significant evolution, galaxies at z$\sim$1 having a larger fraction of \NUV flux within that radius than their peers at z$\sim$0. This could indicate that the star-forming area in them was more compact in the past, but the distribution in relation to that at longer wavelengths (\B) does not change greatly. An evolution of this kind suggests that the stellar discs are growing from the inside outwards.

The difference in central brightness between the centre and \Reff has also evolved, the objects at z$\sim$0 having their central parts brighter in \NUV and especially in \B by a larger factor than the discs, in comparison to galaxies at z$\sim$1. This trend in the \B band seems to be largely due to current bulges/pseudo-bulges being more prominent with respect to the discs than in the past. It remains to be answered whether the similar evolution observed in \NUV is due to a parallel boost of nuclear star formation in recent times, or to the contribution of numerous middle aged stars in those central areas, or to some intermediate situation. The median colour profiles for galaxies at $z\sim$0 (Fig. \ref{Fig:MedProfs}) show significant reddening at small values of \R, but this, of course, does not suffice to state that all \NUV emission in the central parts of galaxies is due to intermediate age populations. A detailed study of the composition of stellar populations, via the analysis of SEDs, in the central areas of discs would be required to clarify this issue.

In this work we have also explored the relation between the radial distribution of \NUV flux (a proxy for SF), and the phenomenon of stellar disc truncations. The mechanism behind these truncations is not fully understood yet, though several models have been proposed. Most of them can be grouped in two groups: a) models related to angular momentum conservation in the proto-galactic cloud (van der Kruit \cite{Kruit87}) or angular momentum cut-off in cooling gas (van den Bosch \cite{Bosch01}); and b) models which appeal to thresholds in star formation (Kennicutt \cite{Kennicutt89}) or star-forming properties of distinct ISM phases (Elmegreen \& Parravano \cite{Elmegreen94}; Schaye \cite{Schaye04}; Elmegreen \& Hunter \cite{Elmegreen06}). More recently, N-body simulations in which both star formation processes and secular dynamics are included, have been devised in an attempt to better fit observational results (e.g. Debattista et al. 2006). In a further elaboration of these ideas, in \Roskar et al. (\cite{Roskar08a}) a model was presented in which the breaks are the result of the interplay between a radial star formation cut-off and a redistribution of stellar mass by secular processes  (see also \Roskar et al. \cite{Roskar08b}). In this model, stars are created inside the break (threshold) radius and then move outwards due to angular momentum exchange in spiral arms. Other recent models explain the occurrence of breaks in the mass density profiles as the result of other secular processes. Foyle et al. (\cite{Foyle08}) presented a model in which angular momentum redistribution inside the break was the cause for it to form. In a different approach, in Bournaud et al. (\cite{Bournaud07}) it was shown that the disruption of large stellar ``clumps'' (with stellar masses in the range $10^8-10^9$ \msun) in primordial discs (observed as ``clump cluster'' galaxies at high redshift) could give way to single or double exponential discs (i.e. discs with a break in their profile).

We have presented results on the fraction of \NUV flux contained within the break radius, \Rbr, for galaxies in the Local and High-z samples. This fraction is relatively high, but there is significant difference between the local and intermediate-redshift results, the median value at $z\sim$0 being $\sim$90\%, while at z$\sim$1 it is $\gtrsim$60\% (in some cases it is as low as $\sim$20\%). The values at intermediate redshifts seem perhaps too low, assuming that this \NUV emission traces the star-forming activity, to fit into  models in which the suppression of star formation outside the break radius is part of the mechanism behind the formation of a truncated disc. Being strict, one would expect a distribution similar to that of the Local sample. But other factors must be taken into account before extracting hasty conclusions from this result. First, the \NUV is not a perfect tracer of recent SF, and so, even if the break could mark the extension of the area where SF takes place, stars outside this break, of intermediate age, could also contribute significantly to the total emission of the galaxy in that band. For example, in the already cited model by \Roskar et al. (\cite{Roskar08a}, \cite{Roskar08b}), it is proposed that a mechanism of radial migration of stars due to secular processes may drive stars away from the zones where they form, supposedly inside the ``break''. Furthermore, from visual inspection we have seen that in some cases there are few, but powerfully emitting regions outside \Rbr which account for reduced fractions of \NUV flux enclosed within \R$<$\Rbr. It could also be possible that there was some contribution to the \NUV emission outside the break due to mergers. In any case, this result does not suffice to decide what is the mechanism behind disc truncations, whether models which rely on angular momentum conservation / cut-off in the proto-galactic cloud, or those which postulate that the break is the signature of a threshold for SF in the disc. At most, it can be said that the fraction of SF outside the break at z$\sim$1 seems to be rather high, even given the limitations of \NUV flux as a tracer for the star-forming activity.

We have seen that the median \B and \NUV profiles have decreased their brightness at \R$=$\Reff by $\sim$60\% and $\sim$80\% respectively between z$\sim$1 and z$\sim$0. Although these variations have been deduced assuming simplistic k-corrections, this result is consistent with a significant decline in the SFR per unit area in stellar discs in the last $\sim$8 Gyr. The decline in \NUV, and in \B, has been similar for galaxies within different ranges of mass (\mstar$<10^{10}$\msun and \mstar$>10^{10}$\msun).

Summarizing, stellar discs, as seen in the \NUV, seem to have undergone significant evolution on several fronts, and this could be succinctly described as follows: a) the surface brightness of the discs has decreased to roughly a 1/5 of the median values at z$\sim$1; b) \NUV discs have grown, at a fixed stellar mass, only moderately (by $\lesssim$20\%); c) the ratio of the effective radius in \NUV to that in \B has increased by $\sim$10\%; d) the \NUV profiles are rather flat in the inner regions, showing signs of ``truncation'' in many cases, as made evident in averaged profiles for the whole samples at different redshifts; e) a significant excess in emission (\NUV and \B) at the centre of the discs has developed over time (between z$\sim$1 and z$\sim$0). What does this information tell us about the distribution of SF? If the relation between \NUV emission and SFR were strictly proportional, it could be said: a) the star-forming activity decays with radius according to a double exponential law, with a larger scale radius (i.e. a shallower slope of the profile) in the inner parts of the discs; b) the SFR per unit area in discs has decreased significantly in the redshift interval 0$<z<$1; in this same period c) the bulk of this star-forming activity has shifted to larger radii within the discs; and d) the SFR in the bulges/nuclii has grown in relevance over that in the discs. Nonetheless, \NUV flux is not a perfect tracer of this activity, even when it provides a fair approximation. Thus, below we discuss how the two factors, age of the stellar populations and dust in the ISM and IGM, that contribute the most to deviations from a simple relation of proportionality between the \NUV luminosity and the SFR affect these conclusions.

The \NUV is, in general, dominated by the contribution of stars with ages $\lesssim$100 Myr. The \NUV images, which show patchy morphologies, give general support to this. But the median \NUV profiles at z$\sim$0 are close to exponential, and show substantial differences from the \B profile (a tracer of the distribution of significantly older stars, with ages $\gtrsim$1 Gyr), but not so ``erratic'' as typical profiles in H$\alpha$, a better tracer of on-going SF. In particular, the central excesses of \NUV emission that we find at z$\sim$0, as discussed above, are suspected to be, at least in part, ``contaminated'' by the contribution of older stars (the signatures of bulges and pseudo-bulges).

The dust is a more involved issue, that requires special attention, as its absorption is more important at shorter wavelengths (see e.g. Boissier et al. \cite{Boissier04}; Popescu et al. \cite{Popescu00}; Tuffs et al. \cite{Tuffs04}; M\"ollenhoff et al. \cite{Mollenhoff06}). We have used the prescription for dust attenuation as a function of radius in local disc galaxies by Boissier et al. \cite{Boissier04} to correct the \Brest and \NUVrest profiles for dust absorption, as is shown in Sec. \ref{Sec:Dust}. This dust absorption is maximum in the innermost parts of the profiles, and is negligible beyond $\sim$2.5-3 \Reff, and so the underlying \Brest and \NUVrest profiles have presumably steeper inner profiles than observed. Furthermore, the absorption is higher in \NUV than in \B. We have explored how this dust-corrections affect the measured parameters, and the stacked intensity and colour profiles. This test indicates several things. Part of the ``truncation'' effect in median \NUV profiles at \R$\sim$\Reff could be due to dust. The valley-like appearance of the colour profiles (\NUV-\B) seems to be intrinsic to the distribution of stellar populations at $z\sim$0, but at intermediate redshifts it dissipates, with the significant reddening in the central parts seemingly being due to the effect dust. Though the dust has a significant effect on most parameters, such as \C, \Reff, fraction of \NUV flux within a fixed radius, it does not, on its own, seem to mimic evolution for some of the parameters (e.g. \Reff(\NUV)), and for those which does, the effect is not great (e.g. \C, fraction of flux within the central 2.5 kpc, $\Delta\mu$), and largely compatible with the trends found before the corrections. Succinctly, it seems that the decrease in SFR surface density is more pronounced within the central 2.5 kpc, relative to outer parts of the disc. This would explain the growth in \Reff(\NUVrest) since $z\sim$1 at a fixed mass, a phenomenon which is persistent after the dust corrections. Furthermore, and if we take these dust corrections for certain, it seems that the central excess of the \B profiles, which we attribute to pseudo-bulges, could be largely in place at $z\sim$1, but it would be outshone by the more widely radially distributed emission from young stars at that time. This last interpretation neglects the effect of stellar mass redistribution within (dynamical evolution) and amongst galaxies (mergers) in the meantime, and so it could be erroneous. 

How can we understand the previous results in terms of the growth of stellar discs? We see that the (\NUVrest-derived) SFR profiles are roughly exponential at $z\sim$1, and since then the SF activity has decreased at all radii, but more so in the inner parts of the discs, leading to a downbending of the SFR profile with the knee at \R$\sim$3 kpc. In some sense, this is compatible with an inside-out scenario for the formation of the disc, but not exactly so. Up to a radius of \R$\sim$8 kpc it seems that there was more star-forming activity in the past, at all radii. But in the inner parts this activity has declined more rapidly since then. The net effect is that the stellar populations must have aged more in the central parts, and so they were formed before the outermost parts, which were being formed at $z\sim$1, a process that is still going on nowadays. In fact, we do not see that the SF has actually migrated to larger radii, within that radial range (imposed by the lower signal-to-noise ratio at large radii at $z\sim$1). Instead, it seems that it is being been shut-off inside-out. Deeper images at all redshifts could provide valuable information with regard to what happens at larger radii.

To end the discussion we comment on some observational biases, namely the change in spatial resolution with redshift, and the differences in the relative depths of the images. We have taken a careful approach to the resolution problem, and different tests (limiting distances to avoid the bias in the selection of the objects in the local sample, and comparing results obtained with images of equal, simulated, resolution) have shown that these results are robust in this regard. More delicate is the issue of the differences in depth, which are substantial, as commented in section \ref{Sec:PhotoDepth}. As we said there, the \GALEX-\NUV images are $\sim$10 times deeper than the \SDSS, and these are $\lesssim$20 times deeper than \GOODS. So, the portrait here presented on the \NUV profiles of disc galaxies in the past must be incomplete, as we do not reach the same depths as with the local data. But some of the results in Local and High-z sample are not consistent with absence of evolutionary effects (even when considering these differences in depth), such as the change in the profiles, with the development of substantial central emission at z$\sim$0, or the significant decrease in surface brightness at \Reff to a 20\% of the value at z$\sim$1. If there were significant, undetected, low surface brightness in the GOODS images this would make the intensity level at \Reff lower at z$\sim$1, and so the difference would be less significant, but that does not convey the fact that the z$\sim$1 profiles are brighter on average than those of galaxies at z$\sim$0, as the differences in the \Reff are not so big ($\R_{eff}^B(z\sim1) \sim \R_{eff}^B(z\sim0$) as to render these differences a mere selection effect. Still, the discs could have had larger \Reff's in the past, as there could be a low surface brightness contribution that may be unnoticed at intermediate redshifts, but can be detected in local objects. For these reasons it would be desirable to have deeper rest-frame \NUV and \B images at z$\sim$1.

\section{Summary \& Conclusions}\label{Sec:conclusions}

We have studied the radial distribution of flux in the rest-frame \NUV and \B bands for a sample of 270 disc galaxies, in the range 0$\lesssim$$z$$\lesssim$1 (i.e. exploring the last $\sim$8 Gyr of cosmic age). The analysis is focused on the evolution in the spatial extension (parameterized by effective radii, \Reff) and concentration (using appropriate parameters) of the rest-frame \NUV flux, which may be taken as an indicator of recent star forming activity. In this way, and to our knowledge, this is the first systematic study of the radial distribution of star formation in disc galaxies extending through a cosmologically significant range of redshifts. Our aim is to provide further constraints on the mechanisms which shape the structure of stellar discs with epoch. These are our conclusions:

   \begin{enumerate}
      \item At a fixed luminosity (\mb=-21 mag) galaxies have larger values of \Reff in the rest-frame \NUV band, \NUVrest, by a factor 1.5$\pm$0.1 at $z=$1 than at $z=$0 (a factor 1.4$\pm$0.1 in the rest-frame \B band, \Brest). 
      
      \item At a given stellar mass (\mstar$=10^{10}$ \msun), the \Reff of the galaxies in \NUVrest has slightly increased, by a factor 1.18$\pm$0.06. The \Reff in \B band, at a given stellar mass, has increased by a smaller factor, 1.10$\pm$0.05 in the same range of redshifts. This is in fair agreement with other observations, which report no evolution in that parameter at longer wavelengths in this range of redshifts (e.g. Barden et al. \cite{Barden05}).
      \item We find no significant evolution in the concentration of \NUV emission in this range of redshifts (using the \C parameter, Kent \cite{Kent85}), nor in the fraction of flux contained in the central parts of the galaxies, when this central region is defined in relation to the Petrosian radius (in \B). Nonetheless, galaxies at lower redshift show a larger difference in surface brightness, $\mu$, from centre to \Reff (i.e. an increase in this difference by $\sim$0.8 mag in \NUV and by $\sim$1.6 mag in \B between z$\sim$0 and z$\sim$1).
      
      \item Galaxies at z$\sim$1 contain a larger fraction of \NUV flux within R$<$2.5 kpc than those at z$\sim$0 ($\sim$30\% vs. $\gtrsim$15\%).
      
      \item The averaged surface brightness profiles in rest-frame \NUV and \B bands for the galaxies under study show: a) the \NUV and \B profiles at z$\sim$1 are quite similar, being somewhat ``flattened'' in the central parts, showing no evidence of a distinct bulge emission. At z$\sim$0 the profiles have a $\sim$20\% and $\sim$40\% of the brightness at z$\sim$1 in \NUV and \B, respectively (measured at the \Brest effective radii in each case). At lower redshifts the profiles show clear signs of bulges or pseudo bulges, more prominently in the \B band, which could be understood as a by-product of progressive accumulation of stellar mass in the inner parts of the galaxies. However, it could be that these bulges were partially outshone at $z\sim$1 by discs with their intense star forming activity. Analysis of dust-corrected \NUV and \B profiles seems to favour the second interpretation, but this is still inconclusive.
      
      \item At both redshifts, the median \NUV profiles are similar to those of stellar discs which show truncations, with the corresponding break roughly at the position of the (\B-band) effective radius. The median colour profiles (\NUV-\B) show an evident ``blue valley'' shape, with a minimum in colour found around \Reff(\Brest), a feature similar to that found in averaged colour profiles of truncated discs in galaxies at $z\sim$1 (Azzollini et al. \cite{ATB08a}) and in local galaxies (Bakos et al. \cite{Bakos08}).
      \item The fraction of \NUV flux enclosed by the ``break'' radius, \Rbr, in truncated galaxies is $\sim$90\% at z$\sim$0 and $\gtrsim$60\% (at 0.5$<z<$1.1), with significant dispersion. This finding seems to favour models for stellar disc truncations which postulate that these are due to a cut-off in the star-forming activity at the ``break'' radius (e.g. Elmegreen \& Parravano \cite{Elmegreen94}; Schaye \cite{Schaye04}; Elmegreen \& Hunter \cite{Elmegreen06}, Debattista et al. \cite{Debattista06}, \Roskar et al. \cite{Roskar08a}), though the lower fraction at higher redshift indicates that this interpretation may not be complete.
\end{enumerate}

Finally, it would be desirable to perform an analysis of stellar populations (e.g. from SED modelling) at different galacto-centric radii which would complement the study of the distribution of \NUV flux, in order to weight the effects of varying dust absorption, and the ageing of underlying stellar populations, and disentangle them from genuine evolutionary changes in the distribution of star formation across stellar discs.

\begin{acknowledgements}
      We are grateful to Dr. Marco Barden for kindly providing us with the GEMS morphological analysis catalogue. We acknowledge the COMBO-17 collaboration (especially Dr. Christian Wolf) for the public provision of a unique database on which this study is partially based. We also want to thank Dr. Carlos J. Delgado for sharing with us his software for easily downloading \SDSS images. Part of our gratitude also goes to the anonymous referee for valuable comments and suggestions along the revision process. This work is based on observations made with several facilities. The NASA / ESA Hubble Space Telescope is operated by the Association of Universities for Research in Astronomy, Inc., under NASA contract NAS5-26555. \GALEX (Galaxy Evolution Explorer) is a NASA Small Explorer, developed in cooperation with the Centre National d\'Etudes Spatiales of France and the Korean Ministry of Science and Technology. The SDSS is managed by the Astrophysical Research Consortium for the Participating Institutions.

Partial support for this work has been provided by projects AYA2004-08251-C02-01 and AYA2007-67625-C02-01 of the Spanish Ministry of Education and Science and P3/86 of the Instituto de Astrof\'isica de Canarias.

\end{acknowledgements}

\appendix{APPENDIX\\\newline
DEGRADATION OF IMAGES TO A COMMON SPATIAL RESOLUTION}
\newline\\
Here we explain the process by which we degrade the images of galaxies in different redshift bins to a common resolution, in order to avoid potential bias due to resolution differences. These transformations are always degradations by definition, in the sense that we apply only transformations in which the final images have coarser sampling (larger angular scale) and a larger value of the \FWHM than the initial images. Otherwise, we would be merely interpolating the images, which is not useful in this context, or would have to deconvolve the images from their PSFs, a process which usually generates artefacts in the images.

First, we interpolate the pixel values of the \GALEX-\NUV images to change their angular scale $\gamma$ from 1.5''/pixel to the 0.396''/pixel value of the \SDSS images. We checked that this does not produce significant alterations in the extracted radial distributions of \NUV flux. We then selected a projected dimension of the \FWHM in kpc, \FWHMkpc, which is larger than the projected size of the FWHM of the \GOODS-\bb images at $z$=1.1, and also larger than the projected size of the relatively large \GALEX-\NUV \FWHM (5''), at the limiting distance of the Local sample, which is 60 Mpc. Once the \FWHMkpc is fixed ($=$1.5 kpc), all images are convolved with a Gaussian kernel, with a certain matching \FWHM ($F_{m}$), so as to produce an image whose final \FWHM ($F_f$), subtends the desired \FWHMkpc at the distance of the object. To compute this $F_m$ we make use of the expression:

\begin{equation}
F_{m} = \sqrt{F_{f}^{2} - F_{o}^{2}} \label{Eq:A1}
\end{equation}

where $F_o$ is the original value of the \FWHM. This expression is valid for Gaussian PSFs, which is our working assumption. Using expression (\ref{Eq:A1}), and convolving the images according to the described scheme, we produce images of objects in the Local and High-z sample which share a common physical size of the PSF, \FWHMkpc. 

But the images still have different physical scales, $\Gamma$; i.e. their pixels still project different lengths, in kpc, at the distances of the targets. We select a certain value of $\Gamma$, that which corresponds to the 1.5'' pixels of the \GALEX images at a distance of 60 Mpc ($\Gamma$=0.5 kpc), which is also larger than the values of this parameter for the galaxies in the High-z sample ($\Gamma\lesssim$0.25 kpc in that sample). Then, all the selected images are re-sampled to match that value of $\Gamma$ by performing a suitable ``binning'' on them (i.e. averaging a certain number of pixels, though this number may be fractional, which is computationally feasible by means of interpolation).

These procedures of convolution and re-sampling were performed using several IRAF tasks and a script coded in Python language.

We have seen that there are also substantial differences with regard to photometric depth amongst the images from the surveys used. There is the possibility to adjust the intensities (scale them) and noise level (adding Gaussian nose) of the images of the galaxies in the Local sample to match the significance level of the images of the High-z sample. But this strategy has also its inconveniences, firstly a) the resulting images would not unambiguously represent the appearance of local galaxies observed at High-z, given the uncertainties in the intensity scaling factor, through the corresponding k-correction (which is reasonable to assume as spatially variant); and b) even more important, the decrease in surface brightness between z$\sim$1 and z$\sim$0 has been so significant, that local objects would be, in many cases, practically undetected in the GOODS-ACS images. This is why we have not attempted that type of simulation in this work.

\end{document}